\documentclass[pre,twocolumn,amsmath,amssymb,floats,superscriptaddress,nofootinbib,10pt]{revtex4-1}

\pdfoutput=1

\AtBeginDocument{%
    \newwrite\bibnotes
    \def\bibnotesext{Notes.bib}
    \immediate\openout\bibnotes=\jobname\bibnotesext
    \immediate\write\bibnotes{@CONTROL{REVTEX41Control}}
    \immediate\write\bibnotes{@CONTROL{%
    apsrev41Control,author="08",editor="1",pages="1",title="0",year="1"}}
     \if@filesw
     \immediate\write\@auxout{\string\citation{apsrev41Control}}%
    \fi
}

\usepackage{graphicx}
\usepackage{dcolumn}
\usepackage{bm}
\usepackage{revsymb}
\usepackage[usenames]{color}
\usepackage{subfigure}
\usepackage{color}
\usepackage{soul}

\usepackage{physics}
\usepackage{amsmath}
\usepackage[usenames]{color}
\usepackage{amsfonts}
\usepackage{graphicx}
\usepackage{dcolumn}
\usepackage{bm}
\usepackage{revsymb}
\usepackage[usenames]{color}
\usepackage{subfigure}
\usepackage{color}
\usepackage{physics}
\usepackage{amsmath}
\usepackage{amssymb,bbm}

\usepackage{xcolor}

\newcommand{\cD}{{\cal D}}

\newcommand{\cL}{{\cal L}}

\newcommand{\bea}{\begin{eqnarray}}
\newcommand{\eea}{\end{eqnarray}}
\newcommand{\rmd} {{\rm d}}
\newcommand{\rme} {{\rm e}}
\newcommand{\rmi} {{\rm i}}

\definecolor{nblue}{RGB}{28,130,185}

\definecolor{cgreen}{RGB}{76,153,0}

\definecolor{myorange}{RGB}{245,156,74}

\usepackage{hyperref}
\hypersetup{
  colorlinks=true,
  citecolor=magenta,
  urlcolor=-myorange
}

\newcommand{\bmP}{\tens{\Phi}}
\newcommand{\bmf}{\tens{f}}

\usepackage{xcolor}
\definecolor{ogreen} {RGB}{71,191,145}

\newcommand{\com}[1]{\textcolor{black}{#1}}

\usepackage{hyperref}
\hypersetup{
  colorlinks=true,
  citecolor=magenta,
  urlcolor=-myorange
}

\newcommand{\DDt}[1]{%
\frac{{\rm D}#1}{{\rm D}t}%
}

\newcommand{\vect}[1]{%
\underline{#1}%
}

\newcommand{\tens}[1]{%
\underline{\underline{#1}}%
}

\newcommand{\quotes}[1]{``#1''}

\begin{document}

\title{%
Probe particles in odd active viscoelastic fluids: how activity and dissipation determine linear stability
}

\makeatletter
\newcommand\thankssymb[1]{\textsuperscript{a}}
\makeatother

\author{Charlie Duclut\thankssymb{1}}
\email{charlie.duclut@curie.fr}
\affiliation{Laboratoire Physico-Chimie Curie, CNRS UMR 168, Institut Curie, Université PSL, Sorbonne Université, 75005, Paris, France
}
\affiliation{Universit\'e Paris Cit\'e, Laboratoire Mati\`ere et Syst\`emes Complexes (MSC), UMR 7057 CNRS,F-75205 Paris,  France}
\affiliation{Max Planck Institute for the Physics of Complex Systems, 01187 Dresden, Germany}

\author{Stefano Bo\thankssymb{1}}
\email{stefano.bo@kcl.ac.uk}
\affiliation{Max Planck Institute for the Physics of Complex Systems, 01187 Dresden, Germany}
\affiliation{Department of Physics, King's College
London, London WC2R 2LS, U.K.}
\author{Ruben Lier\thankssymb{1}}
\email{r.lier@uva.nl}
\affiliation{Institute for Theoretical Physics, University of Amsterdam, 1090 GL Amsterdam, The Netherlands}
\affiliation{Dutch Institute for Emergent Phenomena (DIEP), University of Amsterdam, 1090 GL Amsterdam, The Netherlands}

\author{Jay Armas}
\email{j.armas@uva.nl}
\affiliation{Institute for Theoretical Physics, University of Amsterdam, 1090 GL Amsterdam, The Netherlands}
\affiliation{Dutch Institute for Emergent Phenomena (DIEP), University of Amsterdam, 1090 GL Amsterdam, The Netherlands}

\author{Piotr Sur\'{o}wka}
\email{piotr.surowka@pwr.edu.pl}
\affiliation{Department of Theoretical Physics, Wroc\l{}aw  University  of  Science  and  Technology,  50-370  Wroc\l{}aw,  Poland}

\author{Frank J\"{u}licher}
\email{julicher@pks.mpg.de}
\affiliation{Max Planck Institute for the Physics of Complex Systems, 01187 Dresden, Germany}
\affiliation{Center for Systems Biology Dresden, Pfotenhauerstrasse 108, 01307 Dresden, Germany}
\affiliation{Cluster of Excellence Physics of Life, TU Dresden, 01062 Dresden, Germany}

\renewcommand{\thefootnote}{\alph{footnote}}
\footnotetext[1]{These authors contributed equally.}
\renewcommand{\thefootnote}{\arabic{footnote}}

\begin{abstract}
Odd viscoelastic materials are constrained by fewer symmetries than their even counterparts. The breaking of these symmetries allows these materials to exhibit different features, which have attracted considerable attention in recent years.  
Immersing a bead in such complex fluids allows for probing their physical properties, highlighting signatures of their oddity and exploring the consequences of these broken symmetries.
We present the conditions under which the activity of an odd viscoelastic fluid can give rise to linear instabilities in the motion of the probe particle and unveil how the features of the probe particle dynamics depend on the oddity and activity of the viscoelastic medium in which it is immersed.
\end{abstract}

\maketitle

\tableofcontents

\section*{Introduction}

Chiral two-dimensional materials break parity symmetry and display transport phenomena characterized by tensors that are odd under index exchange. Such odd systems attract considerable interest from a broad range of communities. 
These include soft active matter, statistical physics and biological physics~\cite{epstein2020timereversal,hargus2020time,lier2022passive,han2021fluctuating,hargus2021odd,soni2019odd,reichhardt2019active,markovich2021odd,reichhardt2022active,furthauer2012active,tan2022odd,banerjee2021active,surowka2022odd,kalz2023oscillatory,PhysRevLett.129.090601}, fluid dynamics~\cite{banerjee2017odd,abanov2018odd,ganeshan2017odd,khain2022stokes,hosaka2023hydrodynamics,hosaka2021hydrodynamic,hosaka2023pair}, complex materials~\cite{scheibner2020odd,braverman2021topological,floyd2022signatures,fruchart2022odda,liao2020rectification}, electron fluids \cite{pellegrino2017nonlocal,narozhny2019magnetohydrodynamics,berdyugin2019measuring}, and topological waves \cite{fossati2022odd,green2020topological,souslov2019topological,tauber2019bulkinterface}.
Recent experiments have begun measuring odd systems ~\cite{berdyugin2019measuring,soni2019odd,tan2022odd,bililign2022motile}. One of these odd phenomena is odd elasticity \cite{scheibner2020odd,tan2022odd}. Odd elastic solids can perform net work under quasi-static cycles, implying that they are active materials, as such phenomena can only occur in the presence of a constant energy injection at the microscopic level. Such energy injection, if not properly dissipated, can give rise to linear instabilities \cite{floyd2022signatures,scheibner2020odd}. Complex materials that exhibit elastic properties at short times and fluid-like ones at long times are called viscoelastic. Odd viscoelastic systems were first discussed in \cite{banerjee2021active} in the context of active matter and recently shown to exist also in a passive context, provided that certain thermodynamic constraints are obeyed \cite{lier2022passive}. 

A powerful method of probing the physical properties of a fluid is to observe the motion of a probe particle immersed in it \cite{mason1995optical,mackintosh1999microrheology,squires2010fluid}. This is even more important for active odd viscoelastic materials.
First, because the response contains signatures of the odd nature of the material, such as odd lift forces \com{orthogonal to the probe's motion}~\cite{lier2023lift,hosaka2021nonreciprocal}.
Second, because the viscoelastic nature of the system makes the response in time to instantaneous external perturbations non-trivial~\cite{Gardel2005}.
Third, because activity can give rise to sustained motion of the probe particle and, potentially, instability in the response to perturbations (see, {\it e.g.}, Ref.~\cite{di2010bacterial}).

\begin{figure}[t]
    \centering
    \includegraphics[width=0.9\linewidth]{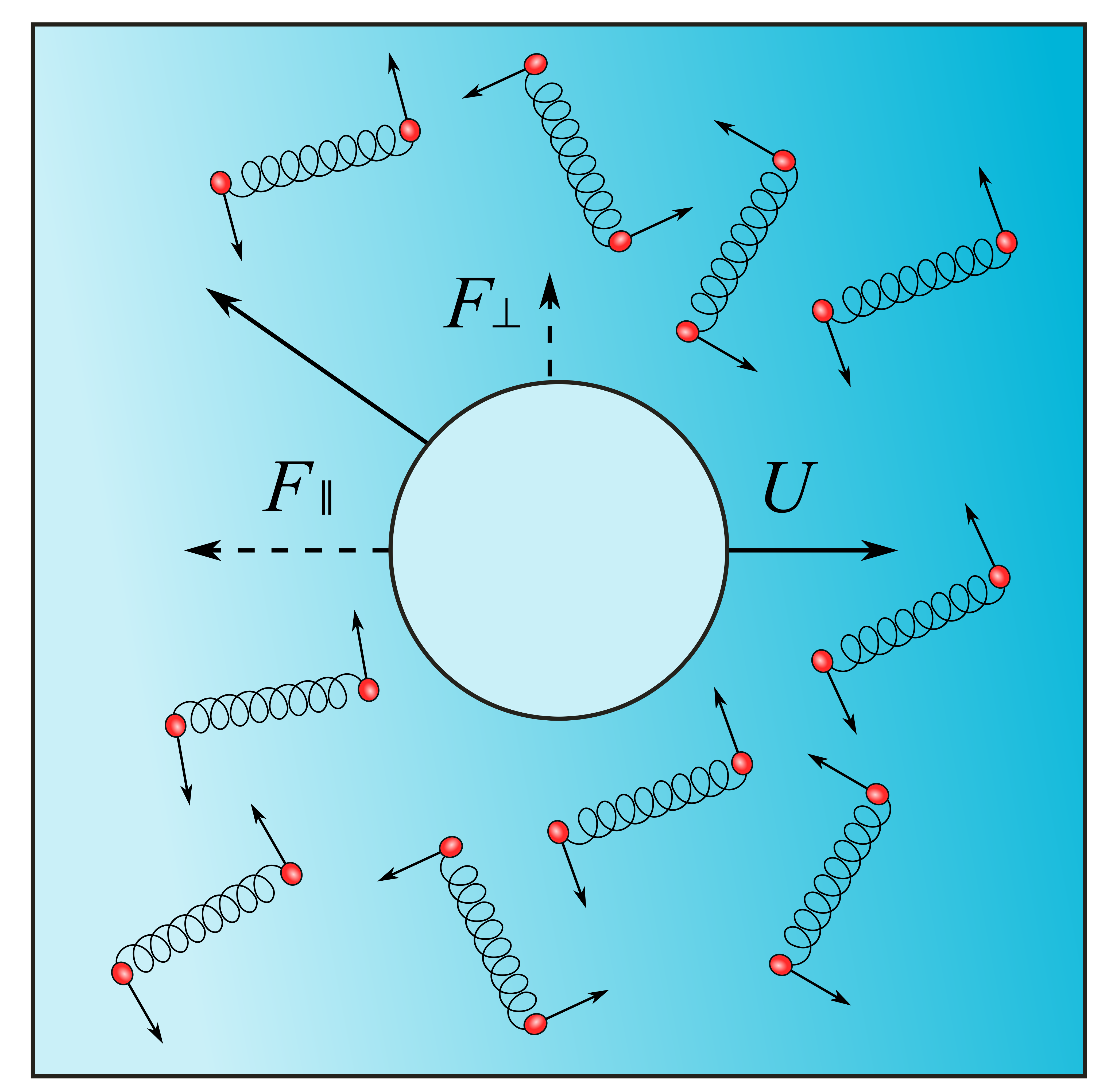}
    \caption{Schematic picture of a probe particle moving with velocity $U$ in an odd viscoelastic fluid, experiencing a drag force $F_{\parallel}$ and a lift force $F_\perp$.}
    \label{fig:enter-label298}
\end{figure}

In this work, we shed light on these three points by looking at the motion of a passive probe particle in an active odd viscoelastic fluid (see Fig.~\ref{fig:enter-label298}).
\com{For symmetry reasons, isotropic odd fluids only exist in two dimensions~\cite{avron1998odd}, and we, therefore, focus on this case in this paper, although our methods could be applied in higher dimensions. Moreover, lift forces vanish in incompressible odd fluids~\cite{ganeshan2017odd} when no-slip boundary conditions hold \cite{lier2024slipinduced}, but can exist in compressible fluids~\cite{lier2023lift}.  To assess odd viscoelastic responses in a generic framework, we therefore consider in the following a weakly compressible fluid, and odd lift responses will also be discussed.}

First, we derive the material properties of such fluid from Onsager's theory and show that activity introduces interesting physics by breaking the Onsager-symmetry and thermodynamic inequalities that otherwise constrain the stress-strain relations.
We then unveil the conditions that grant the stability of the motion of the probe particle when it is embedded in an active fluid. 
After confirming that a probe in a passive system has a stable trajectory, i.e. its velocity decays at long times, we show that, within our linear approach, there exists an intermediate regime where the probe motion through an active system is still stable. For larger activities, a probe that is subjected to an external force at an initial time becomes linearly unstable.
We also focus on the signatures of oddity, which can be seen in the dynamics of the probe. We observe that odd viscoelastic terms promote long-lasting oscillations during the relaxation process after an instantaneous perturbation. 
These findings provide tools to assess the degree of oddity of a material by studying the motion of a probe inserted in it.\\

\section{probe particle in an active odd viscoelastic fluid}
\label{activechiralviscoelasticfluidapp}

    \subsection{Constitutive equations for the fluid}

In this Section, we discuss the material properties of the two-dimensional odd viscoelastic fluid in which the probe particle moves. For this purpose, we start with the total stress tensor $\tens{\sigma}^{\rm tot}$, which enters the momentum balance equation as 
\begin{align} \label{eq:momconservation}
   \partial_t \mathbf{g}  -  \nabla \cdot \tens{\sigma}^{\rm tot}  = \mathbf{f}^{\rm{ext}} \, . 
\end{align}
We have introduced the momentum $\mathbf{g}=\rho \mathbf{v}$ with $\rho$ and $v_i$ the fluid mass density and velocity, respectively, and $\mathbf{f}^{\rm{ext}}$ is an external force density.
Mass density is conserved and obeys $\partial_t \rho + \nabla \cdot (\rho \mathbf{v})=0$.
The total stress $\tens{\sigma}^{\rm tot}$ can be decomposed into
\com{\begin{align}
\tens{\sigma}^{\rm tot}    = - \rho \mathbf{v} \otimes \mathbf{v} - P \mathbbm{1}   + \tens{\sigma}^{\rm s} + \tens{\sigma}^{\rm a} \, ,
\end{align}}
where $\otimes$ is the outer product and $\mathbbm{1}$ the identity tensor. 
\com{The tensor $\tens{\sigma}^{\rm s}$ is the symmetric deviatoric stress, which incorporates viscoelastic contributions and $\tens{\sigma}^{\rm a}$ is the anti-symmetric part of the stress ~\cite{callan-jones2011hydrodynamics}.}

\com{For a \textit{passive} odd viscoelastic fluid, the constitutive equations for the stress $\tens{\sigma}^{\rm s}$ and strain $\tens{u}$ tensors are found to be (see App.~\ref{app:const_eq} for more details, also on $ \tens{\sigma}^{\rm a} $):
\begin{subequations} 
\begin{align}%
    \tens{\sigma}^{\rm s} &=  \bm\nu^{\rm p} : \tens{\sigma}^{\rm el}  + 2\bm\eta^{\rm p}  : \tens{v}  \, , \label{eq_constitutiveEquation_passive_1}  \\
 \DDt{\tens{u}} &= -\bm\gamma^{\rm p}  : \tens{\sigma}^{\rm el} + \bm\nu^{\rm p}  : \tens{v} \, , \label{eq_constitutiveEquation_passive_2}
\end{align}\label{eq_constitutiveEquation_passive}%
\end{subequations}}
where ${\rm D} /{\rm D}t$ is the corotational derivative and $\tens{v}=[ \nabla \vect{v} + (\nabla \vect{v})^\top ]/2$ is the symmetric part of the velocity gradient tensor. The elastic stress $\tens\sigma^{\rm el}$ results from the local strain $\tens{u}$ within the material and is obtained as $\tens\sigma^{\rm el} = \delta f  / \delta \tens{u}$, where $f$ denotes the free energy density.

\com{The objects $\bm\nu^{\rm p} $, $\bm\eta^{\rm p} $ and $\bm\gamma^{\rm p} $ in Eq.~\eqref{eq_constitutiveEquation_passive} are four-tensors 
that are isotropic and symmetric under the exchange of the first two as well as the last two indices, i.e. for a general four-tensor $\bm\beta$ we have $\beta_{ij k l } = \beta_{ji  k l }=   \beta_{ij  lk  }$.}
Such a general tensor can thus be characterized in two dimensions by a shear $\beta_{\rm s}$, bulk $\beta_{\rm b}$ and odd $\beta_{\rm o}$ coefficient (more details on the tensor notation can be found in App.~\ref{sec_rank4_algebra}). The four-tensor \com{$\bm\eta^{\rm p}$} in Eq.~\eqref{eq_constitutiveEquation_passive} corresponds to all viscous corrections, whereas \com{$\bm \gamma^{\rm p}$} incorporates plasticity or strain relaxation. 

\com{As a consequence of Onsager symmetry (see App.~\ref{sec_Onsager_odd} for its derivation for odd materials), the same four-tensor $\bm\nu^{\rm p}$ relates the deviatoric stress to elastic stress and the strain rate to the velocity gradient. 
We emphasize that the odd coefficient $\nu_{\rm o}^{\rm p}$ is responsible for a \textit{passive} and transient odd elasticity~\cite{lier2022passive} (while steady-state odd elasticity cannot be passive~\cite{scheibner2020odd}).   
Finally, to qualify as a passive material, the coefficients entering the four-tensors of Eq.~\eqref{eq_constitutiveEquation_passive} must satisfy \cite{lier2022passive}
\begin{align} \label{eq:passiveconstraint1}
\eta^{\rm p}_{\rm s} , \eta^{\rm p}_{\rm b} , \gamma^{\rm p}_{\rm s} , \gamma^{\rm p}_{\rm b} \geq 0 ~~ , ~~ 2\eta_{\rm s}^{\rm p}\gamma_{\rm s}^{\rm p}\ge(\nu_{\rm o}^{\rm p})^2 \, ,   
\end{align}
as dictated by the positivity of the entropy production rate.
}

\com{In this paper, we go beyond passive systems and include activity. This can be done in Onsager framework, in the spirit of active gel models, by introducing fuel consumption at the microscopic level (see, {\it e.g.}, Refs.~\cite{callan-jones2011hydrodynamics,julicher2018hydrodynamic}). Fuel consumption can be represented by adding a thermodynamic flux, the scalar reaction rate $r$,  and its associated thermodynamic force $\Delta\mu$, which is the chemical potential difference of a chemical reaction, to the description. Since it is an additional thermodynamic force, one should supplement equation~\eqref{eq_constitutiveEquation_passive} with terms proportional to $\Delta\mu$, where, by symmetry, the proportionality factor should be a rank-2 tensor. 
We thus add the active contributions $\Delta\mu \left(  \bm\zeta^{(1)} : \tens{\sigma}^{\rm el} + 2 \bm\zeta^{(2)} : \tens{v} \right)$ to Eq.~\eqref{eq_constitutiveEquation_passive_1} and  $\Delta\mu \left( -  \bm\zeta^{(3)} : \tens{\sigma}^{\rm el} +  \bm\zeta^{(4)} : \tens{v} \right)$ to Eq.~\eqref{eq_constitutiveEquation_passive_2}.
In Onsager framework, the thermodynamic forces (here~$\Delta\mu$,~$\tens{\sigma}^{\rm el}$ and~$\tens{v}$) are considered small, and the expansion of the thermodynamic fluxes as a function of the forces is kept to the lowest order. This means that the $\Delta \mu \tens{\sigma}^{\rm el}$- and $\Delta \mu \tens{v}$-terms are second order, and one would therefore naively think that these terms should be neglected. However, as we will show in the rest of the paper, because these contributions are not bound by passivity constraints, even a small contribution can destabilize the odd viscoelastic model, and therefore to correctly capture the qualitative behavior of the active viscoelastic system we must retain them.} 

\com{Finally, the constitutive equation for the stress $\tens{\sigma}^{\rm s}$ and strain $\tens{u}$ tensors characterizing two-dimensional odd active viscoelastic fluids read:
\begin{subequations} 
\begin{align}%
    \tens{\sigma}^{\rm s} &=  \bm\nu : \tens{\sigma}^{\rm el}  + 2\bm\eta : \tens{v}  \, ,  \\
 \DDt{\tens{u}} &= -\bm\gamma : \tens{\sigma}^{\rm el} + \bm\nu^{\prime } : \tens{v} \, , 
\end{align}\label{eq_constitutiveEquationSimple}%
\end{subequations}
where we have defined the active rank-four tensors $\bm\nu =\bm\nu^{\rm p}+\Delta\mu\bm\zeta^{(1)} $, 
$\bm\eta =\bm\eta^{\rm p}+\Delta\mu\bm\zeta^{(2)}$, 
$\bm\gamma =  \bm\gamma^{\rm p}+ \Delta\mu\bm\zeta^{(3)}$, and 
$\bm\nu^{\prime} =\bm\nu^{  \rm p}+\Delta\mu\bm\zeta^{(4)}$.}

\com{
Note that the effect of activity ($\Delta\mu\neq0$) is two-fold: (i)~it can break the symmetry of the Onsager matrix, which happens whenever $\bm \nu' \neq \bm\nu$.  For this reason,  $ \bm\nu$ and $ \bm\nu^{\prime}$ can be understood as a non-equilibrium and non-reciprocal generalization of classical elasticity; or (ii)~it can break the semi-definiteness of the Onsager matrix. 
}

When discussing the stability of the probe in Sec.~\ref{sec_stability} and~\ref{sec_drag}, we will make use of the passive limit for the generalized tensor objects, which can be imposed by setting\footnote{Note that in this limit, activity has not necessarily disappeared from the system, i.e. active agents might still be present. However, from the point of view of the coarse-grained description, Eq.~\eqref{eq_constitutiveEquationSimple}, the viscoelastic fluid is indistinguishable from a passive system.}
\begin{align}%
 2\eta_{\rm s}\gamma_{\rm s}\ge \nu_{\rm o}^2   ~~ , ~~      \bm\nu =     \bm\nu^{\prime} ~~ , 
    \label{eq:thermoConstraint}
\end{align}
and where $\eta_{\rm s} , \eta_{\rm b} , \gamma_{\rm s} , \gamma_{\rm b} \geq 0$ is always satisfied provided $\Delta \mu$ is small.

  \subsection{Jeffreys model for the odd viscoelastic fluid}

Having in mind the description of a probe particle embedded in a viscoelastic fluid, we now show how Eq.~\eqref{eq_constitutiveEquationSimple} leads to an odd Jeffreys model\com{~\cite{raikher2013brownian,Oswald2009}.} We consider a material with linear elastic properties, and first use the identity
\begin{align} \label{eq:elasticidentity}
    \tens{\sigma}^{\rm el}=2 \bm{G} :  \tens{u} \, , 
\end{align}
where $ \bm{G}$ is the four-tensor of elastic moduli. Since Eq.~\eqref{eq:elasticidentity} is derived from free energy, it is not possible to have an odd elastic modulus \cite{scheibner2020odd} and thus $G_{\rm o}=0$. Note however that odd linear couplings between the stress and the strain are present in Eq.~\eqref{eq_constitutiveEquationSimple}, and correspond to non-equilibrium corrections derived following Onsager's theory.

\begin{figure}[t]
    \centering
    \includegraphics[width=0.9\linewidth]{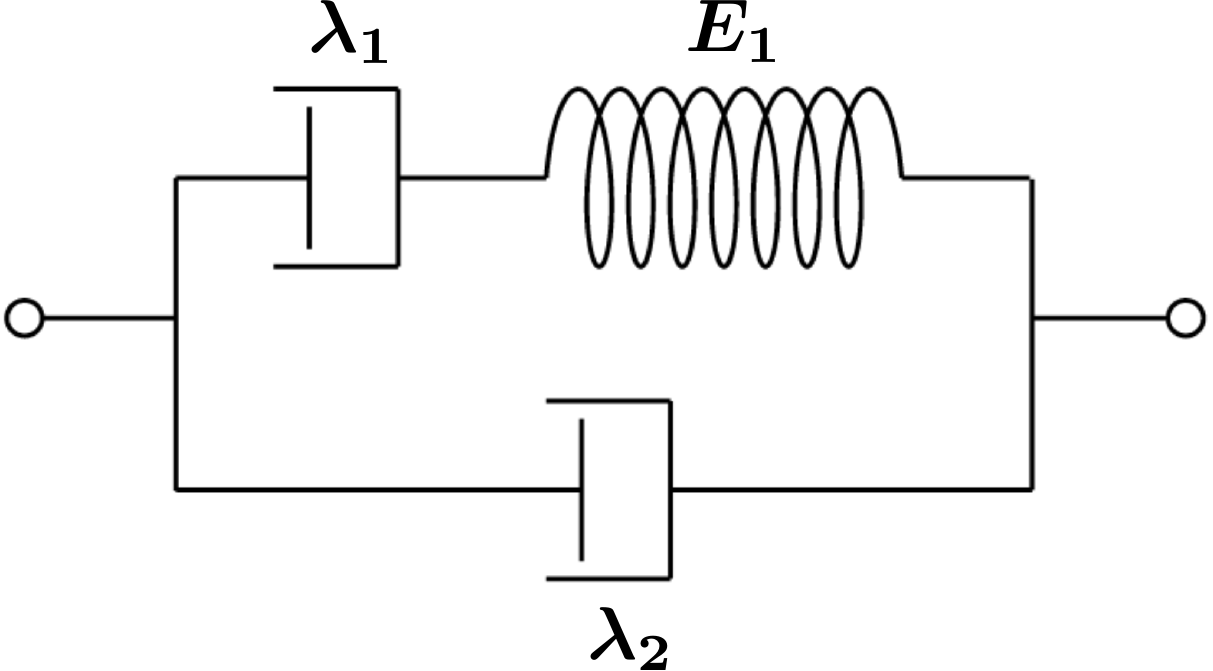}
    \caption{
    \com{
    Circuit representation of the odd Jeffreys model given in Eq.~\eqref{eq_activeOddJeffreys} including dampers~($\bm \lambda_{1,2}$) and an elastic spring~($\bm E_1$). Each element of this representation has odd and even contributions, and can be represented as a rank-4 tensor (see text for details). The relation between the rank-4 tensors appearing in Eq.~\eqref{eq_activeOddJeffreys} and the circuit rank-4 tensors is given by
    $ \bm \lambda_1  = \bm{\eta} $, $ \bm E_1   = \bm{\nu G \nu'} $ and $\bm \lambda_2  = \bm{\nu \gamma^{-1} \nu'} /2$.}
    }
    \label{fig_odd_jeffreys}
\end{figure}

Solving for the strain $\tens{u}$, we find that the constitutive equations~\eqref{eq_constitutiveEquationSimple} can be rewritten as:
\begin{align}
    \left(\bm{I} + \bm \tau_1 \DDt{} \right)  :  \tens{\sigma}^{\rm s} = 2\bm\eta^{\rm eff}\left(\bm{I} + \bm\tau_2 \DDt{} \right) :  \tens{v} \, ,
    \label{eq_activeOddJeffreys}
\end{align}
where $\bm{I}$ is the rank-4 identity and where we have used the fact that the contraction of an odd isotropic rank-4 tensor with a rank-2 tensor commutes with taking the corotational time derivative. We have moreover defined: 
\com{$\bm\tau_1 = \bm\gamma^{-1} \bm{G}^{-1}\bm/2$, $\bm \eta^{\rm eff}= \bm\eta+ \bm\nu \bm\gamma^{-1} \bm\nu'/2$, and $\bm\tau_2 =  (\bm\eta^{\rm eff})^{-1} \bm\tau_1 \bm\eta$,}
where four-tensor algebra is given in App.~\ref{sec_rank4_algebra}.
\com{Equation~\eqref{eq_activeOddJeffreys} describes an odd active Jeffreys model, that is, a viscoelastic fluid with two (odd) dampers and one (odd) elastic spring, as depicted schematically on Fig.~\ref{fig_odd_jeffreys}. Furthermore, one}
consequence of the thermodynamic constraint~\eqref{eq:passiveconstraint1} is that an odd Maxwell model, for which $\bm\tau_2$ in Eq.~\eqref{eq_activeOddJeffreys} vanishes, cannot be passive~\cite{lier2022passive}. Similarly, the odd Kelvin--Voigt limit ($\gamma_{\rm s}\to0$, $\nu_{\rm o}\neq 0$) is also forbidden by thermodynamics in the absence of fuel consumption. 

In Laplace space, dropping the (nonlinear) vorticity contributions to the corotational time derivative, we can write the constitutive relation of the active odd Jeffreys model~\eqref{eq_activeOddJeffreys} as \cite{levine2001response,mason1995optical}
\begin{align}
    \tens{\sigma}^{\rm s}(s) = 2\bm A(s) : \tens{v}(s) \, ,
\label{eq_material_response_function}
\end{align}
where 
$\bm A(s)=\bm \eta+    \left(\bm{I} + s \bm \tau_1 \right)^{-1} \cdot\bm\nu \bm\gamma^{-1}\bm \nu'/2$ \com{contains the shear and odd components}
\begin{subequations}%
\label{eq_Laplace_viscosity_tensor}
\begin{align}%
 & A_{\rm s}(s) =  \eta_{\rm s} +  \frac{  \left( \gamma_{\rm s}+ \frac{s}{2 G_{\rm s}} \right) (  \nu_{\rm s}\nu_{\rm s}' - \nu_{\rm o}\nu_{\rm o}' )+ \gamma_{\rm o} (\nu_{\rm o} \nu_{\rm s}'+\nu_{\rm o}' \nu_{\rm s}) }{ 2[\gamma_{\rm o}^2+(\gamma_{\rm s}+ s / (2 G_{\rm s}) )^2]}   \, ,  \\ 
& A_{\rm o}(s) =  \eta_{\rm o} +   \frac{ \left( \gamma_{\rm s}+ \frac{s}{2 G_{\rm s}} \right) (\nu_{\rm o} \nu_{\rm s}'+\nu_{\rm o}' \nu_{\rm s})  -  \gamma_{\rm o} (  \nu_{\rm s}\nu_{\rm s}' - \nu_{\rm o}\nu_{\rm o}' ) }{ 2[\gamma_{\rm o}^2+(\gamma_{\rm s}+ s/ (2 G_{\rm s}) )^2 ]} \, , \\
& A_{\rm b} (s)  = \eta_{\rm b}  +  \frac{    \nu_{\rm b}\nu_{\rm b}' }{ 2[ s /(2  G_{\rm b})  + \gamma_{\rm b}]  }  \, , 
\end{align}%
\end{subequations}%
where we have introduced the Laplace transform $\mathcal{L}[h(t)](s) =h(s)=\int_0^\infty \rmd t\, \rme^{-st}h(t)$ (using the same notation for the function and its transform). \com{We will refer to the coefficients in Eq.~\eqref{eq_Laplace_viscosity_tensor} as Laplace viscosities, as they can be seen as viscoelastic generalizations of viscosity in the Laplace domain.} 

\com{
To illustrate the physical meaning of the Laplace viscosity tensor $\bm A(s)$, let us consider an even material with $\nu = \nu ' $. In this case, $A_{\rm s}$ turns into 
\begin{align}
\tilde A_{\rm s}(s)=  \eta_{\rm s} +  \frac{   \nu_{\rm s}^2  }{2  \gamma_{\rm s}+ s / G_{\rm s}  }  \, ,
\end{align} 
which corresponds to the usual Jeffrey's model, which is viscous at very short timescales with viscosity $ \lim_{s \rightarrow \infty } \tilde A_{\rm s}(s ) =  \eta_{\rm s}$, elastic on intermediate timescales and again viscous at long timescales with viscosity $\lim_{s \rightarrow 0 } \tilde A_{\rm s}(s) = \eta_{\rm s} +  \frac{   \nu_{\rm s}^2  }{2  \gamma_{\rm s} }$. Note that the only role of activity, in this case, is to modify the values of the different parameters compared to their passive counterparts, while the physics is qualitatively unchanged.} 

\com{From this simple discussion, we understand that the odd visco-elastic Jeffreys' model [Eqs.~\eqref{eq_material_response_function} and~\eqref{eq_Laplace_viscosity_tensor}]: (i)~is also characterized by a viscous, fluid-like behavior at short and long timescales; (ii)~has three characteristic timescales in the shear and odd sectors, and two in the bulk sector, (iii)~can have qualitative differences compared to the even Jeffreys' model since the sign of some effective parameters is not fixed (for instance the term $\nu_{\rm s}\nu_{\rm s}' - \nu_{\rm o}\nu_{\rm o}'$ in $A_{\rm s}$). }

        \subsection{Probe particle}
\label{sec:tracerparticle}

We now study the motion of a probe particle dragged in the viscoelastic fluid with constitutive equation~\eqref{eq_material_response_function}. 
\com{As discussed in the introduction, isotropic odd materials only exist in two dimensions, and we thus focus on this case in the following. 
A natural setting for a two-dimensional viscoelastic fluid is to consider a thin layer of such a fluid at the interface between two bulk (even) fluids, for instance, water and air~\cite{lier2023lift,hosaka2021nonreciprocal}. For simplicity, we consider this layer to be flat and infinitely thin, such that the odd fluid can be described effectively as two-dimensional.}

The probe particle velocity $\mathbf{U}(t)$ corresponding to an applied time-dependent force $\mathbf{F}(t)$ can be computed using Laplace transforms as:
\begin{align} \label{eq:responsematrix}
    U_i  (t)   =    \mathcal{L}^{-1} \left[  \mathbb{M}_{ij } (s )   F_j (s) 
  \right] (t) ~~ , 
\end{align}
where Latin indices correspond to Cartesian coordinates and a summation over repeated indices is implied, and with $\mathcal{L}^{-1} \left[ h (s )  \right] (t ) $ the inverse Laplace transform. The response matrix $\mathbb{M}_{ij } (s ) $ can be understood as a viscoelastic and therefore $s$-dependent generalization of the drag and lift coefficients for a purely viscous fluid. 
\com{
For simplicity, we consider only the translational degrees of freedom of the probe, noticing that in the incompressible limit and for rigid probes there is no odd viscosity-induced torque~\cite{ganeshan2017odd}.
}

To obtain the response matrix $\mathbb{M}_{ij } (s )$ of an odd viscoelastic fluid, we first linearize to first order in $v_i$ and $\delta\rho=\rho-\rho_0$ the momentum and mass conservation equations.
\com{In order to discuss weakly compressible fluids, we need to specify an equation of state for the pressure, which at first order in $\delta\rho$ reads $P = \frac{\delta \rho }{\psi \rho_0}$ where $\psi$ is the the compressibility. 
Note that the incompressible case, for which the pressure is independent of density, is recovered by taking the limit  $\psi\to 0$ in the response matrix $\mathbb{M}_{ij } (s )$.}
Then, using the constitutive equation~\eqref{eq_material_response_function}, we obtain the following coupled linearized equations:
\begin{subequations} \label{eq_Stokes_Laplace}%
\begin{align}%
\begin{split}
\rho_0 s  v_i &= A_{\rm s}(s) \partial_k\partial_k v_i + A_{\rm b}(s) \partial_i \partial_k v_k  \\
& \quad + A_{\rm o}(s) \varepsilon_{ij} \partial_k\partial_k v_j -  \frac{1}{\psi \rho_0 } \partial_i \delta\rho   +  f^{\text{ext}}_i  \, , 
\end{split} \\
s  \delta\rho  &= - \rho_0\partial_k  v_k  \,  . 
\end{align}%
\end{subequations}
Using a spatial Fourier transform  with the convention   $h(\mathbf{x}) = \frac{1}{(2\pi)^2} \int  {\rm d}^2 \mathbf{k} \, h(\mathbf{k}) e^{   \rmi \mathbf{k} \cdot \mathbf{x} }$, Eq.~\eqref{eq_Stokes_Laplace} can then be rewritten in matrix form :
\begin{align}
\mathcal{G}_{ij} v_j = f^{\rm{ext}}_i     \label{eq:matrixform111} ~~ , 
\end{align}
with $\mathcal{G}_{ij}$ given by~\cite{lier2023lift,hosaka2021nonreciprocal}
\begin{align} 
\begin{split}
    \mathcal{G}_{ij}  &= \hat{k}_i  \hat{k}_{j} \left[   s       \rho_0   + \left(  A_{\rm s} (s)  + A_{\rm b} (s)  +  \frac{  1 }{\psi   s  }  \right) k^2          \right]   \\
    \quad  & +    (\delta_{ij}- \hat{k}_i  \hat{k}_{j})  \left[ s    \rho_0    +   A_{\rm s} (s) k^2  \right]   + \varepsilon_{ij}   A_{\rm o} (s)  k^2    \, , 
\end{split}
\label{eq_fullResponseMatrix}
\end{align}
where $k=\sqrt{k_ik_i}$ and $\hat{k}_i = k_i/k$. Having formulated the linearized equations for a given Laplace frequency and Fourier wave vector, we can then use the shell localization approach \cite{levine2001response,weisenborn1984oseen,lier2023lift} to obtain the linear response for a rigid disk-shaped probe particle moving in the viscoelastic fluid. When no-slip boundary conditions hold, the shell localization approach prescribes that, upon identifying the probe particle velocity $U_i (s)$ with $ v_{i} (|\mathbf{x}| =0, s)$, the response matrix is given by (more details in App.~\ref{app:shelllocalization}):
\begin{align}  
\mathbb{M}_{ij } (s )  =   \frac{1}{(2 \pi)^2 }   \int_0^{2 \pi } \!\! {\rm d} \theta  \int_0^{\infty} \!\! {\rm d} k \, k J_0 (a k)  \mathcal{G}^{-1}_{ij}  (\mathbf{k},s)  \, , \label{eq_responseMatrix}
\end{align}
where $\theta$ is the angle of the wave vector, $a$ is the disk radius and $J_0(z)$ is the $0^{\rm th}$ Bessel function of the first kind. 

\com{For concreteness, and before discussing in detail the drag and lift responses of a probe in an odd active viscoelastic fluid, let us first relate this expression, Eq.~\eqref{eq_responseMatrix}, to the well-known drag in an incompressible, even, purely viscous fluid in two dimensions. In this case, the response matrix is diagonal and can be written as $\mathbb{M}_{ij }^{\rm simp} (s ) = M_\parallel^{\rm simp}(s) \delta_{ij}$, where the drag coefficient reads  $M_\parallel^{\rm simp}(s) = \frac{1}{4\pi\eta_{\rm s}} K_0(\sqrt{\tau_0 s})$,
with $\tau_0=\rho_0 a^2/\eta_{\rm s}$ and $K_0$ is the 0$^{\rm th}$ modified Bessel function of the second kind. This expression can then be expanded in series of the Laplace frequency~$s$, leading to:
\begin{align}
    M_\parallel^{\rm simp}(s) \!=\! \frac{-1}{4\pi\eta_{\rm s}}\!\!\left(\!
    \log \frac{ \sqrt{s \tau_0} }{2} + \gamma_{\rm EM}     \!\right) \!+ \mathcal{O}(s \tau_0) ,
\end{align}
where $\gamma_{\rm EM}$ is the Euler-Mascheroni constant. This last expression shows the expected divergence of the drag at vanishing frequency, a signature of the Stokes paradox in two dimensions.
}

\begin{figure*}[t]
	\centering
	\subfigure[ \label{fig_stability_gammas_vs_gammao}]
    {\includegraphics[width=0.45\linewidth]{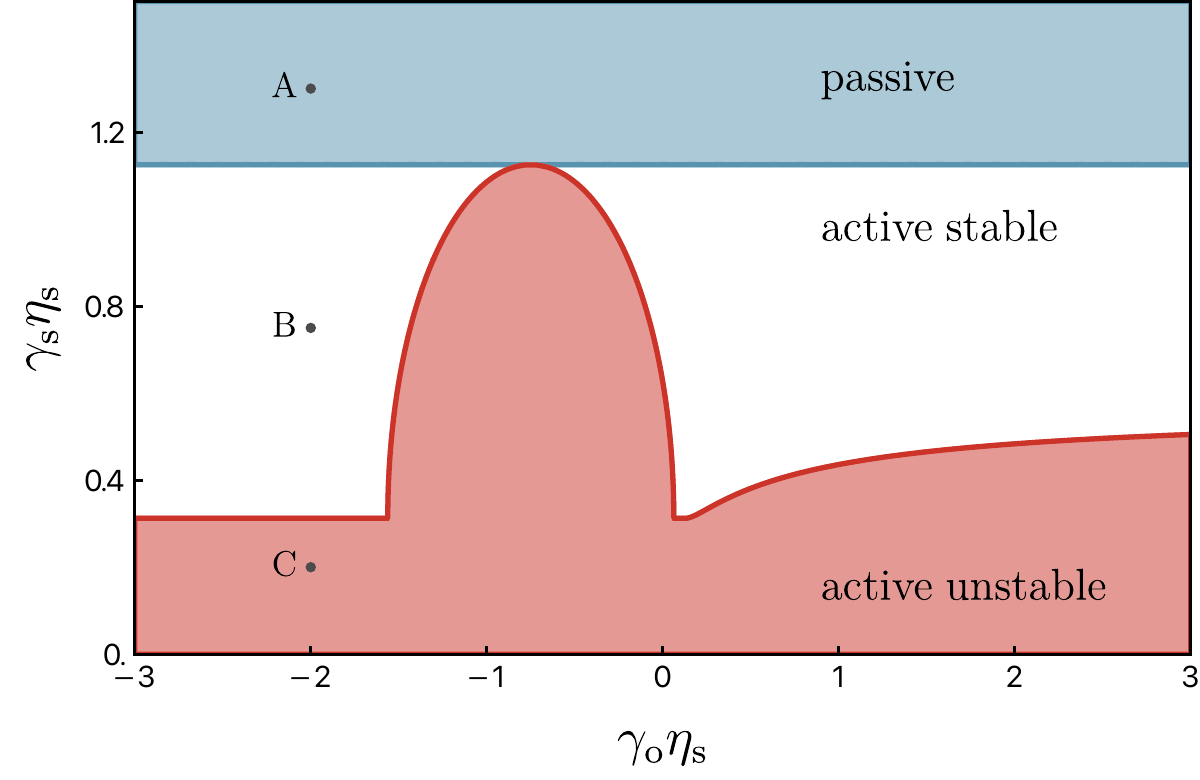}}
    \hfill
	\subfigure[ \label{fig_trajectories}]
    {\includegraphics[width=0.45\linewidth]{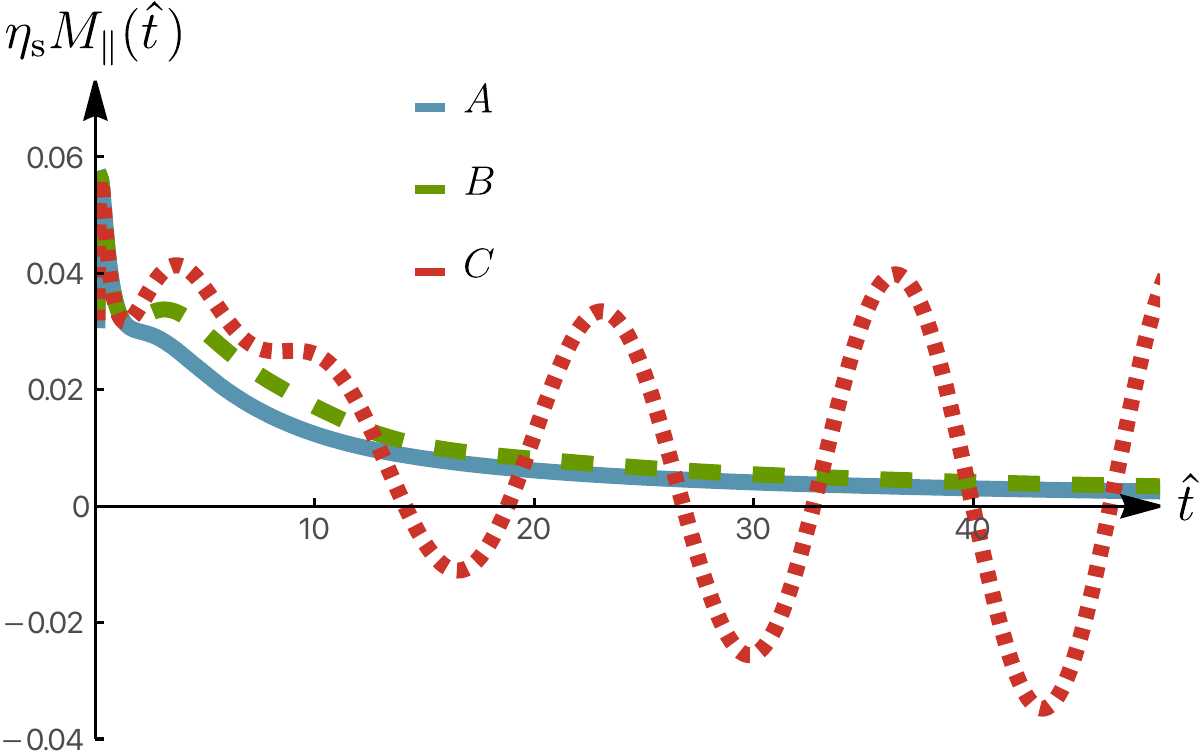}}
	\subfigure[ \label{fig_stability_gammas_vs_nuo}]
    {\includegraphics[width=0.45\linewidth]{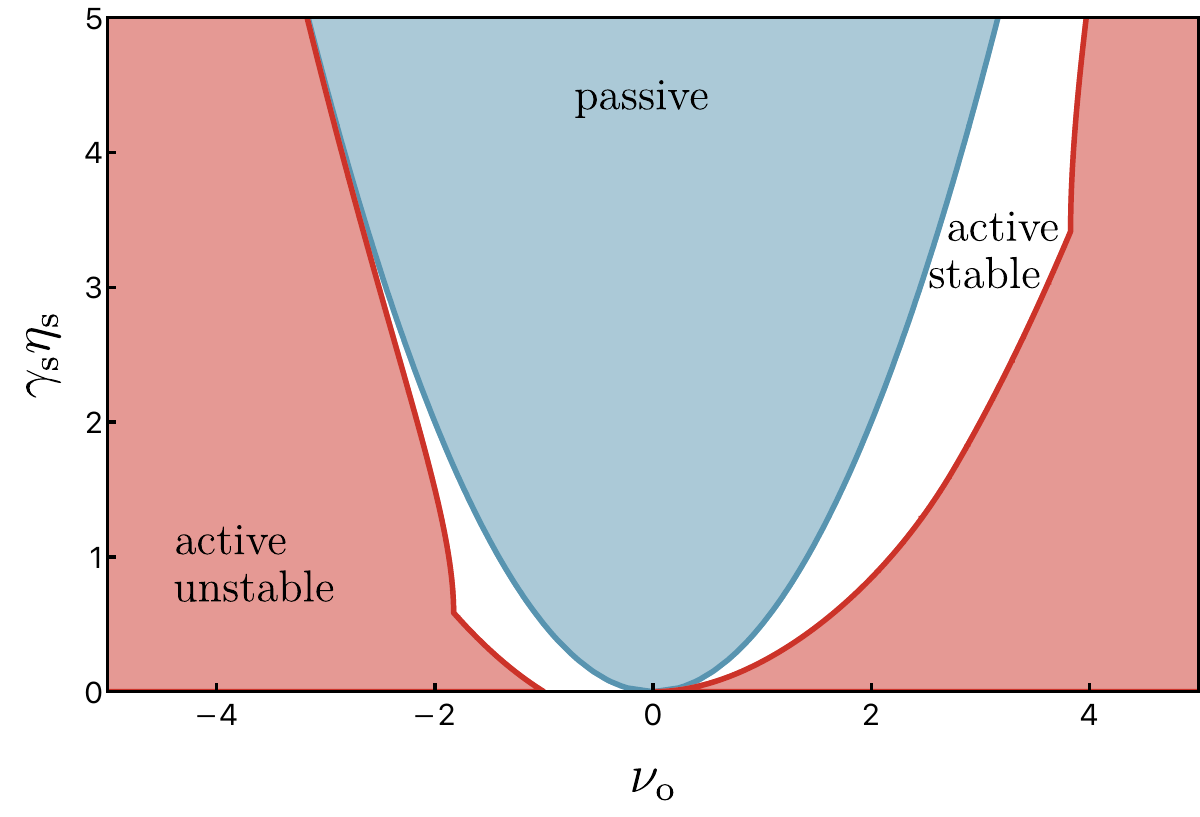}}
    \hfill
	\subfigure[ \label{fig_stability_3d}]
    {\includegraphics[width=0.45\linewidth]{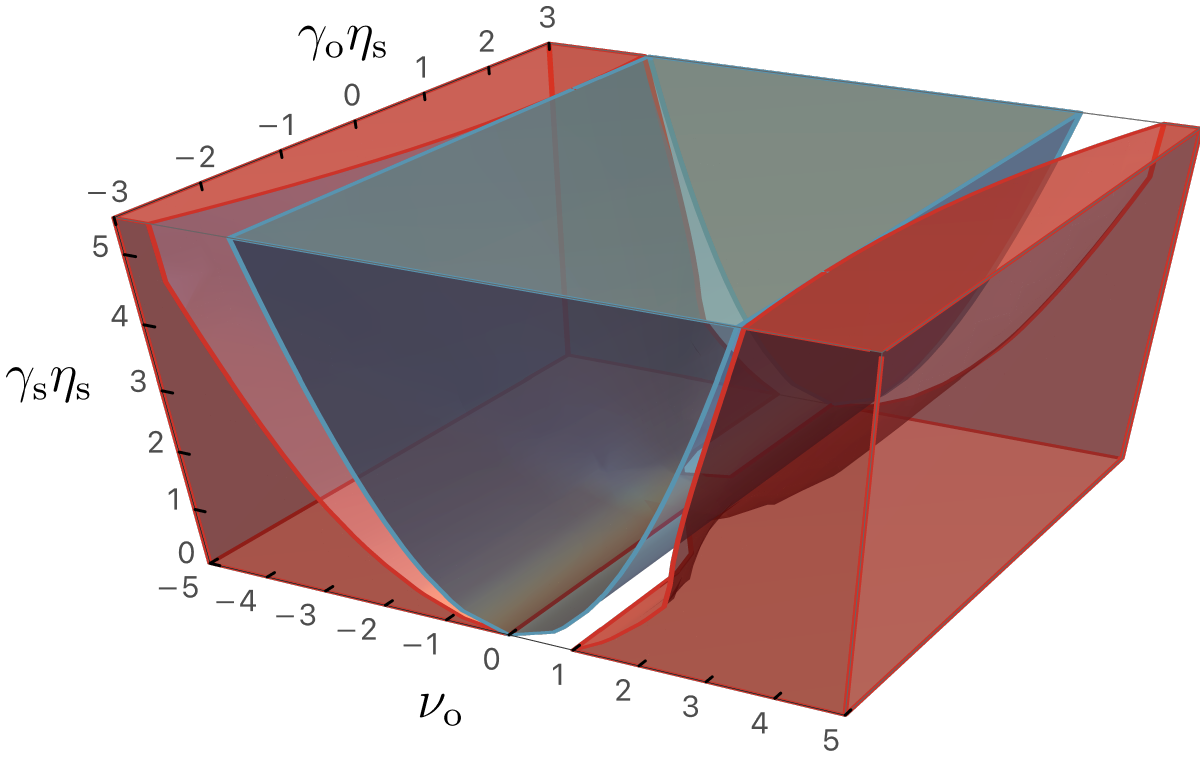}}
	\caption{Stability of a probe in an odd viscoelastic medium. \textbf{(a,c)} stable (white and blue) and linearly unstable (red) regions in the parameter space $(\gamma_{\rm o} \eta_{\rm s},\gamma_{\rm s} \eta_{\rm s} )$ in panel (a) and $(\nu_{\rm o},\gamma_{\rm s} \eta_{\rm s})$ in panel (c). The passive (blue) region indicates where the passivity constraint~\eqref{eq:thermoConstraint} is satisfied. \textbf{(b)} Drag response $\eta_{\rm s} M_\parallel$ as a function of the dimensionless time $\hat{t}=t/\tau_0$ with $\tau_0=a^2\rho_0/\eta_{\rm s}$ for the points $A,B,C$ in the parameter space indicated in panel~(a). \textbf{(d)} Linearly unstable (red) and passive (blue) regions in the three-dimensional parameter space $(\nu_{\rm o},\gamma_{\rm o}  \eta_{\rm s},\gamma_{\rm s} \eta_{\rm s})$. Importantly, the transformation $(\nu_{\rm o},\gamma_{\rm o})\to (-\nu_{\rm o}, -\gamma_{\rm o})$ is a symmetry of the system. Parameters used: for all panels, $\nu_{\rm s}=1$.
	(a) $\nu_{\rm o}=1.5$. 
	(b) $A,B,C$: $\nu_{\rm o}=1.5$,  $\gamma_{\rm o} \eta_{\rm s}=-2$, $G_{\rm s}\tau_0/\eta_{\rm s}=0.25$, and $A$: $\gamma_{\rm s}\eta_{\rm s} =1.3$, $B$: $\gamma_{\rm s} \eta_{\rm s} = 0.75$, $C$: $\gamma_{\rm s} \eta_{\rm s} =0.2$.  
	(c) $\gamma_{\rm o} \eta_{\rm s} = 2$.
	}
	\label{fig_stability_and_thermo}
\end{figure*}

    \section{Linear stability of a probe particle in an odd active viscoelastic fluid}
    \label{sec_stability}

\subsection{General case}
Active viscoelastic fluids can give rise to sustained motion.
Therefore, the response of a probe placed into the medium after an initial perturbation does not necessarily decay to zero. At the level of linear response, such sustained motion is signaled by instability, which appears in the long-time behavior of the probe velocity given by Eq.~\eqref{eq:responsematrix}.
Without loss of generality, \com{we consider an instantaneous perturbation applied at $t=0$, such that $\mathbf{F}(t) \propto \delta(t)$, where $\delta(t)$ is the Dirac delta function}.  The velocity of the probe after this initial perturbation is given by: \com{
\begin{align}
    U_i(t) &= \frac{1}{2 \pi \rmi} \int_{\Gamma -\rmi \infty}^{\Gamma +\rmi \infty} \!\! \rmd s \, \rme^{s t}  \int \frac{\rmd^2 \mathbf{k}}{(2\pi)^2}  m_i (\mathbf{k}, s) J_0 (k) \, ,
\end{align}}%
where \com{$m_i (\mathbf{k}, s) = \mathcal{M}_{ij} (\mathbf{k}, s)F_j$} and $\Gamma$ is a real number \com{such that the integral over $s$ is convergent.}
To discuss stability, we need not focus on the precise expression of the velocity, but we remark the crucial point that $m_i (\mathbf{k}, s)$ is a rational fraction in $s$. This implies that the long-term behavior of the velocity, and, therefore, the probe stability is determined by the sign of the real part of the poles in $s$ of $m_i$:
\begin{align} \label{eq:stabilitycriterion}
    {\rm Re} \left( s_\ell^*(\mathbf{k}) \right) \leq 0 \qquad \forall \ell,\, \forall \mathbf{k}\, ,
\end{align}
where $s_\ell^*(\mathbf{k})$ is a pole of $m_i$, see App.~\ref{sec_appendix_stability} for details. In the general case of an odd compressible viscoelastic fluid, the poles of $m_i$ are roots of a high-order polynomial, which makes further analytical discussion difficult. Nonetheless, Eq.~\eqref{eq:stabilitycriterion} provides a simple and powerful criterion to determine the probe stability that can easily be used numerically. 
In the following, Sec.~\ref{sec:stabilityconstraint}, we discuss the case of an incompressible fluid, where the analytical discussion regarding stability can be continued.

Note finally that in the regions of the parameter space where the inequality~\eqref{eq:stabilitycriterion} is not satisfied, the motion of a probe is not linearly stable: the passive probe is therefore set into motion by the active viscoelastic fluid. In order to obtain physical solutions at longer times, nonlinear corrections must then be added to the description. The discussion of the form \com{ that these nonlinear corrections should take} and their impact on the probe dynamics is left for future work.

\subsection{Stability of probe motion in the incompressible limit}
\label{sec:stabilityconstraint}

To continue our stability analysis while keeping the technicalities to a minimum, we now restrict to the incompressible case. In this simpler case, odd lift effects are absent \cite{abanov2018odd,lier2023lift}, the tensor $\mathcal{M}_{ij}$ is proportional to the identity, and the probe particle stability is determined by the poles of 
\begin{align}
    r(k,s)= \frac{1}{4\pi\eta_{\rm s}} \frac{ak J_0(ak)}{\tau_0 s  + (ak)^2 A_{\rm s}(s)/\eta_{\rm s}} \, ,
\end{align}
with $\tau_0=\rho_0 a^2/\eta_{\rm s}$.
As shown in App.~\ref{sec_appendix_stability}, these poles \com{in the complex plane} correspond to the roots of a third-order polynomial, and stability 
%
can be discussed analytically. We first discuss the simpler case where Onsager symmetry is enforced. 

\medskip
\noindent\textit{Onsager-symmetric fluid.}
We can take $\bm\nu=\bm\nu'$ and stability
%
is ensured iff
\begin{subequations}\label{eq:allconstraints}%
\begin{align}%
    & 4 \gamma_{\rm s} \eta_{\rm s} -\nu_{\rm o}^2+\nu_{\rm s}^2 > 0 \, , \quad  \label{eq:condition1}   \\
    & 2 \eta_{\rm s} \gamma_{\rm o}^2+2 \gamma_{\rm o} \nu_{\rm o} \nu_{\rm s}+\gamma_{\rm s} \left(2 \gamma_{\rm s} \eta_{\rm s}-\nu_{\rm o}^2+\nu_{\rm s}^2\right) >0 \, , \quad  \label{eq:condition2} 
\end{align}%
and 
\begin{align}%
    \Delta <0 \,\, \text{or} \,
    \begin{cases}
        \Delta \geq 0 \, \quad 
    \text{and} \\
        \gamma_{\rm s} \left(\nu_{\rm s}^2-\nu_{\rm o}^2\right)-2 \gamma_{\rm o} \nu_{\rm o} \nu_{\rm s}+8 \eta_{\rm s}\gamma_{\rm s}^2 <0
    \end{cases} \!\!\! \label{eq:condition3}   
\end{align}
with $\Delta= b_1^2-4b_0b_2$, where $b_0= 4 G_{\rm s}^3 \gamma_{\rm s} (\gamma_{\rm s}^2+\gamma_{\rm o}^2)$, 
$b_1 = G_{\rm s}^2 [\gamma_{\rm s}(8\gamma_{\rm s}\eta_{\rm s}-\nu_{\rm o}^2+\nu_{\rm s}^2)-2\gamma_{\rm o}\nu_{\rm o}\nu_{\rm s}]/(\tau_0\eta_{\rm s})$, 
$b_2= G_{\rm s} (4\gamma_{\rm s}\eta_{\rm s}-\nu_{\rm o}^2+\nu_{\rm s}^2)/(\tau_0^2\eta_{\rm s})$.
\end{subequations}%
It is important to note that if the thermodynamics constraint~\eqref{eq:thermoConstraint} holds, all the conditions of Eq.~\eqref{eq:allconstraints} are satisfied and the system is always stable at long-time, as expected. Constraint~\eqref{eq:condition2} coincides with the condition for the stability of a general odd viscoelastic fluid, which we derived in Ref.~\cite{lier2022passive} by considering the stability of perturbative modes. We also note that the constraint~\eqref{eq:condition3} depends on the mass density $\rho_0$ (through $\tau_0$), whereas the other constraints Eqs.~\eqref{eq:condition1} and \eqref{eq:condition2} only depend on viscoelastic coefficients.

For concreteness, we show in Fig.~\ref{fig_stability_and_thermo} stability diagrams of a probe immersed in an odd incompressible viscoelastic fluid. Figures~\ref{fig_stability_gammas_vs_gammao},~\ref{fig_stability_gammas_vs_nuo} and~\ref{fig_stability_3d} display the unstable (red) regions of the parameter space and the passive (blue) regions, where the thermodynamics constraint~\eqref{eq:thermoConstraint} is satisfied. These two regions are disjoint, as expected from thermodynamics. Note that the three-dimensional view of the parameter space $(\nu_{\rm o}, \gamma_{\rm o}\eta_{\rm s},\gamma_{\rm s}\eta_{\rm s})$ of Fig.~\ref{fig_stability_3d} shows the symmetry $(\nu_{\rm o},\gamma_{\rm o})\to (-\nu_{\rm o}, -\gamma_{\rm o})$ of the system.

As intuition would suggest, the friction $\gamma_{\rm s}$, which characterizes the elastic stress relaxation, has a stabilizing role and increasing it always leads to more stable systems. On the other hand, the odd coupling $\nu_{\rm o}$ is destabilizing the system as its magnitude is increased (provided its sign is kept constant). The odd elastic relaxation term $\gamma_{\rm o}$ has a more subtle effect and may have a destabilizing or stabilizing influence [see Fig.~\ref{fig_stability_gammas_vs_gammao}].

Figure~\ref{fig_trajectories} highlights the consequences of the odd active coefficients on the time-dependent  (drag) response $M_\parallel (t)$ experienced by the particle along its velocity. In the linearly stable regions [points $A$ and $B$ in panel~\ref{fig_stability_gammas_vs_gammao}], the drag decays to 0, indicating that the initial perturbation is eventually damped by viscous friction. On the other hand, the parallel response corresponding to parameters in the active unstable region [point $C$ in panel~\ref{fig_stability_gammas_vs_gammao}] increases with time, indicating that the probe motion is accelerated by the active fluid surrounding it. This acceleration must eventually be compensated by nonlinearities that we have not considered in the present analysis. 

\begin{figure*}[t!]
	\centering
    \subfigure[\label{fig_stability_b98u98u}]
    {\includegraphics[width=0.49\linewidth]{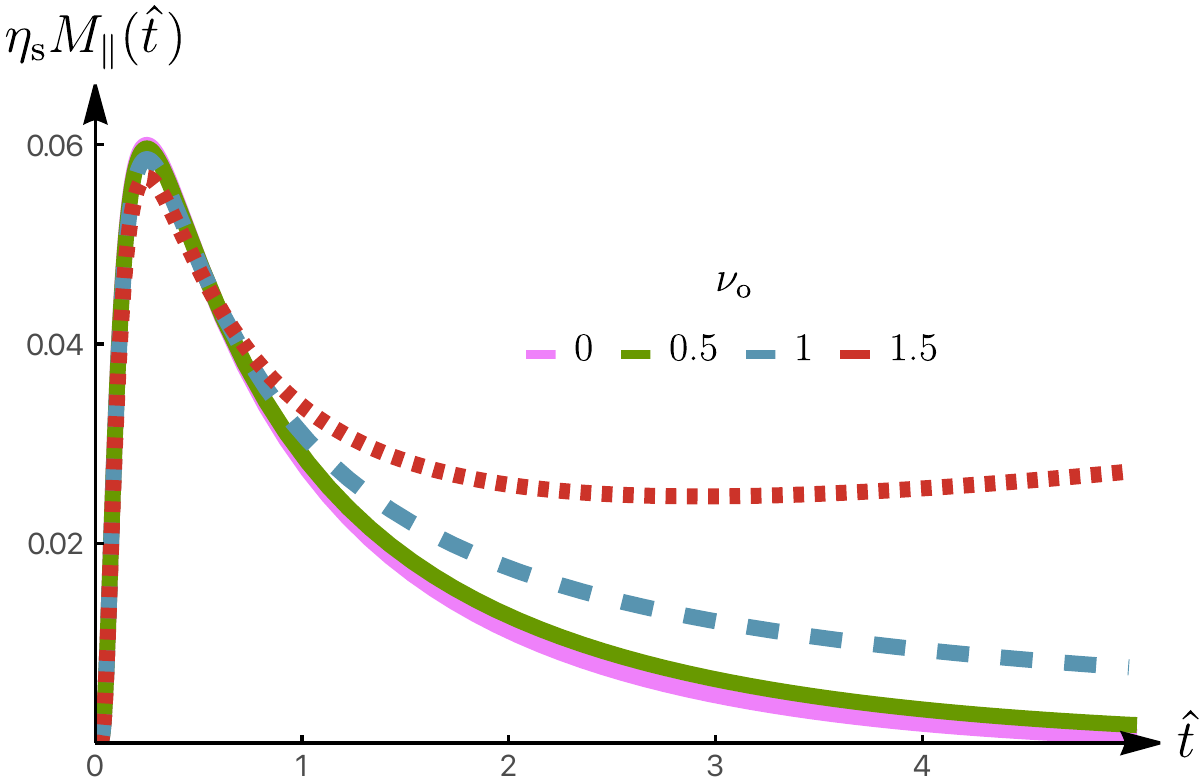}}
    \hfill    	\subfigure[\label{fig_stability_1111}]
    {\includegraphics[width=0.49\linewidth]{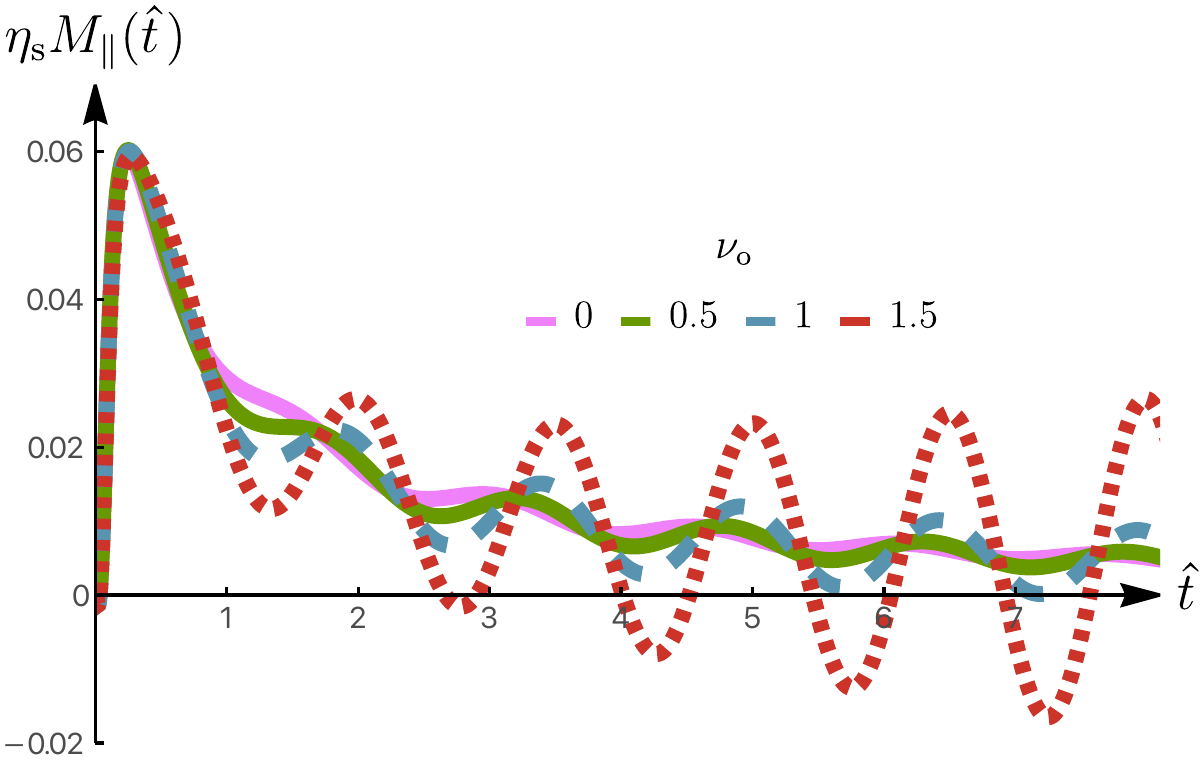}} \\
    \subfigure[\label{fig_stability_119811}]
    {\includegraphics[width=0.49\linewidth]{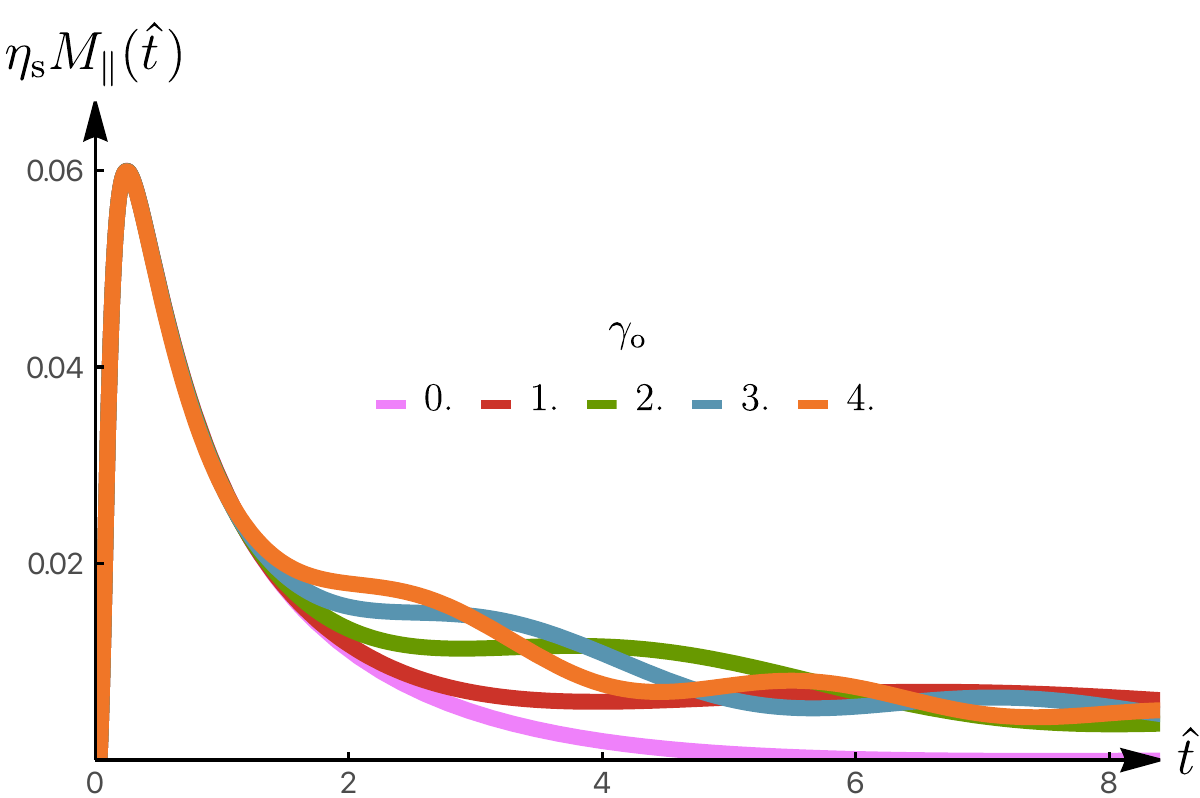}}
     \hfill      \subfigure[\label{fig_stability_771111}]
    {\includegraphics[width=0.49\linewidth]{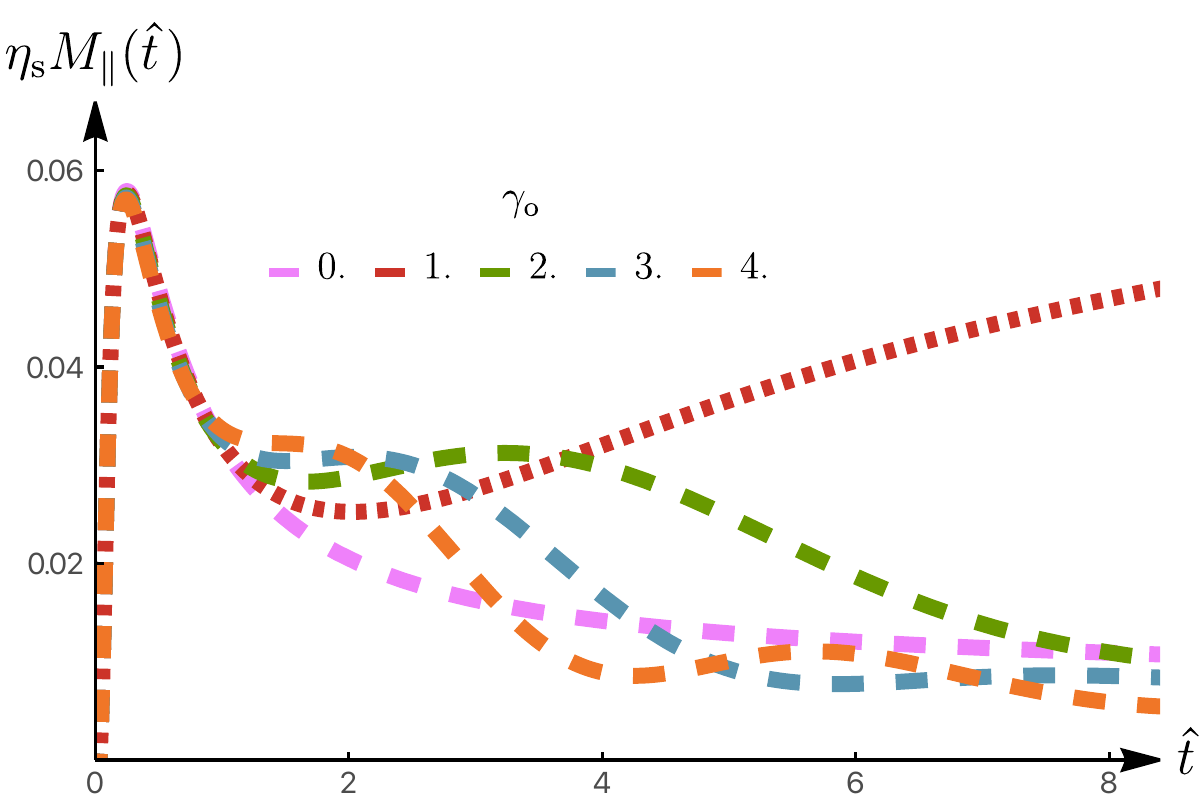}}
	\caption{Drag response $\eta_{\rm s} M_\parallel$ of an odd incompressible viscoelastic fluid as a function of the dimensionless time $\hat{t}=t/\tau_0$ with $\tau_0=a^2\rho_0/\eta_{\rm s}$.  \textbf{(a)} In the absence of the odd elastic stress relaxation ($\gamma_{\rm o} =0$) and for different values of $\nu_{\rm o}$. \textbf{(b)} In the presence of an odd elastic stress relaxation ($\gamma_{\rm o} \eta_{\rm s} =8$) and for different values of $\nu_{\rm o}$.  \textbf{(c)} Drag response for a varied odd elastic stress relaxation $\gamma_{\rm o}$ and a vanishing $\nu_{\rm o}$. \textbf{(d)} Drag response for a varied odd elastic stress relaxation $\gamma_{\rm o}$ and $\nu_{\rm o} =-1.2$.
    In all panels, the solid lines indicate passive fluids for which Eq.~\eqref{eq:thermoConstraint} is satisfied, while dashed lines are used for active fluids. The \com{red dotted lines} are unstable active drag responses for which Eq.~\eqref{eq:allconstraints} is violated. Such responses diverge at long time and nonlinear mechanisms should be included to obtain the corresponding large-time behavior.
    Other parameters are: $\nu_{\rm s}=1$, $\gamma_{\rm s} \eta_{\rm s}=0.3$, $G_{\rm s} \tau_0/ \eta_{\rm s}=0.25$.
	}
	\label{fig_drag_incompressible}
\end{figure*}

\medskip
\com{\noindent\textit{Broken Onsager symmetry.}
We now continue the discussion in the case of an incompressible fluid with broken Onsager symmetry, that is, $\bm\nu'\neq\bm\nu$. We define $\nu'_{\rm s,o}= \tilde{\nu}_{\rm s,o} + e_{\rm s,o}$ and $\nu_{\rm s,o}= \tilde{\nu}_{\rm s,o} - e_{\rm s,o}$, such that the Onsager symmetry is recovered for $e_{\rm s,o}=0$. Breaking Onsager symmetry modifies the conditions for stability, although these new conditions take a form similar to those given in Eq.~\eqref{eq:allconstraints}. They read:
\begin{subequations}\label{eq:allconstraints_brokenOnsager}%
\begin{align}%
    & e_{\rm o}^2 - e_{\rm s}^2 + 4 \gamma_{\rm s} \eta_{\rm s} -\tilde{\nu}_{\rm o}^2+\tilde{\nu}_{\rm s}^2 > 0 ,  \label{eq:condition1_brokenOnsager}   \\
    & 2 \eta_{\rm s} \gamma_{\rm o}^2+2 \gamma_{\rm o} \tilde{\nu}_{\rm o} \tilde{\nu}_{\rm s}+\gamma_{\rm s}\! \left(2 \gamma_{\rm s} \eta_{\rm s}-\tilde{\nu}_{\rm o}^2+\tilde{\nu}_{\rm s}^2\right) + \gamma_- > 0  , \label{eq:condition2_brokenOnsager} 
\end{align}%
where $\gamma_\pm = \gamma_{\rm s} (e_{\rm o}^2 - e_{\rm s}^2)  \pm 2 \gamma_{\rm o} e_{\rm s} e_{\rm o}$,
and 
\begin{align}%
    \Delta \! < \! 0 \,\, \text{or} \,
    \begin{cases}
        \Delta \geq 0 \, \quad 
    \text{and} \\
        \gamma_{\rm s} \!\left(\tilde{\nu}_{\rm s}^2-\tilde{\nu}_{\rm o}^2\right) \!- \! 2 \gamma_{\rm o} \tilde{\nu}_{\rm o} \tilde{\nu}_{\rm s} \!+\! 8 \eta_{\rm s}\gamma_{\rm s}^2 \!+\! \gamma_+ \!\!< \!0
    \end{cases} \!\!\! \label{eq:condition3_brokenOnsager}   
\end{align}
\end{subequations}%
where we have defined $\Delta= b_1^2-4b_0b_2$ with
$b_1 = G_{\rm s}^2 [\gamma_{\rm s}(8\gamma_{\rm s}\eta_{\rm s}-\tilde{\nu}_{\rm o}^2+\tilde{\nu}_{\rm s}^2)-2\gamma_{\rm o}\tilde{\nu}_{\rm o}\tilde{\nu}_{\rm s} + \gamma_+]/(\tau_0\eta_{\rm s})$, 
$b_2= G_{\rm s} (e_{\rm o}^2 - e_{\rm s}^2+4\gamma_{\rm s}\eta_{\rm s}-\tilde{\nu}_{\rm o}^2+\tilde{\nu}_{\rm s}^2)/(\tau_0^2\eta_{\rm s})$.
Importantly, note that $e_{\rm o}$  cannot destabilize the system. This is expected, as this coefficient enters anti-symmetrically in the constitutive equations, which causes its contribution to the entropy production to vanish. In fact, taking $e_{\rm s}=0$ in Eq.~\eqref{eq:allconstraints_brokenOnsager} shows that the constraints are more easily satisfied for $e_{\rm o}\neq0$ and $e_{\rm o}$ even has a stabilizing effect.}

\com{
On the other hand, $e_{\rm s}$ does add to the dissipation rate and therefore a non-vanishing $e_{\rm s}$ can drive the system into an unstable regime, even if the symmetric equivalent of thermodynamic constraint $2\eta_{\rm s}\gamma_{\rm s}\ge \tilde{\nu}_{\rm o}^2$ is satisfied. If $e_{\rm s}\neq0$ and $\gamma_{\rm o}\neq0$, the odd deviation from Onsager symmetry $e_{\rm o}$ can also play a role in destabilizing the system through the term $\gamma_{\rm o} e_{\rm s} e_{\rm o}$ entering Eq.~\eqref{eq:allconstraints_brokenOnsager}.  These results show the complex interplay between oddity and the breaking of Onsager symmetry in determining the stability of a probe.
}

\section{Drag and lift response}\label{sec_drag}

\subsection{Drag response for an incompressible fluid}
\label{sec:incompressible}

In this Section, we consider the response  along the velocity direction [drag, $M_{\rm \parallel} (s)$] and perpendicular to it [lift, $M_{\rm \perp } (s)$], making the decomposition
\label{sec:weaklycompressible}
\begin{align} 
\label{responsecoefficients}
    \mathbb{M}_{ij } (s  ) = M_{\rm \parallel} (s  ) \delta_{ij} - M_{\rm \perp } (s  ) \varepsilon_{ij}  \, . 
\end{align}
For simplicity, we enforce Onsager symmetry in this Section and take $\bm\nu'=\bm\nu$. 
We first consider the incompressible limit, so that $ M_{\rm \perp }^{\rm inc} (s  ) =0 $, whereas the drag response coefficient reads:
\begin{align}
  \label{eq:weaklycompressiblelimit}
 M_{\rm \parallel}^{\rm inc} (s  )    = \frac{ 1 }{4  \pi    A_{\rm s} (s) } K_0 \left( \sqrt{\frac{s \tau_0  }{ A_{\rm s} (s) /\eta_{\rm s}  }} \right)  \, ,
\end{align}
with $\tau_0=\rho_0 a^2/\eta_{\rm s}$ and $K_0(z)$ the $0^{\rm th}$ modified Bessel function of the second kind. 
The corresponding drag response in time domain can be obtained numerically (see App.~\ref{sec_Laplace_transform} for details), and the results are displayed in Fig.~\ref{fig_drag_incompressible}.

\com{In Figs.~\ref{fig_stability_b98u98u} and~\ref{fig_stability_1111}, we first explore the role of the odd elastic coefficient $\nu_{\rm o}$ and of the odd elastic strain relaxation rate $ \gamma_{\rm o}$.}
\com{For a vanishing odd elastic strain relaxation rate $ \gamma_{\rm o}$ [Fig.~\ref{fig_stability_b98u98u}], the drag response does not display oscillations, while a}
nonvanishing $\gamma_{\rm o}$ [Fig.~\ref{fig_stability_1111}] leads to a long-lived oscillatory behavior, despite strain relaxation due to plasticity.  This is a unique property that can signal odd viscoelasticity in experiments. Furthermore, we note that the amplitude of these oscillations is strongly amplified by the odd elasticity coefficient $\nu_{\rm o}$. \com{This finding highlights again the nontrivial interplay between the different odd coefficients, which} was already observed in Sec.~\ref{sec:stabilityconstraint} for the stability constraints. 
\com{It shows that increasing $\gamma_{\rm o} $ and $\nu_{\rm o} $ eventually induces instabilities.}
Specifically, the \com{dotted red line} in Fig.~\ref{fig_stability_b98u98u} (corresponding to $\nu_{\rm o}  \geq \gamma_{\rm s}^2 + \nu_{\rm s}^2 $)  corresponds to a diverging trajectory. Such a non-converging drag response should be regularized by nonlinear corrections. 

In Figs.~\ref{fig_stability_119811} and~\ref{fig_stability_771111}, we compute the drag response for a varied $\gamma_{\rm o} $ for $\nu_{\rm o}=0$ and $\nu_{\rm o}=1$, respectively. We find that an increasing $\gamma_{\rm o} $ leads to an increasing oscillation frequency. In Fig.~\ref{fig_stability_771111}, the choice of $\nu_{\rm o}$ makes the viscoelastic fluid active, and it was shown in Sec.~\ref{sec:stabilityconstraint} that this may induce instabilities for the motion of the probe particle. This is the case for the \com{dotted red line}, which violates the stability constraint of Eq.~\eqref{eq:condition2}.

\subsection{Drag and lift response for a compressible fluid}
\label{sec:weakoddity}

\begin{figure*}[t]
	\centering
    \subfigure[\label{fig_compressible_drag}]
    {\includegraphics[width=0.49\linewidth]{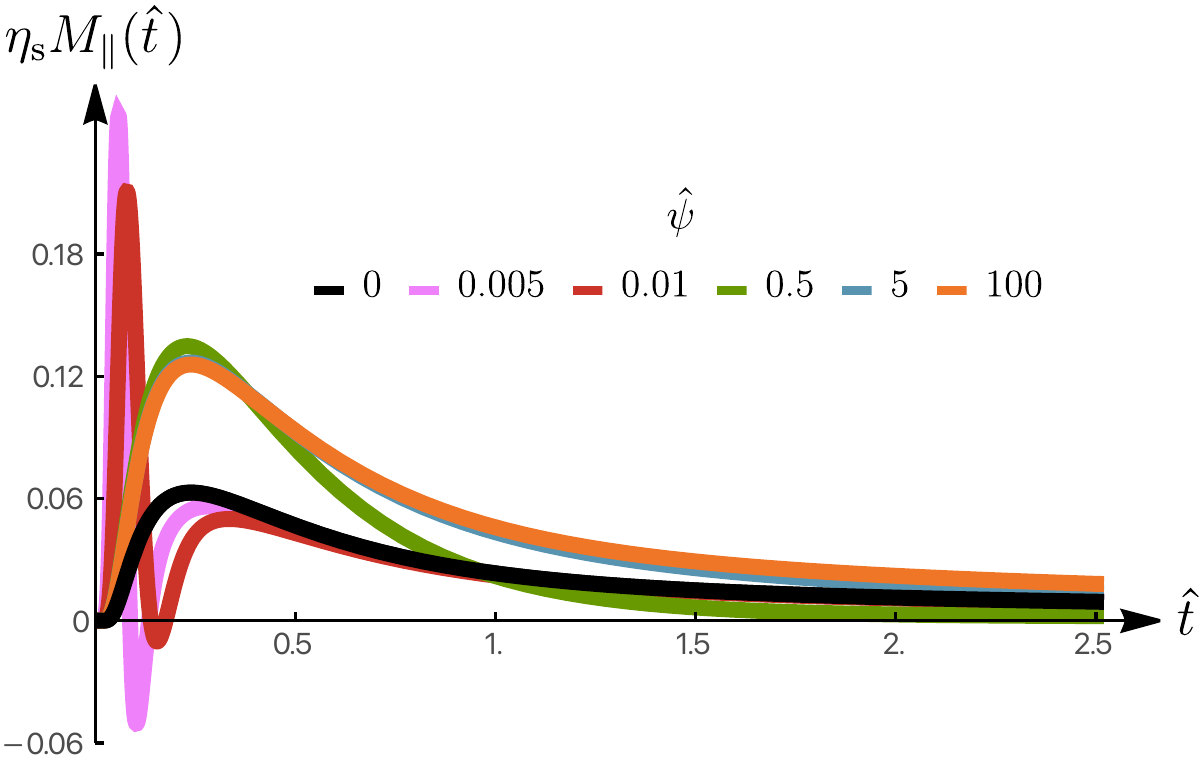}}
    \hfill
    	\subfigure[\label{fig_compressible_lift}]
    {\includegraphics[width=0.49\linewidth]{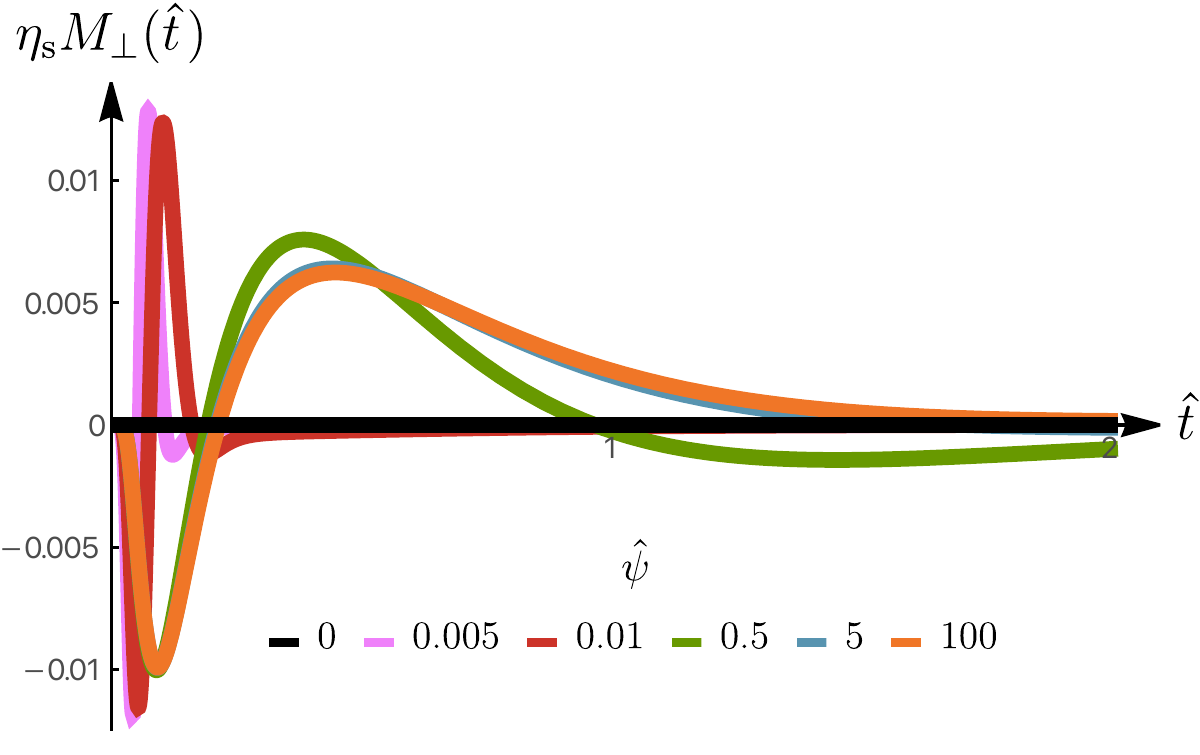}}
	\caption{Drag and lift responses of an odd compressible viscoelastic fluid induced as a function of the dimensionless time $\hat{t}=t/\tau_0$ with $\tau_0=a^2\rho_0/\eta_{\rm s}$.  \textbf{(a)} Drag response $\eta_{\rm s} M_\parallel$ for different values of the dimensionless inverse compressibility $\hat\psi = \psi\eta_{\rm s}^2/(a\rho_0)^2$. \textbf{(b)} Lift response $\eta_{\rm s} M_\perp$ for different values of the dimensionless inverse compressibility $\hat\psi$. Other parameters are: $\nu_{\rm s}=\nu_{\rm b}=1$, $\gamma_{\rm s} \eta_{\rm s}=1$, $G_{\rm s} \tau_0/ \eta_{\rm s}=1$, $\eta_{\rm b}/ \eta_{\rm s}=0$, $\gamma_{\rm b} \eta_{\rm s}=0$, $G_{\rm b} \tau_0/ \eta_{\rm s}=0$, $\eta_{\rm o}/ \eta_{\rm s}=0.1$, $\gamma_{\rm o} \eta_{\rm s}=0.1$, $\nu_{\rm o}=0$.}
	\label{fig_compressible}
\end{figure*}  

Compressibility is a necessary condition for observing lift forces~\cite{ganeshan2017odd,lier2023lift}, a signature of odd fluids. In this Section, we thus discuss how a finite compressibility of the fluid can be considered within our framework.
Although exploring systematically the role of finite compressibility for the response of a probe in an odd viscoelastic medium is beyond the scope of this paper, we illustrate here a few important features of finite compressibility.

To relax the incompressibility constraint while continuing analytic computations, 
we consider a weak oddity approximation, i.e. we consider $\eta_{\rm o},\gamma_{\rm o}$ and $\nu_{\rm o}$ small compared to their shear and bulk counterparts. We introduce $\varepsilon_{\rm o}$ the corresponding small parameter (for instance, $\varepsilon_{\rm o}=\eta_{\rm o}/\eta_{\rm s}$). We discuss here the first nonvanishing correction, although the series expansion can be continued to higher orders.
Starting from Eq.~\eqref{eq_fullResponseMatrix}, we obtain the response coefficients that were defined in Eq.~\eqref{responsecoefficients}:
\begin{subequations} \label{eq:responsematrices}%
\begin{align}%
 M_{ \rm \parallel  } (s  )  &  = M_{ \rm \parallel  }^{\rm inc} (s  ) + \frac{\Theta(s ) K_0\left( \sqrt{\tau_0  s \Theta(s ) }\right)}{ 4 \pi \eta_{\rm s}}   + \mathcal{O} (\varepsilon_{\rm o}^2 )  \,  , \label{eq_compressible_drag} \\ 
\begin{split}%
   M_{\rm \perp } (s  )   &  = 
  \frac{\tau_0 s A_{\rm o}(s)/\eta_{\rm s}}{  \hat\psi^{-1} + \tau_0 s A_{\rm b}(s)/\eta_{\rm s} } \Bigg[ 
   2 M_{ \rm \parallel  }^{\rm inc} (s) \\  &
   -  \frac{\Theta(s) K_0\left( \sqrt{\tau_0  s \Theta(s ) }\right)}{2\pi \eta_{\rm s}}    
   \Bigg]+ \mathcal{O} (\varepsilon_{\rm o}^2 ) \,  ,  
\end{split}%
\end{align}%
where we have defined the dimensionless compressibility~$\hat\psi = \psi\eta_{\rm s}^2/(a\rho_0)^2$ and 
\begin{align}
  \Theta (s )  =  \frac{ \tau_0 s}{  \hat\psi^{-1}  + \tau_0 s   \left(A_{\rm b} (s )+A_{\rm s} (s ) \right)/\eta_{\rm s} } \, . 
\end{align}
\end{subequations}%
As expected, the lift response vanishes in the incompressible limit $\hat\psi\to 0$. Note also the absence of odd corrections to the drag response [Eq.~\eqref{eq_compressible_drag}] at first order. 
 
Figure~\ref{fig_compressible} displays the drag and lift responses for different values of $\hat\psi$,  illustrating the role of a finite compressibility. Increasing the compressibility $\hat\psi$ spreads out the response in time both for the drag and lift coefficients. For small values of the compressibility, we observe a short-time response, eventually converging to the incompressible limit [black line, Fig.~\ref{fig_compressible_drag}]. 
The lift response, which is a hallmark of odd systems and thus vanishes in the absence of the odd coefficients, is shown in Fig.~\ref{fig_compressible_lift}. Note the rapid sign change of the lift coefficient, which becomes more pronounced and symmetrical as the compressibility $\hat\psi$ is decreased, leading eventually to a vanishing lift response in the incompressible limit $\hat\psi\to0$.

\section{Discussion}

In this work, we considered the stability of two-dimensional odd viscoelastic fluids by looking at the motion of probe particles that experience \com{an instantaneous} push at $t=0$. We considered contributions to the constitutive equation that are allowed for passive systems (as found in Ref.~\cite{lier2022passive}), as well as active terms that rely on fuel consumption at a microscopic level.
Such active terms modify the passive picture by breaking the two hallmarks of passive hydrodynamics: Onsager-symmetry and positive-definiteness of the Onsager matrices.
To characterize the effect of activity, we first discussed the stability condition of a probe in a generic, compressible odd viscoelastic fluid, before focusing on the incompressible limit, where the discussion is simpler while the essential ingredients are kept. Importantly, \com{although the lift force vanishes in this limit~\cite{abanov2018odd}}, odd viscoelastic effects manifest themselves spectacularly as they can destabilize the probe motion. 
We find that passive odd viscoelastic fluids, where viscoelastic coefficients are constrained by the Second Law of Thermodynamics,
are always linearly stable. We find the converse statement that active odd viscoelastic fluids are always unstable not to be true.
Specifically, we find that there is an intermediate region where the motion of probe particles is stable, despite the system being active and constantly consuming fuel. In this intermediate regime, there is a non-trivial interplay between the odd elasticity coefficient $\nu_{\rm o}$ and odd plasticity $\gamma_{\rm o}$, which is graphically displayed in Fig.~\ref{fig_stability_and_thermo}. This interplay is even more surprising considering that odd plasticity is a non-dissipative phenomenon and therefore plays no role in determining the passivity of an odd viscoelastic fluid~\cite{lier2022passive}. 
Similarly, the non-dissipative shear elastic coefficient $\nu_{\rm s}$  is involved in stability constraints despite being non-dissipative.

The stability of viscoelastic models was considered in previous works by looking at the stability of perturbative modes. In Ref.~\cite{lier2022passive}, a small wave-vector analysis yielded a constraint identical to Eq.~\eqref{eq:condition2}, which is one of the three constraints obtained in this work for stable probe particle motion that are given in Eq.~\eqref{eq:allconstraints}. 
In Ref.~\cite{floyd2022signatures}, odd viscoelastic materials were studied in the context of the formation of spatial-temporal patterns, whose presence indicates instability. The authors performed a numerical analysis of the perturbative modes to determine their stability for a range of wave vectors.
The numerical findings of Ref.~\cite{floyd2022signatures}, showing that odd elasticity can destabilize the viscoelastic material, whereas shear elasticity, plasticity and viscosity tend to stabilize the viscoelastic material qualitatively agree with the analytic constraints derived in the present work. Such a qualitative finding is also consistent with the analytically derived stability constraint of Eq.~\eqref{eq:condition2} found in Ref.~\cite{lier2022passive}.

In this work, we furthermore considered the effect of several odd viscoelastic coefficients on probe particle motion. The obtained results provide a viscoelastic extension of Ref.~\cite{ganeshan2017odd} for the incompressible case and of Refs.~\cite{hosaka2021nonreciprocal,lier2023lift} for the compressible case, which considered lift and drag force for an odd viscous fluid. When generalizing to viscoelastic fluids, one must consider inhomogeneities in time to see qualitatively different effects, as at vanishing frequencies only viscous effects remain. For this reason, we only considered the motion of a probe particle after a push at $t =0 $. A computational advantage of doing so is that there is no problem arising from the Stokes paradox~\cite{veysey2007simple}, which for the steady case was resolved in Refs.~\cite{hosaka2021nonreciprocal,lier2023lift} by regularizing the integral with momentum relaxation coming from the three-dimensional bulk.
We first considered the incompressible limit, which is a limit that is decoupled from passivity constraints and therefore a useful limit for considering active destabilizing effects. One key qualitative feature that we find is that the relaxing velocity becomes oscillatory for a non-zero odd plasticity coefficient $\gamma_{\rm o}$ and that the amplitude of this oscillation is amplified by the odd elasticity coefficient $\nu_{\rm o}$. This again highlights the non-trivial interplay between these two odd viscoelastic coefficients, and is illustrated in Fig.~\ref{fig_drag_incompressible}.
\com{We also}
discussed the role of compressibility in the response of the probe. In this case, we showed that the odd lift force, transverse to the direction of motion of the probe particle, that exists for viscous odd fluids, becomes time-dependent in viscoelastic systems. We discussed this case focusing on a small oddity limit, where analytical results can be obtained, and the results are displayed in Fig.~\ref{fig_compressible}. In the compressible regime, probes placed in a linearly unstable odd viscoelastic fluid are likely to perform limit cycles in the plane. 
Exploring these limit cycles, and more generally the probe motion in the linearly-unstable regime is an interesting aspect which requires introducing nonlinear terms whose symmetry and time-dependence need to be discussed. We leave this work for future publications.

\com{Lastly, chiral biological systems such as cell assemblies or tissues are good candidates for displaying simultaneously odd, active, and viscoelastic behaviors. In this context, a layer of such cells is likely to exchange momentum with the three-dimensional bulk (for instance, an extra-cellular matrix in the case of an epithelial tissue, or sea water in the case of starfish oocytes~\cite{tan2022odd}). In addition, cell number is usually not conserved in such systems since cell divisions, deaths or extrusions can occur. Although we have not considered finite momentum or density relaxation processes in this paper, taking them into account would be straightforward~\cite {elfring2016surface,stone1990simple} in our formalism. The effects of such terms have been discussed for purely viscous odd fluids in Ref.~\cite{lier2023lift}.}

{\it Acknowledgements.---} 
The authors wish to acknowledge Debarghya Banerjee, Gwynn Elfring, Alexander Morozov and Tzer Han Tan for inspiring discussions. 
C.D. acknowledges the support of the LabEx “Who~Am~I?” (ANR-11-LABX-0071) and of the Université Paris Cité IdEx (ANR-18-IDEX-0001) funded by the French Government through its “Investments for the Future” program.
PS was supported by the DFG through the Leibniz Program and the cluster of excellence ct.qmat (EXC 2147, project-id 39085490) and the National Science Centre (NCN) Sonata Bis grant 2019/34/E/ST3/00405. 

\onecolumngrid

\appendix

\section{Constitutive equations for active odd materials}\label{app:const_eq}

 In this Appendix, we will derive the most general constitutive equations for the conserved currents by constraining entries using the Onsager relations and the Second Law of Thermodynamics. In Sec.~\ref{passivechiralviscoelasticfluid}, we first consider the passive case, following to a large extent the steps performed in Ref.~\cite{callan-jones2011hydrodynamics}. This analysis will yield a result that coincides with Ref.~\cite{lier2022passive} where a general model for passive odd viscoelasticity was constructed. Then we consider the active case in Sec.~\ref{activechiralviscoelasticfluid}. Here, like in Ref.~\cite{julicher2018hydrodynamic}, we add fuel consumption, which allows for a much broader class of non-equilibrium corrections, for which we consider the effect of some of them on the motion of beads in the main text.

\subsection{Passive chiral viscoelastic fluid}
\label{passivechiralviscoelasticfluid}

In order to hydrodynamically describe a viscoelastic fluid we first formulate the conservation laws in absence of external forces and torques. Firstly, there is conservation of mass density $\rho $ and momentum $g_i$
\begin{subequations} \label{eq:conservationn}
\begin{align}
  \partial_t \rho + \partial_{i} (\rho v_i ) 
  & =0    ~~ ,  \\ 
       \partial_t g_i  -  \partial_j \sigma^{\rm tot}_{ij}   &  = 0 ~~  ,  
  \end{align}
where $\sigma^{\rm tot}_{ij}$ is the total stress, $v_i$ is the fluid velocity. Momentum is related to fluid velocity as $g_i = \rho v_i$. In addition, we have conservation of angular momentum, given by
 \begin{align} \label{eq:angularmomentumconservation}
   \partial_t l_{ij} +  \partial_k M_{ijk}  &  = -2 \sigma^{\rm tot}_{[ij]}   ~~ , 
\end{align}    
\end{subequations}
 $l_{ij}$ the density of intrinsic angular momentum and $M_{ijk}$ is the angular momentum flux. Another property that is important for describing the state of viscoelastic fluids the object $u_{ij}$. In the elastic limit, $u_{ij}$ is the elastic strain, however for general viscoelastic fluids $u_{ij}$ can undergo relaxation through plastic deformations \cite{fukuma2011relativistic,azeyanagi2009universal}. To study the evolution of $u_{ij}$ and the behavior the viscoelastic fluid in general, we start with formulating a local free energy~\cite{julicher2018hydrodynamic,callan-jones2011hydrodynamics}:
\begin{align}
    \mathcal{F} = \int \left(\frac{1}{2}\rho v^2 + f(u_{ij}) \right) \dd V \, ,
\end{align}
where $f(u_{ij})$ the local free energy density, which depends on $u_{ij}$. $u_{ij}$ is conjugate to the local elastic stress $\sigma^{\rm el}_{ij}$, such that
$    \sigma^{\rm el}_{ij} = \partial f/\partial u_{ij} $. 
 For an isotropic solid, the local free energy density at leading order in strain is given by
        \begin{align}
    f(u_{ij}) = f_0  +  G_{\rm s} ( \tilde u_{ij} )^2 + \frac{1}{2}  G_{\rm b} (u_{jj})^2   \, .
    \label{eq_hookean}
\end{align}
To derive the constitutive equations for the viscoelastic fluid using the standard hydrodynamic approach, we decompose our stress into 
\com{\begin{align}
\sigma^{\rm tot}_{ij}    = - \rho v_i v_j - P \delta_{ij}   + \sigma^{\rm s}_{ij} + \sigma^{\rm a}_{ij}  \, ,
\end{align}
where $\sigma^{\rm s}_{ij}$ is the deviatoric stress, which is symmetric, and $\sigma^{\rm a}_{ij} $ is the anti-symmetric contribution contribution to the stress.} The pressure $P$ is obtained with the Euler relation
\begin{align}
    P = - f  + \mu n ~~ , 
\end{align}
with $\mu$ being the chemical potential and where the density $n$ is given by $n = \rho /m $. If one considers the change of time of $F $, one finds upon plugging in the conservation laws of Eq.~\eqref{eq:conservationn} that the entropy production rate $\Theta$ is given by \cite{julicher2018hydrodynamic,callan-jones2011hydrodynamics}
\begin{align}
    T\Theta = \sigma^{\rm s}_{ij} v_{ij} - \DDt{u_{ij}} \sigma^{\rm el}_{ij} +  \frac{1}{2} \partial_k \omega_{ij} M_{ijk}  \, ,
    \label{eq_entropyProduction_chiral}
\end{align}
where we introduced the rotation \com{$\omega_{ij} = \frac{1}{2}  ( \partial_{i} v_j- \partial_{j} v_i )  $} as well as the corotational derivative which for a general two-tensor $a_{ij}$ is given by
\begin{align}
  \DDt{ a_{ij}} = \partial_t a_{ij} + v_k \partial_k a_{ij} + \omega_{i k } a_{kj } + \omega_{j k } a_{i k  } ~~ . 
\end{align}
\com{The presence of the corotational derivative in Eq.~\eqref{eq_entropyProduction_chiral} follows from the requirement that the free energy is invariant under rotations \cite{callan-jones2011hydrodynamics}. To have a corotational derivative on the strain field requires the anti-symmetric stress to take the form $\sigma^{\rm a}_{ij}  =    u_{ik} \sigma^{\rm el}_{kj} -   u_{j k} \sigma^{\rm el}_{ki}  $.}
Because of the conservation of angular momentum given by Eq.~\eqref{eq:angularmomentumconservation}, there is no contribution coming from the anti-symmetric stress in the entropy production rate of Eq.~\eqref{eq_entropyProduction_chiral} \cite{salbreux2017mechanics}. Instead, there is only a contribution from the deviatoric angular momentum flux $M_{ijk}$, which will  give contributions to the equations of motion that are two orders higher in gradients, and these contributions will therefore be omitted. Note that because odd viscoelastic fluids are often introduced by chiral agents that draw arbitrary amounts of angular momentum from the environment, the conservation of angular momentum tends not to hold. In the absence of angular momentum conservation, there are ways in which the anti-symmetric stress can get hydrodynamic corrections. However, we will omit these anti-symmetric contributions for simplicity by upholding angular momentum conservation. Having the entropy production rate of Eq.~\eqref{eq_entropyProduction_chiral} at our disposal, we are now ready to construct the Onsager matrix with containing the most general entries in the deviatoric currents. An important difference compared to the isotropic achiral viscoelastic fluids in two dimensions is that, in addition to the identity, the fully anti-symmetric two-dimensional Levi-Civita tensor $\varepsilon_{ij}$ can be used to construct entries in the constitutive equations. As we will discuss below, these tensors can feature components that are odd under the exchange of indices. This can for example give rise to odd viscosity \cite{avron1995viscosity,levay1995berry,jensen2012parityviolating,soni2019odd}. Such terms signal parity-breaking, and for passive two-dimensional systems, one often finds that this breaking of parity is accompanied by a breaking of time-reversal symmetry, as is for example the case for parity-breaking induced by a background magnetic field. We therefore uphold symmetry under the simultaneous transformation of parity and time-reversal when deriving the passive constitutive equations, which modify the Onsager relations. The details of this are provided in App.~\ref{sec_Onsager_odd}.  From the expression of entropy production \eqref{eq_entropyProduction_chiral} and the Onsager relations, we obtain the following constitutive relations:
\begin{subequations}%
\begin{align}%
     \sigma^{\rm s}_{ij} &= 2\eta^{\rm p}_{\rm s}  \tilde v_{ij} + \eta^{\rm p}_{\rm b}v_{kk} \delta_{ij} + 2\eta^{\rm p}_{\rm o}  v^{\rm o}_{ij}+\nu^{\rm p}_{\rm s} \sigma^{\rm el}_{ij} + \frac{1}{2}\nu^{\rm p}_{\rm b} \sigma^{\rm el}_{kk} \delta_{ij}  + \nu^{\rm p}_{\rm o} (\sigma^{\rm el})^{\rm o}_{ij} \, , \label{eq_chiral_sigmaSym} \\
    \DDt{ u_{ij}} &= -\gamma^{\rm p}_{\rm s} \tilde\sigma^{\rm el}_{ij} -\frac{1}{2}  \gamma^{\rm p}_{\rm b} \sigma^{\rm el}_{kk} \delta_{ij}-\gamma^{\rm p}_{\rm o} ( \sigma^{\rm el})^{\rm o}_{ij} + \nu^{\rm p}_{\rm s}   v_{ij} + \frac{1}{2} \nu^{\rm p}_{\rm s} v_{kk} \delta_{ij}  + \nu^{\rm p}_{\rm o}  {v}^{\rm o}_{ij} \, . \label{eq_chiral_dudt}
\end{align} \label{eq_chiral_constitutiveEquation}%
\end{subequations}
What we see in Eq.~\eqref{eq_chiral_constitutiveEquation} are the shear, bulk and odd viscosity given by the $\eta$-terms, as well as $\gamma$-terms representing shear and bulk plasticity, as well as a coefficient that could be called \quotes{odd plasticity}. This odd plasticity was first considered in Ref.~\cite{https://doi.org/10.48550/arxiv.1907.07187}. Then, there are $\nu$-terms which are non-equilibrium corrections representing shear, bulk and odd elasticity, the latter of which was first considered as an active term in Ref.~\cite{scheibner2020odd}. Using the compact notation described in App.~\ref{sec_rank4_algebra}, Eq.~\eqref{eq_chiral_constitutiveEquation} can be rewritten as
\begin{subequations}%
\begin{align}%
    \tens{\sigma}^{\rm s} &= \bm\nu^{\rm p} : \tens{\sigma}^{\rm el}  + 2\bm\eta^{\rm p} : \tens{v} \, , \label{eq_chiral_sigmaSym1} \\
    \DDt{\tens{u}} &= -\bm\gamma^{\rm p} : \tens{\sigma}^{\rm el} + \bm\nu^{\rm p} : \tens{v} \, .  \label{eq_chiral_dudt1}
\end{align} \label{eq_chiral_constitutiveEquationcompact}%
\end{subequations}
To see the dissipative properties of the entries in the constitutive equations of Eq.~\eqref{eq_chiral_constitutiveEquationcompact}, we plug Eq.~\eqref{eq_chiral_constitutiveEquation} back into Eq.~\eqref{eq_entropyProduction_chiral}, which yields
\begin{align}
    T\Theta = {\rm Tr}\left[2\eta^{\rm p}_{\rm s} \tilde{\tens{v}} \cdot \tilde{\tens{v}}+
    2 \nu^{\rm p}_{\rm o}  \left(\tilde{\tens{\sigma}}^{\rm el}\cdot \tens{\varepsilon} \cdot \tilde{\tens{v}}  \right)+
    \gamma^{\rm p}_{\rm s} \tilde{\tens{\sigma}}^{\rm el} \cdot\tilde{\tens{\sigma}}^{\rm el}
    \right]  + 2 \eta_{\rm b}^{\rm p} \Tr[v]^2 + \gamma_{\rm b}^{\rm p} \Tr[v]^2  ~~ .  \label{eq_entropyProduction_chiral_explicit}
\end{align}
Requiring the first term on the right hand side of Eq.~\eqref{eq_entropyProduction_chiral_explicit} to be non-negative is equivalent to requiring that the matrix
\begin{align}
    \begin{pmatrix}    2\eta^{\rm p}_{\rm s} 
 \tens{\mathbbm{1} } & -  \nu^{\rm p}_{\rm o}\tens{\varepsilon}  \\ 
\nu^{\rm p}_{\rm o}\tens{\varepsilon}  & \gamma^{\rm p}_{\rm s} \tens{\mathbbm{1} } 
  \end{pmatrix}
\end{align}
is semi-definite positive, which is satisfied if all its eigenvalues have non-negative real part. As was found in Ref.~\cite{lier2022passive}, this condition is guaranteed if, in addition to $\eta^{\rm p}_{\rm s} , \gamma^{\rm p}_{\rm s}  , \eta^{\rm p}_{\rm b} , \gamma^{\rm p}_{\rm b}  \geq 0 $, we have
\begin{align} \label{eq:passivityconstraint}
2\eta^{\rm p}_{\rm s} \gamma^{\rm p}_{\rm s}\ge\left(\nu^{\rm p}_{\rm o}\right)^2 ~~,   \end{align}
which is the thermodynamic constraint given in Eq.~\eqref{eq:passiveconstraint1}.

\subsection{Active chiral viscoelastic fluid}\label{activechiralviscoelasticfluid}

The passive model we have introduced above can be generalized by considering activity induced by microscopic fuel consumption of microscopic agents \cite{julicher2018hydrodynamic}. The rate of fuel consumption can be represented by $\Delta\mu$, which is the chemical potential difference of a chemical reaction, multiplied by a scalar reaction rate $r$. Because of this fuel consumption, the entropy production rate of Eq.~\eqref{eq_entropyProduction_chiral} is generalized to
\begin{align}
    T\Theta = \sigma^{\rm s}_{ij} v_{ij} - \DDt{u_{ij}} \sigma^{\rm el}_{ij}  +  \frac{1}{2} \partial_k \omega_{ij} M_{ijk}  +  r \Delta \mu  ~~ , 
    \label{eq_entropyProduction_chiral_active}
\end{align}
This new contribution opens up a wide range of new possible entries in the constitutive equations, which are given by 
\begin{subequations}%
\begin{align}%
    \tens{\sigma}^{\rm s} &= \bm\nu^{\rm p} : \tens{\sigma}^{\rm el}  + 2\bm\eta^{\rm p} : \tens{v} + \Delta\mu \left(  \bm\zeta^{(1)} : \tens{\sigma}^{\rm el} + 2 \bm\zeta^{(2)} : \tens{v} \right) \, ,  \\
    \DDt{\tens{u}} &= -\bm\gamma^{\rm p} : \tens{\sigma}^{\rm el} + \bm\nu^{\rm p} : \tens{v} + \Delta\mu \left( -  \bm\zeta^{(3)} : \tens{\sigma}^{\rm el} +  \bm\zeta^{(4)} : \tens{v} \right) \, . 
\end{align} \label{eq_chiral_active_constitutiveEquation_ran9899898Sk4}%
\end{subequations}
The active terms ($\propto \Delta\mu$)  can be absorbed into a redefinition of the parameters so that constitutive equations read:
\begin{subequations} 
\begin{align}%
    \tens{\sigma}^{\rm s} &=  \bm\nu : \tens{\sigma}^{\rm el}  + 2\bm\eta : \tens{v}  \, ,  \\
 \DDt{\tens{u}} &= -\bm\gamma : \tens{\sigma}^{\rm el} + \bm\nu^{\prime} : \tens{v} \, ,
\end{align}\label{eq_constitutiveEquationSimple989898}%
\end{subequations}%
which is Eq.~\eqref{eq_constitutiveEquationSimple} in the main text.
The new  parameters are related to the passive ones of Eq.~\eqref{eq_chiral_active_constitutiveEquation_ran9899898Sk4} by  $\bm\eta=\bm\eta^{\rm p}+\Delta\mu\bm\zeta^{(2)}$, $\bm\gamma=  \bm\gamma^{\rm p}+\Delta\mu\bm\zeta^{(3)}  $, $\bm\nu=  \bm\nu^{\rm p}+\Delta\mu\bm\zeta^{(1)}  $ and $\bm\nu'=  \bm\nu^{\rm p}+\Delta\mu\bm\zeta^{(4)}  $. Note that $\Delta \mu$, as can be seen in Eq.~\eqref{eq_entropyProduction_chiral_active}, is a thermodynamic force \cite{julicher2018hydrodynamic} corresponding to the flux $r$. Because $\Delta \mu$ is a thermodynamic force, it should be a small correction with respect to the local equilibrium. This means that the $\Delta \mu \tens{\sigma}^{\rm el}$- and $\Delta \mu \tens{v}$-terms are second order, and one would therefore naively think that these terms should be neglected. However, these second-order terms play an important qualitative role as the effective coefficients appearing in $\bm\eta,\bm\gamma,\bm\nu,\bm\nu'$ are not bound by the passivity constraint~\eqref{eq:passivityconstraint}. In particular, this allows for the presence of active odd elasticity. 

To make sure that the second law constraint $\Theta \geq 0$ in Eq.~\eqref{eq_entropyProduction_chiral_active} as well as the Onsager relations described in App.~\ref{sec_Onsager_odd} are satisfied, we require the corresponding reaction rate to be
\begin{align}
    r &   =   \Lambda  \Delta \mu   + \Tr[    - \tens{\sigma}^{\rm el} :  \bm\zeta^{(1)} :  \tens{v}  +  2  \tens{v}:  \bm\zeta^{(2)} : \tens{v} +   \tens{\sigma}^{\rm el}: \bm\zeta^{(3)} : \tens{\sigma}^{\rm el}  + \tens{v}  : \bm\zeta^{(4)} :  \tens{\sigma}^{\rm el}  ]  ~~ .  
\end{align}

\section{Compact tensor notation and odd isotropic rank-4 tensor algebra}
\label{sec_rank4_algebra}

In this Appendix, we describe the notation used for describing tensors and their contractions. First, we describe vectors and two-tensors with a single and a double underline, i.e. 
\begin{align}
V_i \rightarrow   \vect{V} ~~ ,~~ T_{ij} \rightarrow \tens{T}  ~~ . 
\end{align}
We define the components of an odd isotropic rank-4 tensor $\bm B$ in the following way:
\begin{align}
    B_{ijkl}= \frac{B_{\rm s}}{2} \left(\delta_{ik} \delta_{jl} + \delta_{il} \delta_{jk} - \delta_{ij} \delta_{kl} \right) 
    + \frac{B_{\rm b}}{2} \delta_{ij} \delta_{kl}
    + \frac{B_{\rm o}}{4}  \left(
    \delta_{ik} \varepsilon_{jl} +
    \delta_{il} \varepsilon_{jk} +
    \delta_{jk} \varepsilon_{il} +
    \delta_{jl} \varepsilon_{ik} 
    \right) \, .
\end{align}
These tensors are symmetric with respect to their first two and last two indices, such that \mbox{$B_{ijkl}= B_{jikl}$} and \mbox{$B_{ijkl} =B_{ijlk}$}. The contraction of an odd isotropic rank-4 tensor $\bm B$ with an arbitrary rank-2 tensor $\tens{b}$ reads:
\begin{align}
    (\bm B : \tens{b} )_{ij} = B_{ijkl} b_{kl} = B_{\rm s} \tilde b_{ij} + \frac{B_{\rm b}}{2} {\rm Tr}(b) \delta_{ij} + B_{\rm o}  \tilde b^{\rm o}_{ij}  \, ,  \label{eq_contraction_rank2}
\end{align}
where
\begin{align}
{\rm Tr}(b)  &   =   b_{jj} \, , \quad \tilde b_{ij}   =  \frac{1}{2} \left( b_{ij} +  b_{ji}\right) - \frac{1}{2}b_{kk}\delta_{ij}    \, , \quad     \tilde b^{\rm o}_{ij}    = \varepsilon_{ik} \tilde{b}_{kj}  \, . 
\end{align}    
One important relation that is used in App.~\ref{app:const_eq} is
\begin{align}
    \tens{a} : \bm A : \tens{a} = A_{\rm s} (\tilde a_{ij} )^2 +  \frac{ 1}{2} A_{\rm b}  ( {\rm Tr}(a) )^2 ~~,\label{eq_full_contraction}  
\end{align}
where we see that the odd contribution drops out due to its anti-symmetric nature. Consistently with the definition~\eqref{eq_contraction_rank2}, we define the contraction $\bm C= \bm A \bm B $ of two tensors $\bm A$ and $\bm B$ in terms of their components as: 
\begin{align}
    C_{ijkl} = A_{ijmn} B_{mnkl} \, . \label{eq_multiplication_convention}
\end{align}
Odd isotropic rank-4 tensors form a commutative group under multiplication since
\begin{align}
    C_{ijkl} = A_{ijmn} B_{mnkl} = \frac{C_{\rm s}}{2} \left(\delta_{ik} \delta_{jl} + \delta_{il} \delta_{jk} - \delta_{ij} \delta_{kl} \right) 
    + \frac{C_{\rm b}}{2} \delta_{ij} \delta_{kl}
    + \frac{C_{\rm o}}{4}  \left(
    \delta_{ik} \varepsilon_{jl} +
    \delta_{il} \varepsilon_{jk} +
    \delta_{jk} \varepsilon_{il} +
    \delta_{jl} \varepsilon_{ik} 
    \right) \, ,
\end{align}
where
\begin{align}
    C_{\rm s} = A_{\rm s}B_{\rm s}- A_{\rm o} B_{\rm o} \, , \quad C_{\rm b} = A_{\rm b} B_{\rm b}  \, , \quad C_{\rm o} = A_{\rm o}B_{\rm s} + A_{\rm s} B_{\rm o} \, ,
\end{align}
where we have used the relation $\varepsilon_{ij}\varepsilon_{kl}=\delta_{ik}\delta_{jl}-\delta_{il}\delta_{jk}$.
The identity tensor $\bm I$ is defined component-wise as:
\begin{align}
    I_{ijkl} = \frac{1}{2} \left(\delta_{ik} \delta_{jl} + \delta_{il} \delta_{jk} \right)  \, ,
\end{align}
and the inverse $\bm N=\bm M^{-1}$ of $\bm M$ satisfies $\bm M^{-1}\bm M= \bm M \bm M^{-1} = \bm I $ and has the following elements:
\begin{align}
    N_{\rm s} = \frac{M_{\rm s}}{M_{\rm s}^2 + M_{\rm o}^2} \, , \quad N_{\rm b}=  \frac{1}{M_{\rm b}}\, , \quad  N_{\rm o} = -\frac{M_{\rm o}}{M_{\rm s}^2 + M_{\rm o}^2} \, .
\end{align}

\section{Onsager relations with odd isotropic rank-4 tensors}
\label{sec_Onsager_odd}

In this appendix, we derive Onsager relations in the case where the state variables   $\bmP^\alpha$ of the system are rank-2 symmetric tensors. We define their conjugate forces as $\bmf^\alpha = - \delta \mathcal{F}/\delta \bmP^\alpha$ where $\mathcal{F}[\bmP]$ is the free energy of the system. The time derivative of the free energy can therefore be written as:
\begin{align}
    \dot{\mathcal{F}}=- \sum_\alpha \dot{\Phi}_{ij}^{\alpha} f_{ij}^\alpha \, ,
\end{align}
where the summation over repeated Latin indices is implicit, while we write explicitly the summation over Greek indices. Considering linear relations between thermodynamics fluxes and forces read:
\begin{align}
    \dot{\bmP}^{\alpha} = \sum_\beta \bm{\cL}^{\alpha\beta} :  \bmf^{\beta} \, , 
\end{align}
where each $\bm{\mathcal{L}}^{\alpha\beta}$ is an odd isotropic rank-4 tensor. Writing explicitly the components, we have:
\begin{align}
    \dot{\Phi}_{ij}^{\alpha} =\sum_\beta \cL_{ijkl}^{\alpha\beta}  f_{kl}^{\beta} \, , \label{eq_Onsager}
\end{align}
We deduce:
\begin{align}
    \dot{\mathcal{F}}=-\sum_{\alpha\beta}(\bm{\cL}^{\alpha\beta} : \bmf^{\beta})_{ij} f_{ij}^\alpha = - \sum_{\alpha\beta} \cL_{ijkl}^{\alpha\beta} f_{ij}^\alpha f_{kl}^\beta \, ,
\end{align}
from which we deduce that the odd part of the diagonal elements of $\bm \cL$ (i.e. the coefficients $\cL^{\alpha\alpha}_{\rm o}$) do not contribute to the rate of change of the free energy as a consequence of Eq.~\eqref{eq_full_contraction}. We now derive the symmetries of the Onsager matrix $\bm{\mathcal{L}}^{\alpha\beta}$ under time reversal. For this purpose, we first introduce the partition function:
\begin{align}
    \mathcal{Z} =  \int \cD \bmP \, \rme^{-\mathcal{F}[\bmP]/k_B T} \, ,
\end{align}
that can be used to compute correlation functions. We have:
\begin{align}
\begin{split}
    \langle  \bmP^\alpha \bmf^\beta\rangle &= \frac{1}{\mathcal{Z}} \int \cD \bmP \, \bmP^\alpha \bmf^\beta \rme^{-\mathcal{F}[\bmP]/k_B T} \, , \\
    &= -\frac{1}{\mathcal{Z}} \int \cD \bmP \, \bmP^\alpha \frac{\delta F}{\delta \bmP^\beta}  \rme^{-\mathcal{F}[\bmP]/k_B T} \, , \\
    &= \frac{k_B T}{\mathcal{Z}} \int \cD \bmP \, \bmP^\alpha  \frac{\delta}{\delta \bmP^\beta} \rme^{-\mathcal{F}[\bmP]/k_B T} \, , \\
    &= -\frac{k_B T}{\mathcal{Z}} \int \cD  \bmP \frac{\delta \bmP^\alpha}{\delta \bmP^\beta} \rme^{-\mathcal{F}[\bmP]/k_B T} \, , \\
    &= - k_B T \delta^{\alpha\beta} \bm{I} \, .
\end{split} \label{eq_correlation_sigmaf}
\end{align}
Note that the same procedure can be used to obtain relations for higher order correlations functions. Considering for simplicity scalar quantities, we have the generic expression:
\begin{align}
    \langle \phi^{\alpha_1} \cdots \phi^{\alpha_N} f^\beta \rangle = -k_B T \left(
    \delta^{\alpha_1\beta} \langle\phi^{\alpha_2} \cdots \phi^{\alpha_N} \rangle +
    \delta^{\alpha_2\beta} \langle \phi^{\alpha_1} \phi^{\alpha_3} \cdots \phi^{\alpha_N} \rangle
    + \cdots
    \right) \, .
\end{align}
The last equality of Eq.~\eqref{eq_correlation_sigmaf} is best derived using indices:
\begin{align}
    \frac{\delta \Phi_{ij}^\alpha}{\delta \Phi_{kl}^\beta} = \frac{1}{2} \left(\delta_{ik} \delta_{jl} + \delta_{il} \delta_{jk} \right)  \delta^{\alpha\beta} = I_{ijkl}  \delta^{\alpha\beta} \, ,
\end{align}
where we have used the fact that the $\bmP^\alpha$ are symmetric tensors. The second to last equality in Eq.~\eqref{eq_correlation_sigmaf} has been obtained using an integration by parts. The boundary terms vanish with the assumption that $F$ diverges at infinity. Using Eqs.~\eqref{eq_Onsager} and~\eqref{eq_correlation_sigmaf} , we can now compute:
\begin{align}
\begin{split}
    \langle \Phi_{ij}^\alpha \dot{\Phi}_{kl}^\beta \rangle &= \langle \Phi_{ij}^\alpha \cL_{klmn}^{\beta\gamma} f_{mn}^\gamma
    \rangle  \, ,\\
    &= \cL_{klmn}^{\beta\gamma} \langle \Phi_{ij}^\alpha f_{mn}^\gamma
    \rangle  \, ,\\
    &= - k_B T \cL_{klij}^{\beta\alpha} \, .
\end{split}
\end{align}
The same correlation function can also be computed using time reversal. We have:
\begin{align}
\begin{split}
    \langle \bmP^\alpha \dot{\bmP}^\beta \rangle &= \underset{\Delta t\to 0}{{\rm lim}} \frac{1}{\Delta t} \left( 
    \langle \bmP^\alpha(t) \bmP^\beta(t+\Delta t) \rangle - \langle \bmP^\alpha(t) \bmP^\beta(t) \rangle 
    \right) \, , \\
    &= \underset{\Delta t\to 0}{{\rm lim}} \frac{1}{\Delta t} \left( 
    \langle \bmP^\alpha(t - \Delta t) \bmP^\beta(t) \rangle - \langle \bmP^\alpha(t) \bmP^\beta(t) \rangle 
    \right) \, ,\\
    &= \underset{\Delta t\to 0}{{\rm lim}} \frac{\epsilon^\alpha \epsilon^\beta }{\Delta t} \left( 
    \overline{\langle \bmP^\alpha(-t + \Delta t) \bmP^\beta(-t) \rangle} - \overline{\langle \bmP^\alpha(-t) \bmP^\beta(-t) \rangle} 
    \right) \, ,\\
    &= \underset{\Delta t\to 0}{{\rm lim}} \frac{\epsilon^\alpha \epsilon^\beta}{\Delta t} \left( 
    \overline{\langle \bmP^\alpha(t + \Delta t) \bmP^\beta(t) \rangle} - \overline{\langle \bmP^\alpha(t) \bmP^\beta(t) \rangle} 
    \right) \, ,\\
    &= \epsilon^\alpha \epsilon^\beta \overline{\langle \dot{\bmP}^\alpha \bmP^\beta \rangle} \, ,
\end{split}
\end{align}
where $\epsilon^\alpha$ is the signature of $\bmP^\alpha$ under time reversal
and the overline $\overline{\cdots}$ indicates that the sign of some parameters (for instance magnetic field) has to be changed. We thus deduce the Onsager matrix symmetry:
\begin{align}
    \cL_{ijkl}^{\alpha\beta} = \epsilon^\alpha \epsilon^\beta  \overline{\cL}_{klij}^{\beta\alpha}  \, .
    \label{eq_OnsagerSymmetry_appendix}
\end{align}
It yields in particular:
\begin{align}
    \cL_{ijkl}^{\alpha\alpha} = \overline{\cL}_{klij}^{\alpha\alpha}  \, ,
\end{align}
which means that an equilibrium odd viscosity needs to change sign under time reversal to be non-vanishing. This is consistent for instance with odd viscosity \cite{avron1995viscosity,levay1995berry,jensen2012parityviolating,soni2019odd}, which is proportional to the magnetic field and therefore changes sign upon time reversal. Assuming a sign change upon time-reversal  for all  odd coefficients thus leads to the following structure for the Onsager matrix:
\begin{align}
     \bm\cL^{\alpha\beta} = 
\begin{pmatrix}
     \bm\eta^{\rm p} & -\bm\nu^{\rm p} \\
     \bm\nu^{\rm p} & \bm\gamma^{\rm p}
\end{pmatrix}^{\alpha\beta} \, ,
\end{align}
where $\bm\eta^{\rm p}$, $\bm\nu^{\rm p}$ and $\bm\gamma^{\rm p}$ are isotropic rank-4 tensors that include odd and even parts. This yields Eq.~\eqref{eq_chiral_constitutiveEquationcompact}.

\section{Shell localization}
\label{app:shelllocalization}

In this Appendix, we explain how shell localization is used to obtain the response matrix of Eq.~\eqref{eq_responseMatrix}. For this, we first invert Eq.~\eqref{eq:matrixform111} as
\begin{align}
   v_i (\mathbf{k} , s) =   \mathcal{G}^{-1}_{ij } (\mathbf{k} , s )  f^{\text{ext}}_j  (\mathbf{k} , s )      ~~  .  \label{eq:velocityforcerelation}
\end{align}
Eq.~\eqref{eq:velocityforcerelation} gives the velocity induced by a force for a given Fourier wave vector and Laplace frequency. We then consider the force applied on a probe particle, which is a rigid disk of radius $a$ located at the origin. Due to the rotational symmetry of the disk, we can decompose the force density as 
\begin{align}
    f^{\text{ext}}_j (s, \mathbf{k}) =  L(k)  F_j (s) ~~ . 
\end{align}
The shell localization method amounts to assuming that the force density is uniformly localized our the surface of the  probe particle \cite{levine2001response,camley2011creeping,mackintosh1991orientational}, i.e. 
\begin{align}
    L (\mathbf{x}) =  \frac{1}{2 \pi a }  \delta ( |\mathbf{x}|- a )  
    \label{eq_forceDensity_real} \, .
\end{align}
The assumption of a uniform distribution in real space has been found to give consistent results for systems where no-slip boundary conditions apply \cite{levine2001response,weisenborn1984oseen,lier2023lift}. Fourier transforming Eq.~\eqref{eq_forceDensity_real} yields 
\begin{align}
L(k) =  J_0  (a k) ~~ ,     
\end{align}
with $J_0(z)$ the $0^{\rm th}$ Bessel function of the first kind. To obtain the bead velocity, we must inverse Fourier transform Eq.~\eqref{eq:velocityforcerelation} to obtain the fluid velocity in real space which is coupled to the probe particle velocity through no-slip boundary conditions. The most accurate way to do this would to relate the probe particle velocity $U_i (s)$ to $ v_{i} (|\mathbf{x}| =a, s)$, which can be done averaging over the boundary line between the fluid and the probe particle \cite{weisenborn1984oseen}. We instead consider the fluid velocity localized at $|\mathbf{x}|=0$, i.e. we relate
\begin{align}
    U_i (s) =  v_{i} (|\mathbf{x}| =0 , s) ~~ , 
\end{align}
as has also been done in Refs.~\cite{camley2011creeping,levine2001response,lier2023lift}. As shown in Ref.~\cite{weisenborn1984oseen}, averaging over velocities at $|x|=a$ would yield an additional factor $J_0(a k)$ in the integrand of Eq.~\eqref{eq_responseMatrix}. However, this modification only leads to minor quantitative effects, and we thus ignore it for mathematical simplicity.

\section{Computing the probe particle stability for  active odd fluids}\label{sec_appendix_stability}

\subsection{General case}

In this Appendix, we detail the stability analysis of 
 a probe particle in an active odd viscoelastic fluid, as discussed in Sec.~\ref{sec_stability}.
 Such stability is determined by the long-term velocity of the probe particle. This velocity follows from Eq.~\eqref{eq:responsematrix}, which can be rewritten as
\begin{align}
    U_i(t) &=  \frac{1}{2\rmi \pi} \int_{\Gamma -\rmi \infty}^{\Gamma +\rmi \infty} \!\! \rmd s \, \rme^{s t}  \int \frac{\rmd^2 \mathbf{k}}{(2\pi)^2}  \mathcal{M}_{ij} (\mathbf{k}, s) L(k)   F_j (s) \, ,
\end{align}
where $\Gamma$ is a real number so that the contour path of integration is in the region of convergence of the integrand.
As discussed in the main text, we consider a force perturbation along the eigendirections of $\mathcal{M}_{ij}$, and furthermore consider an instantaneous perturbation applied at $t=0$, such that $F_j(t)\propto \delta(t)$, or $F_j(s)\propto 1$ in Laplace space. The velocity of the probe after this initial perturbation is thus given by: 
\begin{align}
    U_i(t) &= \frac{1}{2\pi \rmi } \int_{\Gamma -\rmi \infty}^{\Gamma +\rmi \infty} \!\! \rmd s \, \rme^{s t}  \int \frac{\rmd^2 \mathbf{k}}{(2\pi)^2}  m_i (\mathbf{k}, s) L(k) \, ,
\end{align}
where $m_i (\mathbf{k}, s)$ depends on the eigenvalues of $\mathcal{M}_{ij}$ and on the form of the initial perturbation. We will not focus on its precise expression in the following discussion, but the crucial point is that $m_i (\mathbf{k}, s)$ is a rational fraction in $s$. This implies that the inverse Laplace transform can be computed as:
\begin{align}\label{eq:app_stability}
    U_i(t) &=  \sum_\ell \int  \frac{\rmd^2 \mathbf{k}}{(2\pi)^2} \, m_{i,\ell}(\mathbf{k}) L(k) \, \rme^{s_\ell^*(\mathbf{k}) t} \, ,
\end{align}
where the $s_\ell^*(\mathbf{k})$ are the poles in $s$ of the rational fraction $m_i$ and where the $m_{i,\ell}(\mathbf{k})$ are the residues of $m_i$ at $s=s_\ell^*(\mathbf{k})$. As a consequence of this simple form, a sufficient condition for the long-time stability of the probe is directly given by the sign of the real part of the poles of $m_i$:
\begin{align} \label{eq:app_stabilitycriterion}
    {\rm Re} \left( s_\ell^*(\mathbf{k}) \right) \leq 0 \qquad \forall \ell,\, \forall \mathbf{k}\, ,
\end{align}
which is the condition given in the main text.

\subsection{Incompressible limit}

To continue our analysis while keeping the technicalities to a minimum, we now restrict to the incompressible case. In this simpler case, $\mathcal{M}_{ij}$ is proportional to the identity tensor, such that the probe velocity can be written as:
\begin{align}
    U_i(t) &= \frac{1}{2\rmi \pi} \int_{\Gamma -\rmi \infty}^{\Gamma +\rmi \infty} \!\! \rmd s \, \rme^{s t}  \int_0^\infty \frac{\rmd k}{2\pi}  r (k, s) F_i (s)  \, ,
\end{align}
where we have defined
\begin{align}
    r(k,s)= \frac{1}{4\pi\eta_{\rm s}} \frac{ak J_0(ak)}{\tau_0 s  + (ak)^2 A_{\rm s}(s)/\eta_{\rm s}} \, ,
\end{align}
with $\tau_0=\rho_0 a^2/\eta_{\rm s}$, and we only consider instantaneous perturbations ($F_i (s)$ constant). As discussed in the previous section, the long-time stability is then given by the poles in $s$ of $r$. They correspond to the roots of the third-order polynomial $P(s) = s^3 + a_2 s^2 + a_1 s  + a_0$ with
\begin{align}
\begin{split}
    &a_2 = (ak)^2/\tau_0+4\gamma_{\rm s}G_{\rm s} \, ,  
      \\ 
     & a_1 =  G_{\rm s} \left[ 4 G_{\rm s} (\gamma_{\rm o}^2+\gamma_{\rm s}^2) + \frac{(a k)^2}{\tau_0 \eta_{\rm s}} \left(4 \gamma_{\rm s} \eta_{\rm s} -\nu_{\rm o}^2 +\nu_{\rm s}^2\right) \right]\, , \\
    &a_0 = \frac{ 2 G_{\rm s}^2 (ak)^2}{\tau_0 \eta_{\rm s}}  \left[2 \eta_{\rm s} \gamma_{\rm o}^2+2 \gamma_{\rm o} \nu_{\rm o} \nu_{\rm s}+\gamma_{\rm s} \left(2 \gamma_{\rm s} \eta_{\rm s}-\nu_{\rm o}^2+\nu_{\rm s}^2\right)\right] \, ,
\end{split}
\end{align}
where we have considered the case $\nu_{\rm s,o}'=\nu_{\rm s,o}$. The roots $s_\ell^*(k)$ of $P(s)$ all have negative real parts iff $a_{0,1,2}>0$ and $a_2 a_1 - a_0 >0$ according to the Routh--Hurwitz stability criterion. We note that $a_2$ is always positive. The conditions $a_{0,1}>0$ are equivalent to:
\begin{subequations} \label{eq:app_allconstraints}
\begin{align}%
    & 4 \gamma_{\rm s} \eta_{\rm s} -\nu_{\rm o}^2+\nu_{\rm s}^2 > 0 \, , \quad  \label{eq:app_condition1}   \\
    & 2 \eta_{\rm s} \gamma_{\rm o}^2+2 \gamma_{\rm o} \nu_{\rm o} \nu_{\rm s}+\gamma_{\rm s} \left(2 \gamma_{\rm s} \eta_{\rm s}-\nu_{\rm o}^2+\nu_{\rm s}^2\right) >0 \, . \quad  \label{eq:app_condition2} 
\end{align}%
The condition $a_2 a_1 - a_0 >0$ takes the form of a second-order polynomial $Q(X)= b_2 X^2 + b_1 X + b_0$ with $X=(ak)^2$ and 
$b_0= 8 G_{\rm s}^3 \gamma_{\rm s} (\gamma_{\rm s}^2+\gamma_{\rm o}^2)$, 
$b_1 = G_{\rm s}^2 [\gamma_{\rm s}(8\gamma_{\rm s}\eta_{\rm s}-\nu_{\rm o}^2+\nu_{\rm s}^2)-2\gamma_{\rm o}\nu_{\rm o}\nu_{\rm s}]/(\tau_0\eta_{\rm s})$, 
$b_2= G_{\rm s} (4\gamma_{\rm s}\eta_{\rm s}-\nu_{\rm o}^2+\nu_{\rm s}^2)/(2\tau_0^2\eta_{\rm s})$. 
%
%
%
With $\Delta= b_1^2-4b_0b_2$ the discriminant of~$Q$, the condition $a_2 a_1 - a_0 >0$ thus translates to:
\begin{align}
    \Delta <0 \,\, \text{or} \,
    \begin{cases}
        \Delta \geq 0 \, \quad 
    \text{and} \\
        \gamma_{\rm s} \left(\nu_{\rm s}^2-\nu_{\rm o}^2\right)-2 \gamma_{\rm o} \nu_{\rm o} \nu_{\rm s}+8 \eta_{\rm s}\gamma_{\rm s}^2 <0
    \end{cases} \!\!\! .\label{eq:app_condition3}   
\end{align}%
\end{subequations}%
A similar analysis can be conducted for $\nu_{\rm s,o}'\neq \nu_{\rm s,o}$.

\section{Inverse Laplace transform in the incompressible case}

\label{sec_Laplace_transform}

\begin{figure*}[t]
	\centering
	\subfigure[ \label{fig_contour}]
    {\includegraphics[width=0.45\linewidth]{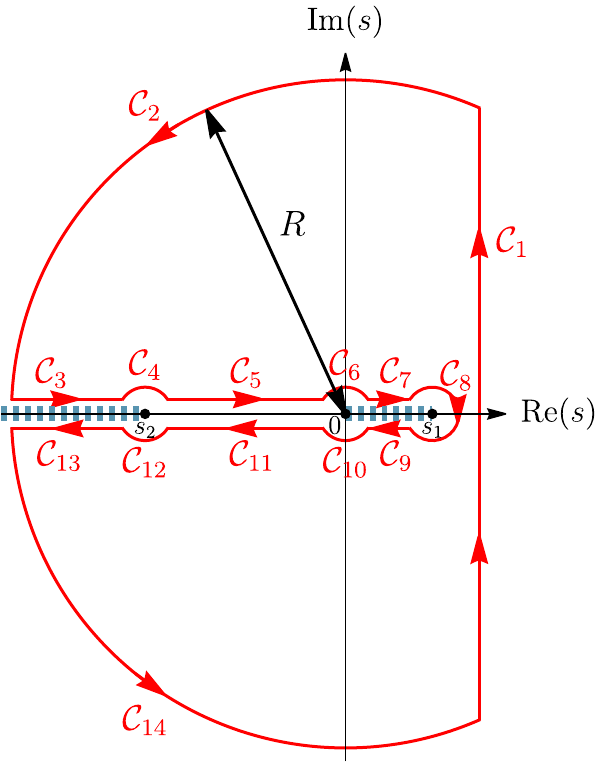}}
    \hfill
	\caption{
	Contour integration in the complex plane used to compute the inverse Laplace transform~\eqref{eq_UparallelAppendix}. The dashed blue lines along the real axis indicate branch cuts of the integrand. In order to compute the integrals $G_i$ along the contours $\mathcal{C}_i$, the radius of the large circle $R$ and the width $2\varepsilon$ between the segments above and below the real axis are sent to $\infty$ and $0$, respectively. 
	}
	\label{fig_contourPlot}
\end{figure*}  

In this Appendix we compute explicitly the velocity of a probe immersed in an odd incompressible visco-elastic fluid. In the incompressible limit, the probe velocity $U_i(t)$ is directly obtained as the inverse Laplace transform of the complex velocity $U_i(s)$:
\begin{subequations}%
\begin{align}%
    U_i(t)=  \frac{1}{2\rmi \pi} \int_{\Gamma -\rmi \infty}^{\Gamma +\rmi \infty} \!\! \rmd s \, \rme^{s t} U_i(s) \, ,
\end{align}
where $U_i(s)$ was obtained in Eq.~\eqref{eq:weaklycompressiblelimit} and we rewrite it here for convenience:
\begin{align} \label{eq:complexVelocityAppendix}
  U_i (s)  = \frac{ K_0 \left( \sqrt{\frac{s \tau_0  }{ A_{\rm s}(s)/\eta_{\rm s}   }} \right)  }{4  \pi   A_{\rm s} (s) } F_i (s)  \, .
\end{align}%
\end{subequations}%
As we will see in the following, the computation of the inverse Laplace transform requires performing contour integration in the complex plane which have to be carefully designed due to the presence of the square root in Eq.~\eqref{eq:complexVelocityAppendix}. Therefore, in order to simplify the following discussion, we consider the case where the coefficient $\gamma_{\rm o}$ vanishes, although a similar result can be obtained in the general case.
For $\gamma_{\rm o}=0$, the complex viscosity $A_{\rm s}$ reads:
\begin{align}
    A_{\rm s}(s)=\eta_{\rm s}\frac{s-s_1}{s-s_2} \, ,
\end{align}
where $s_1=-G_{\rm s}(2\eta_{\rm s}\gamma_{\rm s} - \nu_{\rm o}^2 + \nu_{\rm s}^2)/\eta_{\rm s}$ and $s_2=-2\gamma_{\rm s} G_{\rm s}$. Note that $s_2<0$ while $s_1$ can be either sign, and is positive in the unstable regime according to Eq.~\eqref{eq:condition2}. In the following, we consider the unstable case and have $s_2<0<s_1$. We also introduce $\Delta s = s_2-s_1 <0$ for convenience. The computation in the stable case ($s_1<0$) follows the same steps.

Finally, we consider the velocity after an instantaneous force is applied. We thus take $F_j(t)\propto\delta(t)$ (or $F_j(s)\propto 1$ in Laplace space) and the drag response defined in Eq.~\eqref{responsecoefficients} reads:
\begin{align} \label{eq_UparallelAppendix}
    \eta_{\rm s} M_\parallel(t)=  \frac{1}{4\pi}\int_{\Gamma -\rmi \infty}^{\Gamma +\rmi \infty} \!\! \frac{\rmd s}{2\rmi \pi} \, g(s) \rme^{s t}  \, , \quad g(s) = \frac{s-s_2}{s-s_1} K_0\left( \sqrt{\frac{s(s-s_2)}{s-s_1}} \right) \, ,
\end{align}
This inverse Laplace transform can be computed using residues. We first note that the function $g(s)$ has branch cuts for ${\rm Re} \left[ s(s-s_2)/(s-s_1) \right]<0$, that is for ${\rm Re}(s)<s_2$ and $0<{\rm Re}(s)<s_1$, and is analytic otherwise. We thus define a closed integration contour (see Fig.~\ref{fig_contourPlot}) that avoids the branch cuts and such that $\sum_i G_i =0$ where $G_i = \int_{\mathcal{C}_i} {\rm d} s \, g(s) \rme^{s t}/(2\rmi \pi)$. In the following, we detail the computation of the $G_{i>1}$.

\medskip\noindent\textit{Contours $\mathcal{C}_2$ and  $\mathcal{C}_{14}$.} We set $s=R\rme^{\rmi \theta}$ with $\theta\in[\pi/2,\pi]$ such that:
\begin{align}
    G_2 = \lim_{R\to\infty} \frac{1}{2\rmi \pi}\int_{\pi/2}^{\pi} {\rm d}\theta \, \rmi R \rme^{\rmi \theta} \frac{R \rme^{\rmi \theta}-s_2}{R \rme^{\rmi \theta}-s_1} K_0 \left( \sqrt{R} \rme^{\rmi \theta/2} \sqrt{\frac{R \rme^{\rmi \theta}-s_2}{R \rme^{\rmi \theta}-s_1}} \right) \rme^{R \rme^{\rmi \theta} t} \, ,
\end{align}
which vanishes in the limit $R\to\infty$ because of the last exponential factor in the above equation with real part $R t \cos\theta \leq0$ (for $t>0$ and $\theta\in[\pi/2,\pi]$). Similarly, we find $G_{14} = 0$.

\medskip\noindent\textit{Contours $\mathcal{C}_3$ and  $\mathcal{C}_{13}$.} For $\mathcal{C}_3$, we set $s=s_2+\rme^{\rmi \pi}u$ with $u\in[0,\infty[$, from which we obtain:
\begin{align}
    G_3 &= \frac{1}{2\rmi \pi}\int_{\infty}^{0} {\rm d}u \, \rme^{\rmi \pi} 
    \frac{\rme^{\rmi \pi}u}{\rme^{\rmi \pi}u+s_2-s_1} 
    K_0 \left( \sqrt{\frac{
    (\rme^{\rmi \pi}u+s_2)\rme^{\rmi \pi}u}
    {\rme^{\rmi \pi}u+s_2-s_1}}
    \right) \rme^{(s_2+\rme^{\rmi \pi}u) t} \, , \\
    &= \frac{1}{2\rmi \pi}\int_{0}^{\infty} {\rm d}u \,  
    \frac{u}{u-\Delta s} 
    K_0 \left( \rmi \sqrt{u}\sqrt{\frac{
    (u-s_2)}
    {u-\Delta s}
    } \right) \rme^{s_2 t} \rme^{- u t} \, .
\end{align}
Similarly, setting $s=s_2+\rme^{-\rmi \pi}u$ along $\mathcal{C}_{13}$ yields:
\begin{align}
    G_{13} &= -\frac{1}{2\rmi \pi}\int_{0}^{\infty} {\rm d}u \,  
    \frac{u}{u-\Delta s} 
    K_0 \left( -\rmi \sqrt{u}\sqrt{\frac{
    (u-s_2)}
    {u-\Delta s}
    } \right) \rme^{s_2 t} \rme^{- u t} \, .
\end{align}

\medskip\noindent\textit{Contours $\mathcal{C}_4$ and  $\mathcal{C}_{12}$.} For $\mathcal{C}_4$, we set $s=s_2+\varepsilon\rme^{\rmi \phi}$ with $\phi\in[0,\pi]$, from which we obtain:
\begin{align}
    G_4 &= \lim_{\varepsilon\to0} \frac{1}{2\rmi \pi}\int_{\pi}^{0} {\rm d}\phi \, \varepsilon \rmi \rme^{\rmi \phi} 
    \frac{ \varepsilon\rme^{\rmi \phi}}{\varepsilon\rme^{\rmi \phi}+\Delta s} 
    K_0 \left( \sqrt{\varepsilon} \rme^{\rmi \phi/2} \sqrt{
    \frac{\rme^{\rmi \pi}\varepsilon+s_2}
    {\rme^{\rmi \pi}\varepsilon-s_1}
    }\right) \rme^{(s_2+\varepsilon\rme^{\rmi \phi}) t} \, , \\
    &= \lim_{\varepsilon\to0} \frac{1}{2\rmi \pi}\int_{\pi}^{0} {\rm d}\phi \, 
    \frac{ \varepsilon^2}{\Delta s} 
    K_0 \left( \sqrt{\varepsilon} \rme^{\rmi \phi/2} \sqrt{
    \frac{s_2}
    {-s_1}
    }\right) \rme^{s_2 t} \, , 
\end{align}
which vanishes in the limit $\varepsilon\to0$ since $K_0(x)\underset{{x\to0}}{\sim}-\log x$.
Similar steps yield $G_{12}=0$.

\medskip\noindent\textit{Contours $\mathcal{C}_5$ and  $\mathcal{C}_{11}$.} The contributions stemming from $\mathcal{C}_4$ and $\mathcal{C}_{11}$ cancel each other, since they correspond to an integration in both direction of a holomorphic function along these contours.

\medskip\noindent\textit{Contours $\mathcal{C}_6$ and  $\mathcal{C}_{10}$.} For $\mathcal{C}_6$, we set $s=\varepsilon\rme^{\rmi \phi}$ with $\phi\in[0,\pi]$, from which a computation similar of that of $G_4$ yields:
\begin{align}
    G_5 &= \lim_{\varepsilon\to0} \frac{1}{2\rmi \pi}\int_{\pi}^{0} {\rm d}\phi \, 
    \frac{ s_2}{s_1} 
    \varepsilon K_0 \left( \sqrt{\varepsilon} \rme^{\rmi \phi/2} \sqrt{
    \frac{s_2}
    {s_1}
    }\right) \, , 
\end{align}
which vanishes in the limit $\varepsilon\to0$.
Similar steps yield $G_{10}=0$.

\medskip\noindent\textit{Contours $\mathcal{C}_7$ and  $\mathcal{C}_{9}$.} For $\mathcal{C}_7$, we set $s=s_1+s_1 u\rme^{\rmi \pi}$ with $v\in[0,1]$, from which we obtain:
\begin{align}
    G_7 &= \frac{1}{2\rmi \pi}\int_{0}^{1} {\rm d}v \, 
    \frac{ \Delta s+ s_1 v}{v} 
    K_0 \left( \rme^{-\rmi \pi/2} \sqrt{
    \frac{(v-1)(s_1 v +\Delta s)}
    {v}
    }\right) \rme^{s_1(1-v) t} \, .
\end{align}
Similarly, setting $s=s_2+\rme^{-\rmi \pi}u$ along $\mathcal{C}_{13}$ yields:
\begin{align}
    G_9 &= -\frac{1}{2\rmi \pi}\int_{0}^{1} {\rm d}v \, 
    \frac{ \Delta s+ s_1 v}{v} 
    K_0 \left( \rme^{+\rmi \pi/2} \sqrt{
    \frac{(v-1)(s_1 v +\Delta s)}
    {v}
    }\right) \rme^{s_1(1-v) t} \, .
\end{align}

\medskip\noindent\textit{Contour $\mathcal{C}_8$.} We set $s=s_1+\varepsilon \rme^{\rmi \phi}$ with $\phi\in[-\pi,\pi]$, from which we obtain:
\begin{align}
    G_8 &= \lim_{\varepsilon\to0} \frac{1}{2\rmi \pi}\int_{\pi}^{-\pi} {\rm d}\phi \, (-\Delta s)
    K_0 \left( \frac{1}{\sqrt{\varepsilon}} \rme^{\rmi \phi/2} \sqrt{s_1(-\Delta s)} \right) \rme^{s_1 t} \, , 
\end{align}
which vanishes in the limit $\varepsilon\to0$ since $K_0(x)\underset{{x\to\infty}}{\sim}-\sqrt{\pi/(2x)} \,\rme^{-x}$.

\bigskip

Since $\eta_{\rm s} M_\parallel (t)=G_1 /(4\pi)$ and $G_1 = -\sum_{i>1}G_i = - (G_3+ G_{13} + G_7 + G_9)$, we finally obtain:
\begin{subequations}%
\begin{align}%
    \eta_{\rm s} M_\parallel (t)= \frac{1}{8\pi} \left[
    \int_{0}^{\infty} {\rm d}u\, a(u,t) +  
    \int_{0}^{1} {\rm d}v\, b(v,t)  \right] \, ,
\end{align}
with 
\begin{align}
    a(u,t) &= \frac{u}{u-\Delta s} J_0\left( \sqrt{\frac{u(u-s_2)}{u-\Delta s}} \right) \rme^{s_2t -ut} \, , \\
    b(v,t) &= \frac{\Delta s + s_1 v}{-v} J_0\left( \sqrt{\frac{(v-1)(s_1v + \Delta s)}{v}} \right) \rme^{s_1(1-v)t} \, ,
\end{align}%
\end{subequations}%
and where we have used the identities $K_0(\rmi x)-K_0(-\rmi x)=-\rmi \pi J_0(x)$ for $x>0$.
%
Importantly, we verify that in the long-time limit the behavior of $M_\parallel (t)\underset{{t\to\infty}}{\sim} \rme^{s_1 t}$ is set by the real part of $s_1$ (since $(s_2-u)t$ is always negative), and thus diverges, as expected for our choice $s_1>0$.


\begin{thebibliography}{62}%
\makeatletter
\providecommand \@ifxundefined [1]{%
 \@ifx{#1\undefined}
}%
\providecommand \@ifnum [1]{%
 \ifnum #1\expandafter \@firstoftwo
 \else \expandafter \@secondoftwo
 \fi
}%
\providecommand \@ifx [1]{%
 \ifx #1\expandafter \@firstoftwo
 \else \expandafter \@secondoftwo
 \fi
}%
\providecommand \natexlab [1]{#1}%
\providecommand \enquote  [1]{``#1''}%
\providecommand \bibnamefont  [1]{#1}%
\providecommand \bibfnamefont [1]{#1}%
\providecommand \citenamefont [1]{#1}%
\providecommand \href@noop [0]{\@secondoftwo}%
\providecommand \href [0]{\begingroup \@sanitize@url \@href}%
\providecommand \@href[1]{\@@startlink{#1}\@@href}%
\providecommand \@@href[1]{\endgroup#1\@@endlink}%
\providecommand \@sanitize@url [0]{\catcode `\\12\catcode `\$12\catcode
  `\&12\catcode `\#12\catcode `\^12\catcode `\_12\catcode `\%12\relax}%
\providecommand \@@startlink[1]{}%
\providecommand \@@endlink[0]{}%
\providecommand \url  [0]{\begingroup\@sanitize@url \@url }%
\providecommand \@url [1]{\endgroup\@href {#1}{\urlprefix }}%
\providecommand \urlprefix  [0]{URL }%
\providecommand \Eprint [0]{\href }%
\providecommand \doibase [0]{http://dx.doi.org/}%
\providecommand \selectlanguage [0]{\@gobble}%
\providecommand \bibinfo  [0]{\@secondoftwo}%
\providecommand \bibfield  [0]{\@secondoftwo}%
\providecommand \translation [1]{[#1]}%
\providecommand \BibitemOpen [0]{}%
\providecommand \bibitemStop [0]{}%
\providecommand \bibitemNoStop [0]{.\EOS\space}%
\providecommand \EOS [0]{\spacefactor3000\relax}%
\providecommand \BibitemShut  [1]{\csname bibitem#1\endcsname}%
\let\auto@bib@innerbib\@empty
\bibitem [{\citenamefont {Epstein}\ and\ \citenamefont
  {Mandadapu}(2020)}]{epstein2020timereversal}%
  \BibitemOpen
  \bibfield  {author} {\bibinfo {author} {\bibfnamefont {J.~M.}\ \bibnamefont
  {Epstein}}\ and\ \bibinfo {author} {\bibfnamefont {K.~K.}\ \bibnamefont
  {Mandadapu}},\ }\bibfield  {title} {\enquote {\bibinfo {title} {Time-reversal
  symmetry breaking in two-dimensional nonequilibrium viscous fluids},}\ }\href
  {\doibase 10.1103/PhysRevE.101.052614} {\bibfield  {journal} {\bibinfo
  {journal} {Phys. Rev. E}\ }\textbf {\bibinfo {volume} {101}},\ \bibinfo
  {pages} {052614} (\bibinfo {year} {2020})}\BibitemShut {NoStop}%
\bibitem [{\citenamefont {Hargus}\ \emph {et~al.}(2020)\citenamefont {Hargus},
  \citenamefont {Klymko}, \citenamefont {Epstein},\ and\ \citenamefont
  {Mandadapu}}]{hargus2020time}%
  \BibitemOpen
  \bibfield  {author} {\bibinfo {author} {\bibfnamefont {C.}~\bibnamefont
  {Hargus}}, \bibinfo {author} {\bibfnamefont {K.}~\bibnamefont {Klymko}},
  \bibinfo {author} {\bibfnamefont {J.~M.}\ \bibnamefont {Epstein}}, \ and\
  \bibinfo {author} {\bibfnamefont {K.~K.}\ \bibnamefont {Mandadapu}},\
  }\bibfield  {title} {\enquote {\bibinfo {title} {Time reversal symmetry
  breaking and odd viscosity in active fluids: {{Green}}\textendash{{Kubo}} and
  {{NEMD}} results},}\ }\href {\doibase 10.1063/5.0006441} {\bibfield
  {journal} {\bibinfo  {journal} {J. Chem. Phys.}\ }\textbf {\bibinfo {volume}
  {152}},\ \bibinfo {pages} {201102} (\bibinfo {year} {2020})}\BibitemShut
  {NoStop}%
\bibitem [{\citenamefont {Lier}\ \emph {et~al.}(2022)\citenamefont {Lier},
  \citenamefont {Armas}, \citenamefont {Bo}, \citenamefont {Duclut},
  \citenamefont {J{\"u}licher},\ and\ \citenamefont
  {Sur{\'o}wka}}]{lier2022passive}%
  \BibitemOpen
  \bibfield  {author} {\bibinfo {author} {\bibfnamefont {R.}~\bibnamefont
  {Lier}}, \bibinfo {author} {\bibfnamefont {J.}~\bibnamefont {Armas}},
  \bibinfo {author} {\bibfnamefont {S.}~\bibnamefont {Bo}}, \bibinfo {author}
  {\bibfnamefont {C.}~\bibnamefont {Duclut}}, \bibinfo {author} {\bibfnamefont
  {F.}~\bibnamefont {J{\"u}licher}}, \ and\ \bibinfo {author} {\bibfnamefont
  {P.}~\bibnamefont {Sur{\'o}wka}},\ }\bibfield  {title} {\enquote {\bibinfo
  {title} {Passive odd viscoelasticity},}\ }\href {\doibase
  10.1103/PhysRevE.105.054607} {\bibfield  {journal} {\bibinfo  {journal}
  {Phys. Rev. E}\ }\textbf {\bibinfo {volume} {105}},\ \bibinfo {pages}
  {054607} (\bibinfo {year} {2022})}\BibitemShut {NoStop}%
\bibitem [{\citenamefont {Han}\ \emph {et~al.}(2021)\citenamefont {Han},
  \citenamefont {Fruchart}, \citenamefont {Scheibner}, \citenamefont
  {Vaikuntanathan}, \citenamefont {{de Pablo}},\ and\ \citenamefont
  {Vitelli}}]{han2021fluctuating}%
  \BibitemOpen
  \bibfield  {author} {\bibinfo {author} {\bibfnamefont {M.}~\bibnamefont
  {Han}}, \bibinfo {author} {\bibfnamefont {M.}~\bibnamefont {Fruchart}},
  \bibinfo {author} {\bibfnamefont {C.}~\bibnamefont {Scheibner}}, \bibinfo
  {author} {\bibfnamefont {S.}~\bibnamefont {Vaikuntanathan}}, \bibinfo
  {author} {\bibfnamefont {J.~J.}\ \bibnamefont {{de Pablo}}}, \ and\ \bibinfo
  {author} {\bibfnamefont {V.}~\bibnamefont {Vitelli}},\ }\bibfield  {title}
  {\enquote {\bibinfo {title} {Fluctuating hydrodynamics of chiral active
  fluids},}\ }\href {\doibase 10.1038/s41567-021-01360-7} {\bibfield  {journal}
  {\bibinfo  {journal} {Nat. Phys.}\ }\textbf {\bibinfo {volume} {17}},\
  \bibinfo {pages} {1260} (\bibinfo {year} {2021})}\BibitemShut {NoStop}%
\bibitem [{\citenamefont {Hargus}\ \emph {et~al.}(2021)\citenamefont {Hargus},
  \citenamefont {Epstein},\ and\ \citenamefont {Mandadapu}}]{hargus2021odd}%
  \BibitemOpen
  \bibfield  {author} {\bibinfo {author} {\bibfnamefont {C.}~\bibnamefont
  {Hargus}}, \bibinfo {author} {\bibfnamefont {J.~M.}\ \bibnamefont {Epstein}},
  \ and\ \bibinfo {author} {\bibfnamefont {K.~K.}\ \bibnamefont {Mandadapu}},\
  }\bibfield  {title} {\enquote {\bibinfo {title} {Odd {{Diffusivity}} of
  {{Chiral Random Motion}}},}\ }\href {\doibase 10.1103/PhysRevLett.127.178001}
  {\bibfield  {journal} {\bibinfo  {journal} {Phys. Rev. Lett.}\ }\textbf
  {\bibinfo {volume} {127}},\ \bibinfo {pages} {178001} (\bibinfo {year}
  {2021})}\BibitemShut {NoStop}%
\bibitem [{\citenamefont {Soni}\ \emph {et~al.}(2019)\citenamefont {Soni},
  \citenamefont {Bililign}, \citenamefont {Magkiriadou}, \citenamefont
  {Sacanna}, \citenamefont {Bartolo}, \citenamefont {Shelley},\ and\
  \citenamefont {Irvine}}]{soni2019odd}%
  \BibitemOpen
  \bibfield  {author} {\bibinfo {author} {\bibfnamefont {V.}~\bibnamefont
  {Soni}}, \bibinfo {author} {\bibfnamefont {E.~S.}\ \bibnamefont {Bililign}},
  \bibinfo {author} {\bibfnamefont {S.}~\bibnamefont {Magkiriadou}}, \bibinfo
  {author} {\bibfnamefont {S.}~\bibnamefont {Sacanna}}, \bibinfo {author}
  {\bibfnamefont {D.}~\bibnamefont {Bartolo}}, \bibinfo {author} {\bibfnamefont
  {M.~J.}\ \bibnamefont {Shelley}}, \ and\ \bibinfo {author} {\bibfnamefont
  {W.~T.~M.}\ \bibnamefont {Irvine}},\ }\bibfield  {title} {\enquote {\bibinfo
  {title} {The odd free surface flows of a colloidal chiral fluid},}\ }\href
  {\doibase 10.1038/s41567-019-0603-8} {\bibfield  {journal} {\bibinfo
  {journal} {Nat. Phys.}\ }\textbf {\bibinfo {volume} {15}},\ \bibinfo {pages}
  {1188--1194} (\bibinfo {year} {2019})}\BibitemShut {NoStop}%
\bibitem [{\citenamefont {Reichhardt}\ and\ \citenamefont
  {Reichhardt}(2019)}]{reichhardt2019active}%
  \BibitemOpen
  \bibfield  {author} {\bibinfo {author} {\bibfnamefont {C.}~\bibnamefont
  {Reichhardt}}\ and\ \bibinfo {author} {\bibfnamefont {C.~J.~O.}\ \bibnamefont
  {Reichhardt}},\ }\bibfield  {title} {\enquote {\bibinfo {title} {Active
  microrheology, {{Hall}} effect, and jamming in chiral fluids},}\ }\href
  {\doibase 10.1103/PhysRevE.100.012604} {\bibfield  {journal} {\bibinfo
  {journal} {Phys. Rev. E}\ }\textbf {\bibinfo {volume} {100}},\ \bibinfo
  {pages} {012604} (\bibinfo {year} {2019})}\BibitemShut {NoStop}%
\bibitem [{\citenamefont {Markovich}\ and\ \citenamefont
  {Lubensky}(2021)}]{markovich2021odd}%
  \BibitemOpen
  \bibfield  {author} {\bibinfo {author} {\bibfnamefont {T.}~\bibnamefont
  {Markovich}}\ and\ \bibinfo {author} {\bibfnamefont {T.~C.}\ \bibnamefont
  {Lubensky}},\ }\bibfield  {title} {\enquote {\bibinfo {title} {Odd
  {{Viscosity}} in {{Active Matter}}: {{Microscopic Origin}} and {{3D
  Effects}}},}\ }\href {\doibase 10.1103/PhysRevLett.127.048001} {\bibfield
  {journal} {\bibinfo  {journal} {Phys. Rev. Lett.}\ }\textbf {\bibinfo
  {volume} {127}},\ \bibinfo {pages} {048001} (\bibinfo {year}
  {2021})}\BibitemShut {NoStop}%
\bibitem [{\citenamefont {Reichhardt}\ and\ \citenamefont
  {Reichhardt}(2022)}]{reichhardt2022active}%
  \BibitemOpen
  \bibfield  {author} {\bibinfo {author} {\bibfnamefont {C.~J.~O.}\
  \bibnamefont {Reichhardt}}\ and\ \bibinfo {author} {\bibfnamefont
  {C.}~\bibnamefont {Reichhardt}},\ }\bibfield  {title} {\enquote {\bibinfo
  {title} {Active rheology in odd-viscosity systems},}\ }\href {\doibase
  10.1209/0295-5075/ac2adc} {\bibfield  {journal} {\bibinfo  {journal} {EPL}\
  }\textbf {\bibinfo {volume} {137}},\ \bibinfo {pages} {66004} (\bibinfo
  {year} {2022})}\BibitemShut {NoStop}%
\bibitem [{\citenamefont {F{\"u}rthauer}\ \emph {et~al.}(2012)\citenamefont
  {F{\"u}rthauer}, \citenamefont {Strempel}, \citenamefont {Grill},\ and\
  \citenamefont {J{\"u}licher}}]{furthauer2012active}%
  \BibitemOpen
  \bibfield  {author} {\bibinfo {author} {\bibfnamefont {S.}~\bibnamefont
  {F{\"u}rthauer}}, \bibinfo {author} {\bibfnamefont {M.}~\bibnamefont
  {Strempel}}, \bibinfo {author} {\bibfnamefont {S.~W.}\ \bibnamefont {Grill}},
  \ and\ \bibinfo {author} {\bibfnamefont {F.}~\bibnamefont {J{\"u}licher}},\
  }\bibfield  {title} {\enquote {\bibinfo {title} {Active chiral fluids},}\
  }\href {\doibase 10.1140/epje/i2012-12089-6} {\bibfield  {journal} {\bibinfo
  {journal} {Eur. Phys. J. E}\ }\textbf {\bibinfo {volume} {35}},\ \bibinfo
  {pages} {89} (\bibinfo {year} {2012})}\BibitemShut {NoStop}%
\bibitem [{\citenamefont {Tan}\ \emph {et~al.}(2022)\citenamefont {Tan},
  \citenamefont {Mietke}, \citenamefont {Li}, \citenamefont {Chen},
  \citenamefont {Higinbotham}, \citenamefont {Foster}, \citenamefont {Gokhale},
  \citenamefont {Dunkel},\ and\ \citenamefont {Fakhri}}]{tan2022odd}%
  \BibitemOpen
  \bibfield  {author} {\bibinfo {author} {\bibfnamefont {T.~H.}\ \bibnamefont
  {Tan}}, \bibinfo {author} {\bibfnamefont {A.}~\bibnamefont {Mietke}},
  \bibinfo {author} {\bibfnamefont {J.}~\bibnamefont {Li}}, \bibinfo {author}
  {\bibfnamefont {Y.}~\bibnamefont {Chen}}, \bibinfo {author} {\bibfnamefont
  {H.}~\bibnamefont {Higinbotham}}, \bibinfo {author} {\bibfnamefont {P.~J.}\
  \bibnamefont {Foster}}, \bibinfo {author} {\bibfnamefont {S.}~\bibnamefont
  {Gokhale}}, \bibinfo {author} {\bibfnamefont {J.}~\bibnamefont {Dunkel}}, \
  and\ \bibinfo {author} {\bibfnamefont {N.}~\bibnamefont {Fakhri}},\
  }\bibfield  {title} {\enquote {\bibinfo {title} {Odd dynamics of living
  chiral crystals},}\ }\href {\doibase 10.1038/s41586-022-04889-6} {\bibfield
  {journal} {\bibinfo  {journal} {Nature}\ }\textbf {\bibinfo {volume} {607}},\
  \bibinfo {pages} {287--293} (\bibinfo {year} {2022})}\BibitemShut {NoStop}%
\bibitem [{\citenamefont {Banerjee}\ \emph {et~al.}(2021)\citenamefont
  {Banerjee}, \citenamefont {Vitelli}, \citenamefont {J{\"u}licher},\ and\
  \citenamefont {Sur{\'o}wka}}]{banerjee2021active}%
  \BibitemOpen
  \bibfield  {author} {\bibinfo {author} {\bibfnamefont {D.}~\bibnamefont
  {Banerjee}}, \bibinfo {author} {\bibfnamefont {V.}~\bibnamefont {Vitelli}},
  \bibinfo {author} {\bibfnamefont {F.}~\bibnamefont {J{\"u}licher}}, \ and\
  \bibinfo {author} {\bibfnamefont {P.}~\bibnamefont {Sur{\'o}wka}},\
  }\bibfield  {title} {\enquote {\bibinfo {title} {Active {{Viscoelasticity}}
  of {{Odd Materials}}},}\ }\href {\doibase 10.1103/PhysRevLett.126.138001}
  {\bibfield  {journal} {\bibinfo  {journal} {Phys. Rev. Lett.}\ }\textbf
  {\bibinfo {volume} {126}},\ \bibinfo {pages} {138001} (\bibinfo {year}
  {2021})}\BibitemShut {NoStop}%
\bibitem [{\citenamefont {Sur{\'o}wka}\ \emph {et~al.}(2022)\citenamefont
  {Sur{\'o}wka}, \citenamefont {Souslov}, \citenamefont {J{\"u}licher},\ and\
  \citenamefont {Banerjee}}]{surowka2022odd}%
  \BibitemOpen
  \bibfield  {author} {\bibinfo {author} {\bibfnamefont {P.}~\bibnamefont
  {Sur{\'o}wka}}, \bibinfo {author} {\bibfnamefont {A.}~\bibnamefont
  {Souslov}}, \bibinfo {author} {\bibfnamefont {F.}~\bibnamefont
  {J{\"u}licher}}, \ and\ \bibinfo {author} {\bibfnamefont {D.}~\bibnamefont
  {Banerjee}},\ }\bibfield  {title} {\enquote {\bibinfo {title} {Odd cosserat
  elasticity in active materials},}\ }\href@noop {} {\bibfield  {journal}
  {\bibinfo  {journal} {arXiv preprint arXiv:2210.13606}\ } (\bibinfo {year}
  {2022})}\BibitemShut {NoStop}%
\bibitem [{\citenamefont {Kalz}\ \emph {et~al.}(2023)\citenamefont {Kalz},
  \citenamefont {Vuijk}, \citenamefont {Sommer}, \citenamefont {Metzler},\ and\
  \citenamefont {Sharma}}]{kalz2023oscillatory}%
  \BibitemOpen
  \bibfield  {author} {\bibinfo {author} {\bibfnamefont {E.}~\bibnamefont
  {Kalz}}, \bibinfo {author} {\bibfnamefont {H.~D.}\ \bibnamefont {Vuijk}},
  \bibinfo {author} {\bibfnamefont {J.-U.}\ \bibnamefont {Sommer}}, \bibinfo
  {author} {\bibfnamefont {R.}~\bibnamefont {Metzler}}, \ and\ \bibinfo
  {author} {\bibfnamefont {A.}~\bibnamefont {Sharma}},\ }\href@noop {}
  {\enquote {\bibinfo {title} {Oscillatory force autocorrelations in
  equilibrium odd-diffusive systems},}\ } (\bibinfo {year} {2023}),\ \Eprint
  {http://arxiv.org/abs/2302.01263} {arXiv:2302.01263 [cond-mat.stat-mech]}
  \BibitemShut {NoStop}%
\bibitem [{\citenamefont {Kalz}\ \emph {et~al.}(2022)\citenamefont {Kalz},
  \citenamefont {Vuijk}, \citenamefont {Abdoli}, \citenamefont {Sommer},
  \citenamefont {L\"owen},\ and\ \citenamefont
  {Sharma}}]{PhysRevLett.129.090601}%
  \BibitemOpen
  \bibfield  {author} {\bibinfo {author} {\bibfnamefont {E.}~\bibnamefont
  {Kalz}}, \bibinfo {author} {\bibfnamefont {H.~D.}\ \bibnamefont {Vuijk}},
  \bibinfo {author} {\bibfnamefont {I.}~\bibnamefont {Abdoli}}, \bibinfo
  {author} {\bibfnamefont {J.-U.}\ \bibnamefont {Sommer}}, \bibinfo {author}
  {\bibfnamefont {H.}~\bibnamefont {L\"owen}}, \ and\ \bibinfo {author}
  {\bibfnamefont {A.}~\bibnamefont {Sharma}},\ }\bibfield  {title} {\enquote
  {\bibinfo {title} {Collisions enhance self-diffusion in odd-diffusive
  systems},}\ }\href {\doibase 10.1103/PhysRevLett.129.090601} {\bibfield
  {journal} {\bibinfo  {journal} {Phys. Rev. Lett.}\ }\textbf {\bibinfo
  {volume} {129}},\ \bibinfo {pages} {090601} (\bibinfo {year}
  {2022})}\BibitemShut {NoStop}%
\bibitem [{\citenamefont {Banerjee}\ \emph {et~al.}(2017)\citenamefont
  {Banerjee}, \citenamefont {Souslov}, \citenamefont {Abanov},\ and\
  \citenamefont {Vitelli}}]{banerjee2017odd}%
  \BibitemOpen
  \bibfield  {author} {\bibinfo {author} {\bibfnamefont {D.}~\bibnamefont
  {Banerjee}}, \bibinfo {author} {\bibfnamefont {A.}~\bibnamefont {Souslov}},
  \bibinfo {author} {\bibfnamefont {A.~G.}\ \bibnamefont {Abanov}}, \ and\
  \bibinfo {author} {\bibfnamefont {V.}~\bibnamefont {Vitelli}},\ }\bibfield
  {title} {\enquote {\bibinfo {title} {Odd viscosity in chiral active
  fluids},}\ }\href {\doibase 10.1038/s41467-017-01378-7} {\bibfield  {journal}
  {\bibinfo  {journal} {Nat. Commun.}\ }\textbf {\bibinfo {volume} {8}},\
  \bibinfo {pages} {1573} (\bibinfo {year} {2017})}\BibitemShut {NoStop}%
\bibitem [{\citenamefont {Abanov}\ \emph {et~al.}(2018)\citenamefont {Abanov},
  \citenamefont {Can},\ and\ \citenamefont {Ganeshan}}]{abanov2018odd}%
  \BibitemOpen
  \bibfield  {author} {\bibinfo {author} {\bibfnamefont {A.}~\bibnamefont
  {Abanov}}, \bibinfo {author} {\bibfnamefont {T.}~\bibnamefont {Can}}, \ and\
  \bibinfo {author} {\bibfnamefont {S.}~\bibnamefont {Ganeshan}},\ }\bibfield
  {title} {\enquote {\bibinfo {title} {Odd surface waves in two-dimensional
  incompressible fluids},}\ }\href {\doibase 10.21468/SciPostPhys.5.1.010}
  {\bibfield  {journal} {\bibinfo  {journal} {SciPost Phys.}\ }\textbf
  {\bibinfo {volume} {5}},\ \bibinfo {pages} {010} (\bibinfo {year}
  {2018})}\BibitemShut {NoStop}%
\bibitem [{\citenamefont {Ganeshan}\ and\ \citenamefont
  {Abanov}(2017)}]{ganeshan2017odd}%
  \BibitemOpen
  \bibfield  {author} {\bibinfo {author} {\bibfnamefont {S.}~\bibnamefont
  {Ganeshan}}\ and\ \bibinfo {author} {\bibfnamefont {A.~G.}\ \bibnamefont
  {Abanov}},\ }\bibfield  {title} {\enquote {\bibinfo {title} {Odd viscosity in
  two-dimensional incompressible fluids},}\ }\href {\doibase
  10.1103/PhysRevFluids.2.094101} {\bibfield  {journal} {\bibinfo  {journal}
  {Phys. Rev. Fluids}\ }\textbf {\bibinfo {volume} {2}},\ \bibinfo {pages}
  {094101} (\bibinfo {year} {2017})}\BibitemShut {NoStop}%
\bibitem [{\citenamefont {Khain}\ \emph {et~al.}(2022)\citenamefont {Khain},
  \citenamefont {Scheibner}, \citenamefont {Fruchart},\ and\ \citenamefont
  {Vitelli}}]{khain2022stokes}%
  \BibitemOpen
  \bibfield  {author} {\bibinfo {author} {\bibfnamefont {T.}~\bibnamefont
  {Khain}}, \bibinfo {author} {\bibfnamefont {C.}~\bibnamefont {Scheibner}},
  \bibinfo {author} {\bibfnamefont {M.}~\bibnamefont {Fruchart}}, \ and\
  \bibinfo {author} {\bibfnamefont {V.}~\bibnamefont {Vitelli}},\ }\bibfield
  {title} {\enquote {\bibinfo {title} {Stokes flows in three-dimensional fluids
  with odd and parity-violating viscosities},}\ }\href {\doibase
  10.1017/jfm.2021.1079} {\bibfield  {journal} {\bibinfo  {journal} {J. Fluid
  Mech.}\ }\textbf {\bibinfo {volume} {934}},\ \bibinfo {pages} {A23} (\bibinfo
  {year} {2022})}\BibitemShut {NoStop}%
\bibitem [{\citenamefont {Hosaka}\ \emph
  {et~al.}(2023{\natexlab{a}})\citenamefont {Hosaka}, \citenamefont
  {Golestanian},\ and\ \citenamefont
  {{Daddi-Moussa-Ider}}}]{hosaka2023hydrodynamics}%
  \BibitemOpen
  \bibfield  {author} {\bibinfo {author} {\bibfnamefont {Y.}~\bibnamefont
  {Hosaka}}, \bibinfo {author} {\bibfnamefont {R.}~\bibnamefont {Golestanian}},
  \ and\ \bibinfo {author} {\bibfnamefont {A.}~\bibnamefont
  {{Daddi-Moussa-Ider}}},\ }\bibfield  {title} {\enquote {\bibinfo {title}
  {Hydrodynamics of an odd active surfer in a chiral fluid},}\ }\href@noop {}
  {\bibfield  {journal} {\bibinfo  {journal} {arXiv:2303.11836}\ } (\bibinfo
  {year} {2023}{\natexlab{a}})},\ \Eprint {http://arxiv.org/abs/2303.11836}
  {arxiv:2303.11836 [cond-mat, physics:physics]} \BibitemShut {NoStop}%
\bibitem [{\citenamefont {Hosaka}\ \emph
  {et~al.}(2021{\natexlab{a}})\citenamefont {Hosaka}, \citenamefont {Komura},\
  and\ \citenamefont {Andelman}}]{hosaka2021hydrodynamic}%
  \BibitemOpen
  \bibfield  {author} {\bibinfo {author} {\bibfnamefont {Y.}~\bibnamefont
  {Hosaka}}, \bibinfo {author} {\bibfnamefont {S.}~\bibnamefont {Komura}}, \
  and\ \bibinfo {author} {\bibfnamefont {D.}~\bibnamefont {Andelman}},\
  }\bibfield  {title} {\enquote {\bibinfo {title} {Hydrodynamic lift of a
  two-dimensional liquid domain with odd viscosity},}\ }\href {\doibase
  10.1103/PhysRevE.104.064613} {\bibfield  {journal} {\bibinfo  {journal}
  {Phys. Rev. E}\ }\textbf {\bibinfo {volume} {104}},\ \bibinfo {pages}
  {064613} (\bibinfo {year} {2021}{\natexlab{a}})}\BibitemShut {NoStop}%
\bibitem [{\citenamefont {Hosaka}\ \emph
  {et~al.}(2023{\natexlab{b}})\citenamefont {Hosaka}, \citenamefont
  {Andelman},\ and\ \citenamefont {Komura}}]{hosaka2023pair}%
  \BibitemOpen
  \bibfield  {author} {\bibinfo {author} {\bibfnamefont {Y.}~\bibnamefont
  {Hosaka}}, \bibinfo {author} {\bibfnamefont {D.}~\bibnamefont {Andelman}}, \
  and\ \bibinfo {author} {\bibfnamefont {S.}~\bibnamefont {Komura}},\
  }\bibfield  {title} {\enquote {\bibinfo {title} {Pair dynamics of active
  force dipoles in an odd-viscous fluid},}\ }\href {\doibase
  10.1140/epje/s10189-023-00265-y} {\bibfield  {journal} {\bibinfo  {journal}
  {Eur. Phys. J. E}\ }\textbf {\bibinfo {volume} {46}},\ \bibinfo {pages} {18}
  (\bibinfo {year} {2023}{\natexlab{b}})}\BibitemShut {NoStop}%
\bibitem [{\citenamefont {Scheibner}\ \emph {et~al.}(2020)\citenamefont
  {Scheibner}, \citenamefont {Souslov}, \citenamefont {Banerjee}, \citenamefont
  {Sur{\'o}wka}, \citenamefont {Irvine},\ and\ \citenamefont
  {Vitelli}}]{scheibner2020odd}%
  \BibitemOpen
  \bibfield  {author} {\bibinfo {author} {\bibfnamefont {C.}~\bibnamefont
  {Scheibner}}, \bibinfo {author} {\bibfnamefont {A.}~\bibnamefont {Souslov}},
  \bibinfo {author} {\bibfnamefont {D.}~\bibnamefont {Banerjee}}, \bibinfo
  {author} {\bibfnamefont {P.}~\bibnamefont {Sur{\'o}wka}}, \bibinfo {author}
  {\bibfnamefont {W.~T.~M.}\ \bibnamefont {Irvine}}, \ and\ \bibinfo {author}
  {\bibfnamefont {V.}~\bibnamefont {Vitelli}},\ }\bibfield  {title} {\enquote
  {\bibinfo {title} {Odd elasticity},}\ }\href {\doibase
  10.1038/s41567-020-0795-y} {\bibfield  {journal} {\bibinfo  {journal} {Nat.
  Phys.}\ }\textbf {\bibinfo {volume} {16}},\ \bibinfo {pages} {475} (\bibinfo
  {year} {2020})}\BibitemShut {NoStop}%
\bibitem [{\citenamefont {Braverman}\ \emph {et~al.}(2021)\citenamefont
  {Braverman}, \citenamefont {Scheibner}, \citenamefont {VanSaders},\ and\
  \citenamefont {Vitelli}}]{braverman2021topological}%
  \BibitemOpen
  \bibfield  {author} {\bibinfo {author} {\bibfnamefont {L.}~\bibnamefont
  {Braverman}}, \bibinfo {author} {\bibfnamefont {C.}~\bibnamefont
  {Scheibner}}, \bibinfo {author} {\bibfnamefont {B.}~\bibnamefont
  {VanSaders}}, \ and\ \bibinfo {author} {\bibfnamefont {V.}~\bibnamefont
  {Vitelli}},\ }\bibfield  {title} {\enquote {\bibinfo {title} {Topological
  {{Defects}} in {{Solids}} with {{Odd Elasticity}}},}\ }\href {\doibase
  10.1103/PhysRevLett.127.268001} {\bibfield  {journal} {\bibinfo  {journal}
  {Phys. Rev. Lett.}\ }\textbf {\bibinfo {volume} {127}},\ \bibinfo {pages}
  {268001} (\bibinfo {year} {2021})}\BibitemShut {NoStop}%
\bibitem [{\citenamefont {Floyd}\ \emph {et~al.}(2022)\citenamefont {Floyd},
  \citenamefont {Dinner},\ and\ \citenamefont
  {Vaikuntanathan}}]{floyd2022signatures}%
  \BibitemOpen
  \bibfield  {author} {\bibinfo {author} {\bibfnamefont {C.}~\bibnamefont
  {Floyd}}, \bibinfo {author} {\bibfnamefont {A.~R.}\ \bibnamefont {Dinner}}, \
  and\ \bibinfo {author} {\bibfnamefont {S.}~\bibnamefont {Vaikuntanathan}},\
  }\bibfield  {title} {\enquote {\bibinfo {title} {Signatures of odd dynamics
  in viscoelastic systems: From spatiotemporal pattern formation to odd
  rheology},}\ }\href@noop {} {\bibfield  {journal} {\bibinfo  {journal}
  {arXiv:2210.01159}\ } (\bibinfo {year} {2022})},\ \Eprint
  {http://arxiv.org/abs/2210.01159} {arxiv:2210.01159} \BibitemShut {NoStop}%
\bibitem [{\citenamefont {Fruchart}\ \emph {et~al.}(2022)\citenamefont
  {Fruchart}, \citenamefont {Han}, \citenamefont {Scheibner},\ and\
  \citenamefont {Vitelli}}]{fruchart2022odda}%
  \BibitemOpen
  \bibfield  {author} {\bibinfo {author} {\bibfnamefont {M.}~\bibnamefont
  {Fruchart}}, \bibinfo {author} {\bibfnamefont {M.}~\bibnamefont {Han}},
  \bibinfo {author} {\bibfnamefont {C.}~\bibnamefont {Scheibner}}, \ and\
  \bibinfo {author} {\bibfnamefont {V.}~\bibnamefont {Vitelli}},\ }\bibfield
  {title} {\enquote {\bibinfo {title} {The odd ideal gas: {{Hall}} viscosity
  and thermal conductivity from non-{{Hermitian}} kinetic theory},}\
  }\href@noop {} {\bibfield  {journal} {\bibinfo  {journal} {arXiv:2202.02037}\
  } (\bibinfo {year} {2022})},\ \Eprint {http://arxiv.org/abs/2202.02037}
  {arxiv:2202.02037} \BibitemShut {NoStop}%
\bibitem [{\citenamefont {Liao}\ \emph {et~al.}(2020)\citenamefont {Liao},
  \citenamefont {Irvine},\ and\ \citenamefont
  {Vaikuntanathan}}]{liao2020rectification}%
  \BibitemOpen
  \bibfield  {author} {\bibinfo {author} {\bibfnamefont {Z.}~\bibnamefont
  {Liao}}, \bibinfo {author} {\bibfnamefont {W.~T.~M.}\ \bibnamefont {Irvine}},
  \ and\ \bibinfo {author} {\bibfnamefont {S.}~\bibnamefont {Vaikuntanathan}},\
  }\bibfield  {title} {\enquote {\bibinfo {title} {Rectification in
  {{Nonequilibrium Parity Violating Metamaterials}}},}\ }\href {\doibase
  10.1103/PhysRevX.10.021036} {\bibfield  {journal} {\bibinfo  {journal} {Phys.
  Rev. X}\ }\textbf {\bibinfo {volume} {10}},\ \bibinfo {pages} {021036}
  (\bibinfo {year} {2020})}\BibitemShut {NoStop}%
\bibitem [{\citenamefont {Pellegrino}\ \emph {et~al.}(2017)\citenamefont
  {Pellegrino}, \citenamefont {Torre},\ and\ \citenamefont
  {Polini}}]{pellegrino2017nonlocal}%
  \BibitemOpen
  \bibfield  {author} {\bibinfo {author} {\bibfnamefont {F.~M.~D.}\
  \bibnamefont {Pellegrino}}, \bibinfo {author} {\bibfnamefont
  {I.}~\bibnamefont {Torre}}, \ and\ \bibinfo {author} {\bibfnamefont
  {M.}~\bibnamefont {Polini}},\ }\bibfield  {title} {\enquote {\bibinfo {title}
  {Nonlocal transport and the {{Hall}} viscosity of two-dimensional
  hydrodynamic electron liquids},}\ }\href {\doibase
  10.1103/PhysRevB.96.195401} {\bibfield  {journal} {\bibinfo  {journal} {Phys.
  Rev. B}\ }\textbf {\bibinfo {volume} {96}},\ \bibinfo {pages} {195401}
  (\bibinfo {year} {2017})}\BibitemShut {NoStop}%
\bibitem [{\citenamefont {Narozhny}\ and\ \citenamefont
  {Sch{\"u}tt}(2019)}]{narozhny2019magnetohydrodynamics}%
  \BibitemOpen
  \bibfield  {author} {\bibinfo {author} {\bibfnamefont {B.~N.}\ \bibnamefont
  {Narozhny}}\ and\ \bibinfo {author} {\bibfnamefont {M.}~\bibnamefont
  {Sch{\"u}tt}},\ }\bibfield  {title} {\enquote {\bibinfo {title}
  {Magnetohydrodynamics in graphene: {{Shear}} and {{Hall}} viscosities},}\
  }\href {\doibase 10.1103/PhysRevB.100.035125} {\bibfield  {journal} {\bibinfo
   {journal} {Phys. Rev. B}\ }\textbf {\bibinfo {volume} {100}},\ \bibinfo
  {pages} {035125} (\bibinfo {year} {2019})}\BibitemShut {NoStop}%
\bibitem [{\citenamefont {Berdyugin}\ \emph {et~al.}(2019)\citenamefont
  {Berdyugin}, \citenamefont {Xu}, \citenamefont {Pellegrino}, \citenamefont
  {Krishna~Kumar}, \citenamefont {Principi}, \citenamefont {Torre},
  \citenamefont {Ben~Shalom}, \citenamefont {Taniguchi}, \citenamefont
  {Watanabe}, \citenamefont {Grigorieva}, \citenamefont {Polini}, \citenamefont
  {Geim},\ and\ \citenamefont {Bandurin}}]{berdyugin2019measuring}%
  \BibitemOpen
  \bibfield  {author} {\bibinfo {author} {\bibfnamefont {A.~I.}\ \bibnamefont
  {Berdyugin}}, \bibinfo {author} {\bibfnamefont {S.~G.}\ \bibnamefont {Xu}},
  \bibinfo {author} {\bibfnamefont {F.~M.~D.}\ \bibnamefont {Pellegrino}},
  \bibinfo {author} {\bibfnamefont {R.}~\bibnamefont {Krishna~Kumar}}, \bibinfo
  {author} {\bibfnamefont {A.}~\bibnamefont {Principi}}, \bibinfo {author}
  {\bibfnamefont {I.}~\bibnamefont {Torre}}, \bibinfo {author} {\bibfnamefont
  {M.}~\bibnamefont {Ben~Shalom}}, \bibinfo {author} {\bibfnamefont
  {T.}~\bibnamefont {Taniguchi}}, \bibinfo {author} {\bibfnamefont
  {K.}~\bibnamefont {Watanabe}}, \bibinfo {author} {\bibfnamefont {I.~V.}\
  \bibnamefont {Grigorieva}}, \bibinfo {author} {\bibfnamefont
  {M.}~\bibnamefont {Polini}}, \bibinfo {author} {\bibfnamefont {A.~K.}\
  \bibnamefont {Geim}}, \ and\ \bibinfo {author} {\bibfnamefont {D.~A.}\
  \bibnamefont {Bandurin}},\ }\bibfield  {title} {\enquote {\bibinfo {title}
  {Measuring {{Hall}} viscosity of graphene's electron fluid},}\ }\href
  {\doibase 10.1126/science.aau0685} {\bibfield  {journal} {\bibinfo  {journal}
  {Science}\ }\textbf {\bibinfo {volume} {364}},\ \bibinfo {pages} {162--165}
  (\bibinfo {year} {2019})}\BibitemShut {NoStop}%
\bibitem [{\citenamefont {Fossati}\ \emph {et~al.}(2022)\citenamefont
  {Fossati}, \citenamefont {Scheibner}, \citenamefont {Fruchart},\ and\
  \citenamefont {Vitelli}}]{fossati2022odd}%
  \BibitemOpen
  \bibfield  {author} {\bibinfo {author} {\bibfnamefont {M.}~\bibnamefont
  {Fossati}}, \bibinfo {author} {\bibfnamefont {C.}~\bibnamefont {Scheibner}},
  \bibinfo {author} {\bibfnamefont {M.}~\bibnamefont {Fruchart}}, \ and\
  \bibinfo {author} {\bibfnamefont {V.}~\bibnamefont {Vitelli}},\ }\bibfield
  {title} {\enquote {\bibinfo {title} {Odd elasticity and topological waves in
  active surfaces},}\ }\href@noop {} {\bibfield  {journal} {\bibinfo  {journal}
  {arXiv:2210.03669}\ } (\bibinfo {year} {2022})},\ \Eprint
  {http://arxiv.org/abs/2210.03669} {arxiv:2210.03669} \BibitemShut {NoStop}%
\bibitem [{\citenamefont {Green}\ \emph {et~al.}(2020)\citenamefont {Green},
  \citenamefont {Armas}, \citenamefont {{de Boer}},\ and\ \citenamefont
  {Giomi}}]{green2020topological}%
  \BibitemOpen
  \bibfield  {author} {\bibinfo {author} {\bibfnamefont {R.}~\bibnamefont
  {Green}}, \bibinfo {author} {\bibfnamefont {J.}~\bibnamefont {Armas}},
  \bibinfo {author} {\bibfnamefont {J.}~\bibnamefont {{de Boer}}}, \ and\
  \bibinfo {author} {\bibfnamefont {L.}~\bibnamefont {Giomi}},\ }\bibfield
  {title} {\enquote {\bibinfo {title} {Topological waves in passive and active
  fluids on curved surfaces: A unified picture},}\ }\href@noop {} {\bibfield
  {journal} {\bibinfo  {journal} {arXiv:2011.12271}\ } (\bibinfo {year}
  {2020})},\ \Eprint {http://arxiv.org/abs/2011.12271} {arxiv:2011.12271}
  \BibitemShut {NoStop}%
\bibitem [{\citenamefont {Souslov}\ \emph {et~al.}(2019)\citenamefont
  {Souslov}, \citenamefont {Dasbiswas}, \citenamefont {Fruchart}, \citenamefont
  {Vaikuntanathan},\ and\ \citenamefont {Vitelli}}]{souslov2019topological}%
  \BibitemOpen
  \bibfield  {author} {\bibinfo {author} {\bibfnamefont {A.}~\bibnamefont
  {Souslov}}, \bibinfo {author} {\bibfnamefont {K.}~\bibnamefont {Dasbiswas}},
  \bibinfo {author} {\bibfnamefont {M.}~\bibnamefont {Fruchart}}, \bibinfo
  {author} {\bibfnamefont {S.}~\bibnamefont {Vaikuntanathan}}, \ and\ \bibinfo
  {author} {\bibfnamefont {V.}~\bibnamefont {Vitelli}},\ }\bibfield  {title}
  {\enquote {\bibinfo {title} {Topological {{Waves}} in {{Fluids}} with {{Odd
  Viscosity}}},}\ }\href {\doibase 10.1103/PhysRevLett.122.128001} {\bibfield
  {journal} {\bibinfo  {journal} {Phys. Rev. Lett.}\ }\textbf {\bibinfo
  {volume} {122}},\ \bibinfo {pages} {128001} (\bibinfo {year}
  {2019})}\BibitemShut {NoStop}%
\bibitem [{\citenamefont {Tauber}\ \emph {et~al.}(2019)\citenamefont {Tauber},
  \citenamefont {Delplace},\ and\ \citenamefont
  {Venaille}}]{tauber2019bulkinterface}%
  \BibitemOpen
  \bibfield  {author} {\bibinfo {author} {\bibfnamefont {C.}~\bibnamefont
  {Tauber}}, \bibinfo {author} {\bibfnamefont {P.}~\bibnamefont {Delplace}}, \
  and\ \bibinfo {author} {\bibfnamefont {A.}~\bibnamefont {Venaille}},\
  }\bibfield  {title} {\enquote {\bibinfo {title} {A bulk-interface
  correspondence for equatorial waves},}\ }\href {\doibase
  10.1017/jfm.2019.233} {\bibfield  {journal} {\bibinfo  {journal} {J. Fluid
  Mech.}\ }\textbf {\bibinfo {volume} {868}},\ \bibinfo {pages} {R2} (\bibinfo
  {year} {2019})}\BibitemShut {NoStop}%
\bibitem [{\citenamefont {Bililign}\ \emph {et~al.}(2022)\citenamefont
  {Bililign}, \citenamefont {Balboa~Usabiaga}, \citenamefont {Ganan},
  \citenamefont {Poncet}, \citenamefont {Soni}, \citenamefont {Magkiriadou},
  \citenamefont {Shelley}, \citenamefont {Bartolo},\ and\ \citenamefont
  {Irvine}}]{bililign2022motile}%
  \BibitemOpen
  \bibfield  {author} {\bibinfo {author} {\bibfnamefont {E.~S.}\ \bibnamefont
  {Bililign}}, \bibinfo {author} {\bibfnamefont {F.}~\bibnamefont
  {Balboa~Usabiaga}}, \bibinfo {author} {\bibfnamefont {Y.~A.}\ \bibnamefont
  {Ganan}}, \bibinfo {author} {\bibfnamefont {A.}~\bibnamefont {Poncet}},
  \bibinfo {author} {\bibfnamefont {V.}~\bibnamefont {Soni}}, \bibinfo {author}
  {\bibfnamefont {S.}~\bibnamefont {Magkiriadou}}, \bibinfo {author}
  {\bibfnamefont {M.~J.}\ \bibnamefont {Shelley}}, \bibinfo {author}
  {\bibfnamefont {D.}~\bibnamefont {Bartolo}}, \ and\ \bibinfo {author}
  {\bibfnamefont {W.~T.}\ \bibnamefont {Irvine}},\ }\bibfield  {title}
  {\enquote {\bibinfo {title} {Motile dislocations knead odd crystals into
  whorls},}\ }\href@noop {} {\bibfield  {journal} {\bibinfo  {journal} {Nature
  Physics}\ }\textbf {\bibinfo {volume} {18}},\ \bibinfo {pages} {212--218}
  (\bibinfo {year} {2022})}\BibitemShut {NoStop}%
\bibitem [{\citenamefont {Mason}\ and\ \citenamefont
  {Weitz}(1995)}]{mason1995optical}%
  \BibitemOpen
  \bibfield  {author} {\bibinfo {author} {\bibfnamefont {T.~G.}\ \bibnamefont
  {Mason}}\ and\ \bibinfo {author} {\bibfnamefont {D.~A.}\ \bibnamefont
  {Weitz}},\ }\bibfield  {title} {\enquote {\bibinfo {title} {Optical
  {{Measurements}} of {{Frequency-Dependent Linear Viscoelastic Moduli}} of
  {{Complex Fluids}}},}\ }\href {\doibase 10.1103/PhysRevLett.74.1250}
  {\bibfield  {journal} {\bibinfo  {journal} {Phys. Rev. Lett.}\ }\textbf
  {\bibinfo {volume} {74}},\ \bibinfo {pages} {1250--1253} (\bibinfo {year}
  {1995})}\BibitemShut {NoStop}%
\bibitem [{\citenamefont {MacKintosh}\ and\ \citenamefont
  {Schmidt}(1999)}]{mackintosh1999microrheology}%
  \BibitemOpen
  \bibfield  {author} {\bibinfo {author} {\bibfnamefont {F.}~\bibnamefont
  {MacKintosh}}\ and\ \bibinfo {author} {\bibfnamefont {C.}~\bibnamefont
  {Schmidt}},\ }\bibfield  {title} {\enquote {\bibinfo {title}
  {Microrheology},}\ }\href@noop {} {\bibfield  {journal} {\bibinfo  {journal}
  {Current opinion in colloid \& interface science}\ }\textbf {\bibinfo
  {volume} {4}},\ \bibinfo {pages} {300--307} (\bibinfo {year}
  {1999})}\BibitemShut {NoStop}%
\bibitem [{\citenamefont {Squires}\ and\ \citenamefont
  {Mason}(2010)}]{squires2010fluid}%
  \BibitemOpen
  \bibfield  {author} {\bibinfo {author} {\bibfnamefont {T.~M.}\ \bibnamefont
  {Squires}}\ and\ \bibinfo {author} {\bibfnamefont {T.~G.}\ \bibnamefont
  {Mason}},\ }\bibfield  {title} {\enquote {\bibinfo {title} {Fluid
  {{Mechanics}} of {{Microrheology}}},}\ }\href {\doibase
  10.1146/annurev-fluid-121108-145608} {\bibfield  {journal} {\bibinfo
  {journal} {Annu. Rev. Fluid Mech.}\ }\textbf {\bibinfo {volume} {42}},\
  \bibinfo {pages} {413--438} (\bibinfo {year} {2010})}\BibitemShut {NoStop}%
\bibitem [{\citenamefont {Lier}\ \emph {et~al.}(2023)\citenamefont {Lier},
  \citenamefont {Duclut}, \citenamefont {Bo}, \citenamefont {Armas},
  \citenamefont {J{\"u}licher},\ and\ \citenamefont
  {Sur{\'o}wka}}]{lier2023lift}%
  \BibitemOpen
  \bibfield  {author} {\bibinfo {author} {\bibfnamefont {R.}~\bibnamefont
  {Lier}}, \bibinfo {author} {\bibfnamefont {C.}~\bibnamefont {Duclut}},
  \bibinfo {author} {\bibfnamefont {S.}~\bibnamefont {Bo}}, \bibinfo {author}
  {\bibfnamefont {J.}~\bibnamefont {Armas}}, \bibinfo {author} {\bibfnamefont
  {F.}~\bibnamefont {J{\"u}licher}}, \ and\ \bibinfo {author} {\bibfnamefont
  {P.}~\bibnamefont {Sur{\'o}wka}},\ }\bibfield  {title} {\enquote {\bibinfo
  {title} {Lift force in odd compressible fluids},}\ }\href {\doibase
  10.1103/PhysRevE.108.L023101} {\bibfield  {journal} {\bibinfo  {journal}
  {Phys. Rev. E}\ }\textbf {\bibinfo {volume} {108}},\ \bibinfo {pages}
  {L023101} (\bibinfo {year} {2023})}\BibitemShut {NoStop}%
\bibitem [{\citenamefont {Hosaka}\ \emph
  {et~al.}(2021{\natexlab{b}})\citenamefont {Hosaka}, \citenamefont {Komura},\
  and\ \citenamefont {Andelman}}]{hosaka2021nonreciprocal}%
  \BibitemOpen
  \bibfield  {author} {\bibinfo {author} {\bibfnamefont {Y.}~\bibnamefont
  {Hosaka}}, \bibinfo {author} {\bibfnamefont {S.}~\bibnamefont {Komura}}, \
  and\ \bibinfo {author} {\bibfnamefont {D.}~\bibnamefont {Andelman}},\
  }\bibfield  {title} {\enquote {\bibinfo {title} {Nonreciprocal response of a
  two-dimensional fluid with odd viscosity},}\ }\href {\doibase
  10.1103/PhysRevE.103.042610} {\bibfield  {journal} {\bibinfo  {journal}
  {Phys. Rev. E}\ }\textbf {\bibinfo {volume} {103}},\ \bibinfo {pages}
  {042610} (\bibinfo {year} {2021}{\natexlab{b}})}\BibitemShut {NoStop}%
\bibitem [{\citenamefont {Gardel}\ \emph {et~al.}(2005)\citenamefont {Gardel},
  \citenamefont {Valentine},\ and\ \citenamefont {Weitz}}]{Gardel2005}%
  \BibitemOpen
  \bibfield  {author} {\bibinfo {author} {\bibfnamefont {M.}~\bibnamefont
  {Gardel}}, \bibinfo {author} {\bibfnamefont {M.}~\bibnamefont {Valentine}}, \
  and\ \bibinfo {author} {\bibfnamefont {D.}~\bibnamefont {Weitz}},\ }\enquote
  {\bibinfo {title} {Microrheology},}\ in\ \href {\doibase
  10.1007/3-540-26449-3_1} {\emph {\bibinfo {booktitle} {Microscale Diagnostic
  Techniques}}},\ \bibinfo {editor} {edited by\ \bibinfo {editor}
  {\bibfnamefont {K.~S.}\ \bibnamefont {Breuer}}}\ (\bibinfo  {publisher}
  {Springer Berlin Heidelberg},\ \bibinfo {address} {Berlin, Heidelberg},\
  \bibinfo {year} {2005})\ pp.\ \bibinfo {pages} {1--49}\BibitemShut {NoStop}%
\bibitem [{\citenamefont {Di~Leonardo}\ \emph {et~al.}(2010)\citenamefont
  {Di~Leonardo}, \citenamefont {Angelani}, \citenamefont {Dell’Arciprete},
  \citenamefont {Ruocco}, \citenamefont {Iebba}, \citenamefont {Schippa},
  \citenamefont {Conte}, \citenamefont {Mecarini}, \citenamefont {De~Angelis},\
  and\ \citenamefont {Di~Fabrizio}}]{di2010bacterial}%
  \BibitemOpen
  \bibfield  {author} {\bibinfo {author} {\bibfnamefont {R.}~\bibnamefont
  {Di~Leonardo}}, \bibinfo {author} {\bibfnamefont {L.}~\bibnamefont
  {Angelani}}, \bibinfo {author} {\bibfnamefont {D.}~\bibnamefont
  {Dell’Arciprete}}, \bibinfo {author} {\bibfnamefont {G.}~\bibnamefont
  {Ruocco}}, \bibinfo {author} {\bibfnamefont {V.}~\bibnamefont {Iebba}},
  \bibinfo {author} {\bibfnamefont {S.}~\bibnamefont {Schippa}}, \bibinfo
  {author} {\bibfnamefont {M.~P.}\ \bibnamefont {Conte}}, \bibinfo {author}
  {\bibfnamefont {F.}~\bibnamefont {Mecarini}}, \bibinfo {author}
  {\bibfnamefont {F.}~\bibnamefont {De~Angelis}}, \ and\ \bibinfo {author}
  {\bibfnamefont {E.}~\bibnamefont {Di~Fabrizio}},\ }\bibfield  {title}
  {\enquote {\bibinfo {title} {Bacterial ratchet motors},}\ }\href@noop {}
  {\bibfield  {journal} {\bibinfo  {journal} {Proceedings of the National
  Academy of Sciences}\ }\textbf {\bibinfo {volume} {107}},\ \bibinfo {pages}
  {9541--9545} (\bibinfo {year} {2010})}\BibitemShut {NoStop}%
\bibitem [{\citenamefont {Avron}(1998)}]{avron1998odd}%
  \BibitemOpen
  \bibfield  {author} {\bibinfo {author} {\bibfnamefont {J.~E.}\ \bibnamefont
  {Avron}},\ }\bibfield  {title} {\enquote {\bibinfo {title} {Odd
  {{Viscosity}}},}\ }\href {\doibase 10.1023/A:1023084404080} {\bibfield
  {journal} {\bibinfo  {journal} {J. Stat. Phys.}\ }\textbf {\bibinfo {volume}
  {92}},\ \bibinfo {pages} {543--557} (\bibinfo {year} {1998})}\BibitemShut
  {NoStop}%
\bibitem [{\citenamefont {Lier}(2024)}]{lier2024slipinduced}%
  \BibitemOpen
  \bibfield  {author} {\bibinfo {author} {\bibfnamefont {R.}~\bibnamefont
  {Lier}},\ }\href@noop {} {\enquote {\bibinfo {title} {Slip-induced odd
  viscous flow past a cylinder},}\ } (\bibinfo {year} {2024}),\ \Eprint
  {http://arxiv.org/abs/2402.02628} {arXiv:2402.02628 [physics.flu-dyn]}
  \BibitemShut {NoStop}%
\bibitem [{\citenamefont {{Callan-Jones}}\ and\ \citenamefont
  {J{\"u}licher}(2011)}]{callan-jones2011hydrodynamics}%
  \BibitemOpen
  \bibfield  {author} {\bibinfo {author} {\bibfnamefont {A.~C.}\ \bibnamefont
  {{Callan-Jones}}}\ and\ \bibinfo {author} {\bibfnamefont {F.}~\bibnamefont
  {J{\"u}licher}},\ }\bibfield  {title} {\enquote {\bibinfo {title}
  {Hydrodynamics of active permeating gels},}\ }\href {\doibase
  10.1088/1367-2630/13/9/093027} {\bibfield  {journal} {\bibinfo  {journal}
  {New J. Phys.}\ }\textbf {\bibinfo {volume} {13}},\ \bibinfo {pages} {093027}
  (\bibinfo {year} {2011})}\BibitemShut {NoStop}%
\bibitem [{\citenamefont {J{\"u}licher}\ \emph {et~al.}(2018)\citenamefont
  {J{\"u}licher}, \citenamefont {Grill},\ and\ \citenamefont
  {Salbreux}}]{julicher2018hydrodynamic}%
  \BibitemOpen
  \bibfield  {author} {\bibinfo {author} {\bibfnamefont {F.}~\bibnamefont
  {J{\"u}licher}}, \bibinfo {author} {\bibfnamefont {S.~W.}\ \bibnamefont
  {Grill}}, \ and\ \bibinfo {author} {\bibfnamefont {G.}~\bibnamefont
  {Salbreux}},\ }\bibfield  {title} {\enquote {\bibinfo {title} {Hydrodynamic
  theory of active matter},}\ }\href {\doibase 10.1088/1361-6633/aab6bb}
  {\bibfield  {journal} {\bibinfo  {journal} {Rep. Prog. Phys.}\ }\textbf
  {\bibinfo {volume} {81}},\ \bibinfo {pages} {076601} (\bibinfo {year}
  {2018})}\BibitemShut {NoStop}%
\bibitem [{\citenamefont {Raikher}\ \emph {et~al.}(2013)\citenamefont
  {Raikher}, \citenamefont {Rusakov},\ and\ \citenamefont
  {Perzynski}}]{raikher2013brownian}%
  \BibitemOpen
  \bibfield  {author} {\bibinfo {author} {\bibfnamefont {Y.~L.}\ \bibnamefont
  {Raikher}}, \bibinfo {author} {\bibfnamefont {V.~V.}\ \bibnamefont
  {Rusakov}}, \ and\ \bibinfo {author} {\bibfnamefont {R.}~\bibnamefont
  {Perzynski}},\ }\bibfield  {title} {\enquote {\bibinfo {title} {Brownian
  motion in a viscoelastic medium modelled by a jeffreys fluid},}\ }\href@noop
  {} {\bibfield  {journal} {\bibinfo  {journal} {Soft Matter}\ }\textbf
  {\bibinfo {volume} {9}},\ \bibinfo {pages} {10857--10865} (\bibinfo {year}
  {2013})}\BibitemShut {NoStop}%
\bibitem [{\citenamefont {Oswald}(2009)}]{Oswald2009}%
  \BibitemOpen
  \bibfield  {author} {\bibinfo {author} {\bibfnamefont {P.}~\bibnamefont
  {Oswald}},\ }\href@noop {} {\emph {\bibinfo {title} {{Rheophysics: The
  Deformation and Flow of Matter}}}}\ (\bibinfo  {publisher} {Cambridge
  university press},\ \bibinfo {year} {2009})\BibitemShut {NoStop}%
\bibitem [{\citenamefont {Levine}\ and\ \citenamefont
  {Lubensky}(2001)}]{levine2001response}%
  \BibitemOpen
  \bibfield  {author} {\bibinfo {author} {\bibfnamefont {A.~J.}\ \bibnamefont
  {Levine}}\ and\ \bibinfo {author} {\bibfnamefont {T.~C.}\ \bibnamefont
  {Lubensky}},\ }\bibfield  {title} {\enquote {\bibinfo {title} {Response
  function of a sphere in a viscoelastic two-fluid medium},}\ }\href {\doibase
  10.1103/PhysRevE.63.041510} {\bibfield  {journal} {\bibinfo  {journal} {Phys.
  Rev. E}\ }\textbf {\bibinfo {volume} {63}},\ \bibinfo {pages} {041510}
  (\bibinfo {year} {2001})}\BibitemShut {NoStop}%
\bibitem [{\citenamefont {Weisenborn}\ and\ \citenamefont
  {Mazur}(1984)}]{weisenborn1984oseen}%
  \BibitemOpen
  \bibfield  {author} {\bibinfo {author} {\bibfnamefont {A.}~\bibnamefont
  {Weisenborn}}\ and\ \bibinfo {author} {\bibfnamefont {P.}~\bibnamefont
  {Mazur}},\ }\bibfield  {title} {\enquote {\bibinfo {title} {The {{Oseen}}
  drag on a circular cylinder revisited},}\ }\href {\doibase
  10.1016/0378-4371(84)90111-0} {\bibfield  {journal} {\bibinfo  {journal}
  {Physica A: Statistical Mechanics and its Applications}\ }\textbf {\bibinfo
  {volume} {123}},\ \bibinfo {pages} {191--208} (\bibinfo {year}
  {1984})}\BibitemShut {NoStop}%
\bibitem [{\citenamefont {Veysey}\ and\ \citenamefont
  {Goldenfeld}(2007)}]{veysey2007simple}%
  \BibitemOpen
  \bibfield  {author} {\bibinfo {author} {\bibfnamefont {J.}~\bibnamefont
  {Veysey}}\ and\ \bibinfo {author} {\bibfnamefont {N.}~\bibnamefont
  {Goldenfeld}},\ }\bibfield  {title} {\enquote {\bibinfo {title} {Simple
  viscous flows: {{From}} boundary layers to the renormalization group},}\
  }\href {\doibase 10.1103/RevModPhys.79.883} {\bibfield  {journal} {\bibinfo
  {journal} {Rev. Mod. Phys.}\ }\textbf {\bibinfo {volume} {79}},\ \bibinfo
  {pages} {883--927} (\bibinfo {year} {2007})}\BibitemShut {NoStop}%
\bibitem [{\citenamefont {Elfring}\ \emph {et~al.}(2016)\citenamefont
  {Elfring}, \citenamefont {Leal},\ and\ \citenamefont
  {Squires}}]{elfring2016surface}%
  \BibitemOpen
  \bibfield  {author} {\bibinfo {author} {\bibfnamefont {G.~J.}\ \bibnamefont
  {Elfring}}, \bibinfo {author} {\bibfnamefont {L.~G.}\ \bibnamefont {Leal}}, \
  and\ \bibinfo {author} {\bibfnamefont {T.~M.}\ \bibnamefont {Squires}},\
  }\bibfield  {title} {\enquote {\bibinfo {title} {Surface viscosity and
  {{Marangoni}} stresses at surfactant laden interfaces},}\ }\href {\doibase
  10.1017/jfm.2016.96} {\bibfield  {journal} {\bibinfo  {journal} {J. Fluid
  Mech.}\ }\textbf {\bibinfo {volume} {792}},\ \bibinfo {pages} {712--739}
  (\bibinfo {year} {2016})}\BibitemShut {NoStop}%
\bibitem [{\citenamefont {Stone}(1990)}]{stone1990simple}%
  \BibitemOpen
  \bibfield  {author} {\bibinfo {author} {\bibfnamefont {H.~A.}\ \bibnamefont
  {Stone}},\ }\bibfield  {title} {\enquote {\bibinfo {title} {A simple
  derivation of the time-dependent convective-diffusion equation for surfactant
  transport along a deforming interface},}\ }\href {\doibase 10.1063/1.857686}
  {\bibfield  {journal} {\bibinfo  {journal} {Physics of Fluids A: Fluid
  Dynamics}\ }\textbf {\bibinfo {volume} {2}},\ \bibinfo {pages} {111--112}
  (\bibinfo {year} {1990})}\BibitemShut {NoStop}%
\bibitem [{\citenamefont {Fukuma}\ and\ \citenamefont
  {Sakatani}(2011)}]{fukuma2011relativistic}%
  \BibitemOpen
  \bibfield  {author} {\bibinfo {author} {\bibfnamefont {M.}~\bibnamefont
  {Fukuma}}\ and\ \bibinfo {author} {\bibfnamefont {Y.}~\bibnamefont
  {Sakatani}},\ }\bibfield  {title} {\enquote {\bibinfo {title} {Relativistic
  viscoelastic fluid mechanics},}\ }\href {\doibase 10.1103/PhysRevE.84.026316}
  {\bibfield  {journal} {\bibinfo  {journal} {Phys. Rev. E}\ }\textbf {\bibinfo
  {volume} {84}},\ \bibinfo {pages} {026316} (\bibinfo {year}
  {2011})}\BibitemShut {NoStop}%
\bibitem [{\citenamefont {Azeyanagi}\ \emph {et~al.}(2009)\citenamefont
  {Azeyanagi}, \citenamefont {Fukuma}, \citenamefont {Kawai},\ and\
  \citenamefont {Yoshida}}]{azeyanagi2009universal}%
  \BibitemOpen
  \bibfield  {author} {\bibinfo {author} {\bibfnamefont {T.}~\bibnamefont
  {Azeyanagi}}, \bibinfo {author} {\bibfnamefont {M.}~\bibnamefont {Fukuma}},
  \bibinfo {author} {\bibfnamefont {H.}~\bibnamefont {Kawai}}, \ and\ \bibinfo
  {author} {\bibfnamefont {K.}~\bibnamefont {Yoshida}},\ }\bibfield  {title}
  {\enquote {\bibinfo {title} {Universal description of viscoelasticity with
  foliation preserving diffeomorphisms},}\ }\href {\doibase
  10.1016/j.physletb.2009.10.027} {\bibfield  {journal} {\bibinfo  {journal}
  {Physics Letters B}\ }\textbf {\bibinfo {volume} {681}},\ \bibinfo {pages}
  {290--295} (\bibinfo {year} {2009})}\BibitemShut {NoStop}%
\bibitem [{\citenamefont {Salbreux}\ and\ \citenamefont
  {J{\"u}licher}(2017)}]{salbreux2017mechanics}%
  \BibitemOpen
  \bibfield  {author} {\bibinfo {author} {\bibfnamefont {G.}~\bibnamefont
  {Salbreux}}\ and\ \bibinfo {author} {\bibfnamefont {F.}~\bibnamefont
  {J{\"u}licher}},\ }\bibfield  {title} {\enquote {\bibinfo {title} {Mechanics
  of active surfaces},}\ }\href {\doibase 10.1103/PhysRevE.96.032404}
  {\bibfield  {journal} {\bibinfo  {journal} {Physical Review E}\ }\textbf
  {\bibinfo {volume} {96}},\ \bibinfo {pages} {032404} (\bibinfo {year}
  {2017})}\BibitemShut {NoStop}%
\bibitem [{\citenamefont {Avron}\ \emph {et~al.}(1995)\citenamefont {Avron},
  \citenamefont {Seiler},\ and\ \citenamefont {Zograf}}]{avron1995viscosity}%
  \BibitemOpen
  \bibfield  {author} {\bibinfo {author} {\bibfnamefont {J.~E.}\ \bibnamefont
  {Avron}}, \bibinfo {author} {\bibfnamefont {R.}~\bibnamefont {Seiler}}, \
  and\ \bibinfo {author} {\bibfnamefont {P.~G.}\ \bibnamefont {Zograf}},\
  }\bibfield  {title} {\enquote {\bibinfo {title} {Viscosity of {{Quantum Hall
  Fluids}}},}\ }\href {\doibase 10.1103/PhysRevLett.75.697} {\bibfield
  {journal} {\bibinfo  {journal} {Phys. Rev. Lett.}\ }\textbf {\bibinfo
  {volume} {75}},\ \bibinfo {pages} {697--700} (\bibinfo {year}
  {1995})}\BibitemShut {NoStop}%
\bibitem [{\citenamefont {L{\'e}vay}(1995)}]{levay1995berry}%
  \BibitemOpen
  \bibfield  {author} {\bibinfo {author} {\bibfnamefont {P.}~\bibnamefont
  {L{\'e}vay}},\ }\bibfield  {title} {\enquote {\bibinfo {title} {Berry phases
  for {{Landau Hamiltonians}} on deformed tori},}\ }\href {\doibase
  10.1063/1.531066} {\bibfield  {journal} {\bibinfo  {journal} {J. Math.
  Phys.}\ }\textbf {\bibinfo {volume} {36}},\ \bibinfo {pages} {2792--2802}
  (\bibinfo {year} {1995})}\BibitemShut {NoStop}%
\bibitem [{\citenamefont {Jensen}\ \emph {et~al.}(2012)\citenamefont {Jensen},
  \citenamefont {Kaminski}, \citenamefont {Kovtun}, \citenamefont {Meyer},
  \citenamefont {Ritz},\ and\ \citenamefont
  {Yarom}}]{jensen2012parityviolating}%
  \BibitemOpen
  \bibfield  {author} {\bibinfo {author} {\bibfnamefont {K.}~\bibnamefont
  {Jensen}}, \bibinfo {author} {\bibfnamefont {M.}~\bibnamefont {Kaminski}},
  \bibinfo {author} {\bibfnamefont {P.}~\bibnamefont {Kovtun}}, \bibinfo
  {author} {\bibfnamefont {R.}~\bibnamefont {Meyer}}, \bibinfo {author}
  {\bibfnamefont {A.}~\bibnamefont {Ritz}}, \ and\ \bibinfo {author}
  {\bibfnamefont {A.}~\bibnamefont {Yarom}},\ }\bibfield  {title} {\enquote
  {\bibinfo {title} {Parity-violating hydrodynamics in 2 + 1 dimensions},}\
  }\href {\doibase 10.1007/JHEP05(2012)102} {\bibfield  {journal} {\bibinfo
  {journal} {J. High Energ. Phys.}\ }\textbf {\bibinfo {volume} {2012}},\
  \bibinfo {pages} {102} (\bibinfo {year} {2012})}\BibitemShut {NoStop}%
\bibitem [{\citenamefont
  {Son}(2019)}]{https://doi.org/10.48550/arxiv.1907.07187}%
  \BibitemOpen
  \bibfield  {author} {\bibinfo {author} {\bibfnamefont {D.~T.}\ \bibnamefont
  {Son}},\ }\href {\doibase 10.48550/ARXIV.1907.07187} {\enquote {\bibinfo
  {title} {Chiral metric hydrodynamics, kelvin circulation theorem, and the
  fractional quantum hall effect},}\ } (\bibinfo {year} {2019})\BibitemShut
  {NoStop}%
\bibitem [{\citenamefont {Camley}\ and\ \citenamefont
  {Brown}(2011)}]{camley2011creeping}%
  \BibitemOpen
  \bibfield  {author} {\bibinfo {author} {\bibfnamefont {B.~A.}\ \bibnamefont
  {Camley}}\ and\ \bibinfo {author} {\bibfnamefont {F.~L.~H.}\ \bibnamefont
  {Brown}},\ }\bibfield  {title} {\enquote {\bibinfo {title} {Beyond the
  creeping viscous flow limit for lipid bilayer membranes: {{Theory}} of
  single-particle microrheology, domain flicker spectroscopy, and long-time
  tails},}\ }\href {\doibase 10.1103/PhysRevE.84.021904} {\bibfield  {journal}
  {\bibinfo  {journal} {Phys. Rev. E}\ }\textbf {\bibinfo {volume} {84}},\
  \bibinfo {pages} {021904} (\bibinfo {year} {2011})}\BibitemShut {NoStop}%
\bibitem [{\citenamefont {MacKintosh}\ and\ \citenamefont
  {Lubensky}(1991)}]{mackintosh1991orientational}%
  \BibitemOpen
  \bibfield  {author} {\bibinfo {author} {\bibfnamefont {F.~C.}\ \bibnamefont
  {MacKintosh}}\ and\ \bibinfo {author} {\bibfnamefont {T.~C.}\ \bibnamefont
  {Lubensky}},\ }\bibfield  {title} {\enquote {\bibinfo {title} {Orientational
  order, topology, and vesicle shapes},}\ }\href {\doibase
  10.1103/PhysRevLett.67.1169} {\bibfield  {journal} {\bibinfo  {journal}
  {Phys. Rev. Lett.}\ }\textbf {\bibinfo {volume} {67}},\ \bibinfo {pages}
  {1169} (\bibinfo {year} {1991})}\BibitemShut {NoStop}%
\end{thebibliography}
\end{document}